\documentclass[preprint,trackchanges]{aastex62}

\usepackage{amsmath}
\usepackage{natbib}
\usepackage{multirow} 
\hypersetup{linkcolor=magenta, citecolor=cyan, filecolor=yellow, urlcolor=blue}

\received{2019 May 7}
\revised{2019 September 3}
\accepted{2019 September 5}
\published{2019 November 1}

\shorttitle{Thermal Components in Gamma-ray Bursts}
\shortauthors{Li.}

\begin{document}

\title{Thermal Components in Gamma-ray Bursts. I. 
\\ How Do They Affect Non-Thermal Spectral Parameters?}

\author[0000-0002-1343-3089]{Liang Li}
\affiliation{ICRANet, Piazza della Repubblica 10, I-65122 Pescara, Italy}

\correspondingauthor{Liang Li}
\email{liang.li@icranet.org}

\begin{abstract}

The spectral components of the prompt emission of gamma-ray bursts (GRBs) mainly consist of two possible origins: synchrotron (nonthermal) and photosphere (thermal). The typical spectral properties of GRBs can be modeled by a dominant nonthermal component (a Band-like function or cutoff power law), while some of them have an additional thermal component (a Planck-like function). In this paper, we investigate the effects of thermal components on the nonthermal spectral parameters. We focus on eight {\it Fermi} Gamma-ray Burst Monitor bursts of which the spectra deviate from a Band-only function, and the thermal components are significant. We sort them into thermal-subdominant Group I (e.g. GRB 110721A) and thermal-dominant Group II (e.g., GRB 090902B). 
Several interesting results are found assuming the spectral component is totally attributed to the nonthermal component: (i) the low-energy photon index $\alpha$ becomes harder; (ii) the peak energy $E_{\rm c}$ is significantly smaller and lies between the peak temperature of blackbody component and the peak energy of the cutoff power law + blackbody (CPL+BB) model; (iii) total flux $F$, is generally the same; (iv) the changes ($\Delta \alpha$ and $\Delta E_{\rm c}$) are positively correlated with the ratio between the thermal flux and total flux; and (v) parameter relations ($F-\alpha$, $F-E_{\rm c}$ and $E_{\rm c}$-$\alpha$) also changed prominently. The GRBs in both groups show the same results. Our analysis indicates that the thermal component is important, and it significantly affects the spectral parameters and the consequential physical interpretations.

\end{abstract}

\keywords{gamma-ray burst: general – methods: data analysis – radiation mechanisms: thermal}

\section{Introduction} \label{sec:intro}

The synchrotron emission producing nonthermal photons \citep{1994ApJ...432..181M, 1994ApJ...430L..93R, 1998MNRAS.296..275D, 2000ApJ...543..722L, 2009A&A...498..677B, 2011A&A...526A.110D, 2011ApJ...726...90Z, 2013ApJ...769...69B,  2014MNRAS.445.3892B, 2015ApJ...805..163D, 2018MNRAS.476.1785B, 2018ApJS..234....3G, 2019ApJ...882..184L} and the photosphere of the fireball \citep{1994MNRAS.270..480T, 2005ApJ...628..847R,2006ApJ...652..482P, 2008A&A...480..305G, 2008ApJ...682..463P, 2009ApJ...703.1044B, 2010MNRAS.407.1033B, 2011ApJ...737...68B, 2009ApJ...700L.141L, 2010ApJ...725.1137L, 2010PThPh.124..667I, 2011ApJ...732...49P, 2011MNRAS.415.3693R,  2011MNRAS.415.1663T,  2013MNRAS.436L..54A, 2013ApJ...767..139B, 2013MNRAS.428.2430L, 2013ApJ...772...11R, 2014NewA...27...30R, 2014ApJ...785..112D, 2014IJMPD..2330003V, 2017MNRAS.468.3202B, 2017IJMPD..2630018P, 2018ApJ...860...72M, 2019ApJ...882...26M} generating the Comptonized quasi-thermal photons are two leading candidates of the radiation mechanism to explain the gamma-ray burst (GRB) prompt emission. The observed GRB spectrum in the keV-MeV energy range is usually described by a nonthermal phenomenological Band function \citep{1993ApJ...413..281B} in both the time-integrated (also called `time-averaged') and the time-resolved spectra. The Band function is considered to be a phenomenological representation of the nonthermal synchrotron model. It involves the low-energy and the high-energy segments, in which the photon indices are $\alpha$ (the low-energy index, with a typical value of about -1.0), and $\beta$ (the high-energy index, with a typical value of about -2.2), respectively. These two segments are smoothly connected at a peak energy, $E_{\rm p}$ (with a typical value $\sim$ 250 keV), in the case of the $\nu F_{\nu}$ spectrum. Although GRBs generally exhibit a nonthermal Band-like spectrum, the thermal component also plays a very important role, at least in some GRBs. Previous studies suggested that the GRB spectra may contain three basic components \citep[e.g.,][]{2011ApJ...730..141Z}: (i) the nonthermal component generally fitted with the Band function (broad spectral peak); (ii) the thermal component generally fitted with Planck's blackbody (BB) function (a narrow spectral peak); and (iii) an additional power-law spectral component extending to the high energy range that is present in some \textit{Fermi}-Large Area Telescope GRBs. The observed spectrum of a GRB can be a combination of two of these components or a combination of all the three components simultaneously in one GRB \citep{2015ApJ...807..148G}.

Observationally, some GRBs have confirmed an additional subdominant BB component embedded into the left shoulder of the Band shape \citep[e.g, GRB 110721A;][]{2012ApJ...757L..31A}.
Thermal-dominant GRBs are rare. The time-resolved spectral analysis based on the Burst And Transient Source Experiment (BATSE) on board the Compton Gamma Ray Observatory (CGRO) revealed that a few bright bursts exhibit thermal domination during the entire burst duration, e.g., GRB 930214, GRB 941023, GRB 951228 \citep{2004ApJ...614..827R}, and GRB 990413 \citep{2006MNRAS.370L..33B}, or only at the beginning of the burst, see e.g., GRB 910807, GRB 910927, GRB 911119, GRB 970111, and GRB 980326 \citep{2003A&A...406..879G}.  
These results were confirmed later by the observations of Gamma-ray Burst Monitor (GBM) on board the {\it Fermi Gamma-ray Space Telescope} (see the detailed discussion in \S 2).

Due to the rarity and the dimness of the thermal component, the common spectral analysis often adopts only the nonthermal model, which inevitably brings bias to those GRBs containing evident thermal emission. One interesting question therefore arises: {\it how does the thermal component affect the nonthermal spectral parameters?} In this work, we are going to address this issue by analyzing eight bursts with significant thermal components. We fit these GRBs with and without the inclusion of a thermal component and then compare their $\alpha$, $E_{\rm p}$ evolution, and the relationship of $\alpha-E_{\rm p}$, $F-E_{\rm p}$, and $F-\alpha$, respectively.

This paper is organized as follows. 
In \S 2, we perform the sample selection, data reduction, and sample definition.
In \S 3, we present the results, including the global parameter evolution and parameter relations.
The conclusions are presented in \S 4.  
Throughout the paper, the standard $\Lambda$-CDM cosmology with the parameters of $H_{0}= 67.4$ ${\rm km s^{-1}}$ ${\rm Mpc^{-1}}$, $\Omega_{M}=0.315$, and $\Omega_{\Lambda}=0.685$ is adopted \citep{2018arXiv180706209P}. The convention $Q=10^{x}Q_{x}$ is adopted in cgs units.

\section{Methodology} \label{sec:Data}

\subsection{Sample Selection} 

Compared to the BATSE (25-1800 keV), the {\it Fermi}-GBM provides a broader energy range (8 keV-40 MeV), which can fully assess all the current GRB spectral models \cite[e.g.,][]{2007MNRAS.379...73G}. We, therefore, carry out this task by working with {\it Fermi} GBM data. The {\it Fermi}-GBM \citep{2009ApJ...702..791M} contains 12 sodium iodide (NaI, 8 keV-1 MeV) detectors (n0 to n9, na and nb) and two bismuth germanate (BGO, 200 keV-40 MeV) detectors (b0 and b1). The time tagged event (TTE) and the standard response files provided by the GBM team are used. We collect the data from all the triggered NaI detectors and one BGO detector, of which the selection is dependent on the sequence of NaI detectors, b0 if n0 to n5, and b1 otherwise.

The {\it Fermi}-GBM GRBs with thermal components reported in previous studies are systematically searched; 13 cases are obtained in our sample (see Table 1). 
These bursts are reported to have an additional thermal component that significantly improves the spectral fitting.
Here, we first review the spectral characteristics of these bursts reported in the literature. \cite{2011ApJ...727L..33G} reported that the spectrum of GRB 100724B is dominated by the typical Band function, while it also includes a statistically highly significant thermal contribution. Likewise, this two-component scenario is also reported by a few other bursts.
\cite{2014ApJ...784...17B} reported five bursts, GRB 081224 \citep{2008GCN..8723....1W, 2016MNRAS.456.2157I}, GRB 090719A \citep{2009GCN..9691....1V, 2016MNRAS.456.2157I}, GRB 100707A \citep{2010GCN.10944....1W, 2016MNRAS.456.2157I}, GRB 110721A \citep{2012ApJ...757L..31A, 2013MNRAS.433.2739I}, and GRB 110920A \citep{2016MNRAS.456.2157I}, in which an additional BB component was required for the synchrotron components (Band+BB model) to produce acceptable spectral fits.
\cite{2013MNRAS.432.3237G} analyzed the time-resolved spectrum of GRB 100507 and reported that the burst has a BB spectrum for the entire duration ($\sim$ 30 s) of the prompt emission.

A BB spectrum in the prompt gamma-ray emission was found in a long burst, GRB 101219B, which was analyzed by \cite{2015ApJ...800L..34L}, making it the second such burst observed by {\it Fermi}-GBM.
\cite{2013ApJ...770...32G, 2017ApJ...846..138G} argued that both the time-integrated and time-resolved spectral analyses in GRB 120323A can be described by a two-component scenario similarly, in which thermal emission is observed simultaneously with a nonthermal component. 
Recently, a very bright GRB, 190114C, was reported in \cite{2019arXiv190107505W}, adding a BB was found to improves the fitting over the cutoff power law (CPL) only. Besides, \cite{2019ApJS..242...16L} recently analyzed the time-resolved spectra of four {\it Fermi} bright bursts (GRB 140206B, GRB 140329B, GRB 150330A, and GRB 160625B, see also \citealt{2018NatAs...2...69Z}) and found that the thermal component is detected only at the beginning of the burst, and followed up by a nonthermal synchrotron component. Note that a few bursts may show a thermal dominate form, which is represented by narrower spectra, with the most prominent case being GRB 090902B \citep{2010ApJ...709L.172R}. The emission of GRB 160107A \citep{2018PASJ...70....6K} is reported to come from the jet photosphere combined with nonthermal emission described by a single power law. GRB 081221 can be described with multi-color BB spectra \citep{2013ApJ...768..187B, 2018ApJ...866...13H}. The best spectral fitting in GRB 110920A is a power-law plus two BBs fitting \citep{2015MNRAS.450.1651I}, see also \cite{2012grb..confE..12M, 2013ApJ...778....3S}.

\subsection{Data Reduction} 

The time-resolved spectral analyses in previous works mainly make use of the frequentist approach. In this paper, we performed a Bayesian analysis using the package called the Multi-Mission Maximum Likelihood Framework (3ML; \citealt{2015arXiv150708343V}) as the main tool to carry out the time-resolved spectral analyses for {\it Fermi}-GBM data \citep[see also,][]{2018arXiv181007313Y, 2019ApJS..242...16L, 2019arXiv190809240L}. 

The background is selected by fitting the pre- and post-source intervals with an order of 0-4 polynomial for one of the bright NaIs in photon counts. The optimal order of the polynomial is determined by applying a likelihood ratio test.
This optimal polynomial is then used to fit each of the 128 energy channels (TTE) to estimate the background model. The background photon counts for each channel are obtained by integrating the optimal polynomial over source interval. 
The maximum likelihood-based statistics are used, i.e., the so-called Pgstat, given by a Poisson (observation; \citealt{1979ApJ...228..939C})-Gaussian (background) profile likelihood.
For the Poisson-Gaussian profile likelihood to be valid, we bin the spectra at least one background count per spectral bin. 
The source is selected as the time interval of $T_{90}$ for each burst. To perform the time-resolved spectral analysis, we first rebin the TTE data by using the Bayesian Blocks method (BBlocks; \citealt{2013ApJ...764..167S}) with a false alarm probability of $p_{0}$=0.01 to the TTE light curve of the brightest NaI detector. All of the other used detectors are then binned in matching time bins. This can be done by selecting the detector with the highest significance during the source interval for the case where more than one NaI detector is triggered. We then use the brightest one for the BBlocks and background fitting. To better study the time-resolved spectral evolution, we select the number of spectra with statistical significance S$\geq$20 to be at least greater than 4 for each burst\footnote{The detail derivation of significance $S$, as well as the relationship with signal-to-noise ratio S/N, can be found in \cite{2018ApJS..236...17V} for a Poisson (sources)-Gaussian (backgrounds) profile and in \cite{1983ApJ...272..317L} for a Poisson (sources)-Poisson (backgrounds) profile.}. The sample then reduced to eight bursts with this criterion. The spectra with S$<$20 usually have a huge error from the fitting.

In Table 1, we listed 13 bursts satisfying all aforementioned criteria (Column 1), along with their $T_{90}$ (Column 2),  the fluence (column 3) between 10 and 1000 keV, the detectors used (Column 4), the source (Column 5) and background (Column 6) intervals, the number of the time bins of $S>20$/total using the BBlocks analysis across the source (Column 7), and the best model suggested in the literature (Column 8) and its reference (Column 9).

Since thermal components are generally observed in the left shoulder of the Band spectrum in the low-energy region (below $E_{\rm p}$), its presence does not affect the high-energy $\beta$ index (above $E_{\rm p}$). Therefore, we uniformly employ the CPL model to replace the Band function to perform all the spectral fitting throughout the paper. 
The spectral analysis is performed by fitting Bayesian statistical models using the Markov Chain Monte Carlo (MCMC) technique. The parameters of our model in the Monte Carlo fit allow in the following range index (PL model): [-5, 1]; $kT$ (keV, BB model): [1, 10$^{3}$]; $\alpha$ (CPL model): [-5, 1]; and $E_{\rm c}$ (keV, CPL model): [1, 10$^{4}$]. The informative priors are adopted by the typical spectral parameters from the $Fermi$-GBM catalog: $\alpha \sim \mathcal{N} (\mu=-1.,\sigma=0.5)$; $E_{\rm c} \sim \mathcal{N} (\mu=200,\sigma=300)$; and $\beta \sim \mathcal{N} (\mu=-2.2,\sigma=0.5)$. The priors are used and multiplied to the likelihood, which combines the model and the observed data, yielding a posterior distribution of the parameters. Uncertainties are given at the 68\% (1$\sigma$) Bayesian credible level calculated from the last 80\% of the MCMC chain for 10000 iterations. To select the best model from two different given models, the deviance information criterion (DIC) is adopted in this paper, which is defined as DIC=-2log[$p$(data$\mid\hat{\theta}$)]+2$p_{\rm DIC}$, where $\hat{\theta}$ is the posterior mean of the parameters, and $p_{\rm DIC}$ is the effective number of parameters. The preferred model is the model with the lowest DIC score. Here, we define $\Delta$DIC=DIC(CPL+BB)-DIC(CPL). If $\Delta$DIC is negative, it indicates better CPL+BB. 

\subsection{Sample Definition} 

Following the above methodology, we select the time bins with $S\geq$20 as\footnote{Our selected criteria only consider the number of spectra with $S>$20 at least greater than 4 for a further analysis.} having enough photons to perform the precise spectral fitting. We first fit the data with a canonical CPL model. Then, an additional BB component is added to check whether it improves the fitting statistics (DIC). 
The $\Delta$DIC score is greater than 10 for all time bins in our samples after the BB component is introduced (CPL+BB fit over CPL-alone fitting). Some of them are even as high as hundreds,  indicating a firm detection of a BB component. Our results confirm the thermal component of these 13 GRBs as reported in the previous articles.

After the Bayesian analysis, we finally sort these {\it Fermi} bursts into two groups depending on whether the thermal emissions are subdominant (adhering to the left shoulder of the Band shape) or dominate (narrower) components that appear in the observed spectra. 
\begin{itemize}
\item  Group I: bursts for which the thermal components are typically subdominant. The group is made of six GRBs: GRB 081224, GRB 110721A, GRB 090719, GRB 100724B, and GRB 190114C. The temporal evolution of the parameters ($\alpha$, $E_{\rm c}$ and $F$) is shown in Figure \ref{Alpha}, \ref{Ec} and \ref{Flux}, and the parameter relations ($F-\alpha$, $F-E_{\rm c}$ and $E_{\rm c}-\alpha$) are displayed in Figure \ref{FluxAlpha}, \ref{FluxEp}, and \ref{EpAlpha}.
\item Group II: bursts for which the thermal components are dominant. Two GRBs are in this group: GRB 090902B and GRB 160107. The temporal evolution of the parameters ($\alpha$ or the power-law index, and $F$) are shown in Figure \ref{AlphaII}, and the parameter relations (the $F-\alpha$ or $F-$ index) are displayed in Figure \ref{FluxAlphaII}.
\end{itemize}

To account for the difference of two models and two groups, we define:
\begin{equation}
\rm \Delta DIC^{I}= DIC (CPL+BB)-DIC (CPL),
\end{equation}
\begin{equation}
\rm \Delta DIC^{II}= DIC (PL+BB)-DIC (CPL),
\end{equation}
\begin{equation}
\rm \Delta \alpha^{I} = \alpha (\rm CPL+BB)-\alpha (CPL),
\end{equation}
\begin{equation}
\rm \Delta \alpha^{II} = \alpha (\rm PL+BB)-\alpha (CPL),
\end{equation}
\begin{equation}
\Delta E_{\rm c} = E_{\rm c} (\rm CPL+BB)-{\it E_{\rm c}} (CPL),
\end{equation}
\begin{equation}
\rm Ratio^{I}={\it F_{\rm BB}}(CPL+BB)/{\it F_{\rm obs}}(CPL+BB),
\end{equation}
\begin{equation}
\rm Ratio^{II}={\it F_{\rm BB}}(PL+BB)/{\it F_{\rm obs}}(PL+BB).
\end{equation}
Their values for each burst are presented from Tables \ref{Table:081224} to Table \ref{Table:160107}.

\section{Results} \label{sec:Results}

The time-resolved spectral fitting results for each selected burst of Group I are listed from Tables \ref{Table:081224} to Table \ref{Table:190114C}, and the results of Group II are listed from Tables \ref{Table:090902B} and \ref{Table:160107}. For each table, we list the time interval (Column 1), $S$ (Column 2), parameters of the CPL-alone fitting ($\alpha$ and $E_{\rm c}$: Column 3 - Column 4), parameters of the CPL plus BB fitting ($\alpha$, $E_{\rm c}$ and $kT$: Column 5 - Column 7) or parameters of a power-law plus BB fitting (power-law index and $kT$: Column 5 - Column 6), and the difference of parameters ($\Delta \alpha^{\rm I}$ or $\Delta \alpha^{\rm II}$, $\Delta E_{\rm c}$, Ratio$^{\rm I}$ or Ratio$^{\rm II}$, and $\Delta$DIC$^{\rm I}$ or $\Delta$DIC$^{\rm II}$) between these two models (Group I: Column 8 - Column 11 and Group II: Column 7- Column 9). 

\subsection{Parameter Evolutions}

The evolutionary analysis is performed on the bins that satisfy the aforementioned criteria ($\Delta$DIC$<$ 0 and $S>$20). We present temporal evolution of the spectral parameters ($\alpha$ and $E_{\rm c}$) and the energy flux (erg cm$^{-2}$s$^{-1}$) from Figures \ref{Alpha}-\ref{Flux} for the Group I bursts, along with their GBM light curves.

The comparison of parameters show that $\alpha$ is systemically harder\footnote{Note that a harder $\alpha$ is defined as $\Delta$$\alpha$$>$0 ($\alpha$ index increases) and a smaller $E_{\rm c}$ is defined as $\Delta$$E_{\rm c}$$<$0 ($E_{\rm c}$ value decreases).} (Figure \ref{Alpha}), and $E_{\rm c}$ is significantly smaller (Figure \ref{Ec}) in the CPL model than that in the CPL+BB model. There is no obvious difference for the temporal evolution of the energy flux fitted by different models for all the cases (Figure \ref{Flux}). Besides, GRB 081224, GRB 090719, GRB 100707, and GRB 190114C present very hard spectra, in which the majority of the $\alpha$ indices are beyond the synchrotron limit \citep{1998ApJ...506L..23P}. This is in agreement with the previous results where the hard indices are consistent with the thermal origin. Furthermore, $\alpha$ generally can be softer after adding a thermal component, even crossing the synchrotron limit in GRB 190114C. For this type of the spectral evolution, we find most of them exhibit the hard-to-soft pattern, both in $E_{\rm c}$- and $\alpha$-evolution. 

To compare the results between different bursts and to investigate the difference quantitatively, we further calculate the difference of the parameters ($\Delta \alpha$ and $\Delta E_{\rm c}$), as shown in Figure \ref{Dis}. 
To have a global view on the difference, we plot the distributions of parameters (top panels of Figure \ref{Dis}). For $\Delta \alpha$, one has $\Delta \alpha$=-0.14$\pm$0.28; For $\Delta E_{\rm c}$, we get log$\Delta E_{\rm c}$=2.42$\pm$0.68. These results indicate that the spectral parameters are prominently changed. The distribution of the energy flux shows two peaks, with the first dominant one around 3$\times$10$^{-6}$, and the second smaller one peaks at $\sim$10$^{-4}$. Both the CPL and CPL+BB models exhibit similar behaviors. The majority of $\Delta \alpha$ is below 0 and $\Delta E_{\rm c}$ is above 0, which are consistent with the above results (see bottom panels of Figure \ref{Dis}). To determine the prominence of the thermal component, we also calculate the ratio ($F_{\rm BB}$/$F_{\rm tot}$) between the thermal flux and the total flux (see bottom panels of Figure \ref{Dis}). We find the ratio varies from burst to burst, and within a burst, with the range from $\sim$ 0.2 \% to $\sim$ 70\%. The results indicate that some bursts (e.g. GRB 100724B) present no obvious evolution of the ratio, while other bursts present dramatically evolution within the burst (e.g. GRB 100707).

To denote the interesting results shown above, we use a cartoon picture (Figure \ref{Cartoon}) to vividly demonstrate why $\alpha$ hardens and $E_{\rm c}$ gets smaller if ignoring the contribution of the thermal component---namely, the difference of $\alpha$ and $E_{\rm c}$ between the CPL+BB to CPL-alone fittings. We take the typical observed parameters \cite[e.g.,][]{2000ApJS..126...19P, 2009ApJ...702.1211R} and the same normalization: $E_{\rm p}$=1000 keV, $\alpha$=-1, $kT$=150 keV.
As seen in the picture, the BB component is strong. It raises the $\alpha$ shape and shifts the peak energy ($E_{\rm c}$) to the left. The stronger the BB emission, the harder the $\alpha$ and the smaller the $E_{\rm c}$ become. Typically, the $E_{\rm c}$ values of the CPL model fall in between the peak of BB emission ($kT \times$ 2.8) and $E_{\rm c}$ of the CPL+BB model. If the BB emission is extremely strong, $E_{\rm c}$ can be even smaller than the peak of BB emission, but must be greater than the temperature of the BB.

In order to test our cartoon model (Figure \ref{Cartoon}), one time bin with the biggest $\Delta$DIC (or the highest thermal ratio) is used to present the observed spectral variation between the models. The results are shown in Figure \ref{Spectrum}. The BB temperature $kT$; the $E_{\rm c}$ of CPL; and CPL+BB for GRB 110721A are 31.24 keV, 398.09 keV, and 943.05 keV, respectively. The peak of BB emission, $E$(BB), is $\sim$ 87.47 keV (2.8 $\times$ $kT$). Therefore, we find $E$(BB)$<E_{\rm c}$(CPL)$<E_{\rm c}$(CPL+BB). The $\alpha$ index derived from the CPL model in this time bin is -0.90, which is harder than that from the CPL+BB model (-1.08). 

Likewise, the same results are also found in other cases. 
For GRB 081224, $E$(BB)=53.78 keV$<$$E_{\rm c}$(CPL)=222.28 keV$<$$E_{\rm c}$(CPL+BB)=287.38 keV; $\alpha$(CPL)=-0.57 harder than $\alpha$(CPL+BB)=-0.66.
For GRB 090719, $E$(BB)=86.36 keV$<$$E_{\rm c}$(CPL)=173.84 keV$<$$E_{\rm c}$(CPL+BB)=205.83 keV; and $\alpha$(CPL)=-0.41 harder than $\alpha$(CPL+BB)=-0.48. 
For GRB 100707, $E$(BB)=233.21 keV$>$$E_{\rm c}$(CPL)=\newline213.50 keV$<$$E_{\rm c}$(CPL+BB)=700.53 keV; $\alpha$(CPL)=0.39 is harder than $\alpha$(CPL+BB)=-0.16. Note that here, $E_{\rm c}$ is smaller than the peak of the BB temperature. This is because the thermal component in this time bin is very strong, with the ratio of $\sim$ 30\%, which is much higher than the typical value of the observation. 
For GRB 100724B, $E$(BB)=73.84 keV$<$$E_{\rm c}$(CPL)=444.63 keV$<$$E_{\rm c}$(CPL+BB)=554.95 keV; $\alpha$(CPL)=-0.81 is harder than $\alpha$(CPL+BB)=-0.84. 
For GRB 190114C, $E$(BB)=312.59 keV$<$$E_{\rm c}$(CPL)=520.86 keV$<$$E_{\rm c}$(CPL+BB)=717.56 keV; $\alpha$(CPL)=-0.49 is  harder than $\alpha$(CPL+BB)=-0.71. 

\subsection{Global Parameter Relations}

Based on our findings that the addition of the BB component changes the parameters, like $\alpha$ and $E_{\rm c}$, while the flux remains constant, we want to check how those parameters affect the relations of $F-\alpha$, $F-E_{\rm c}$, and $E_{\rm c}-\alpha$ usually found in GRBs. We assume that the parameters---namely $\alpha$, $E_{\rm c}$, and flux---are related with each other by power-laws\footnote{Here, note that $F-\alpha$ and $E_{\rm c}-\alpha$ relations are denoted in semi-log space and $F-E_{\rm c}$ in log-log space.} \citep[e.g.,][]{2019ApJ...884..109L}. 
For the $F-\alpha$ relation, since $F$ remains about constant and $\alpha$ becomes harder, the relation will shifts toward the right (see the arrows marked in Figure \ref{FluxAlpha}) but with almost the same slope. For the $F-E_{\rm c}$ relation, since $F$ remains about constant and $E_{\rm c}$ tends to be smaller, the relation will moves toward the left (see the arrows marked in Figure \ref{FluxEp}) still with a same slope. For the $E_{\rm c}-\alpha$ relation, since both $E_{\rm c}$ and $\alpha$ will change, so the case could be more complicated. Generally speaking, the relation will move toward the top left direction (see the arrows marked in Figure \ref{EpAlpha}), and the slope could also be changed with all possibilities (greater, less, and the same). On the other hand, stronger thermal components will results in greater $\alpha$ and $E_{\rm c}$ changes, leading to more significant changes of the global parameter relations.

Figures \ref{FluxAlpha} to \ref{EpAlpha} show three plots of parameter pairs (also see the middle panels of Figure \ref{Dis} for the global sample): $F-\alpha$, $F-E_{\rm c}$ and $E_{\rm c}-\alpha$. It is observed that in the high-significance cases, $S>$20. The majority of cases in both $F-\alpha$ and $F-E_{\rm c}$ exhibit monotonous positive relations, but the relations are shifted. The $E_{\rm c}-\alpha$ relation displays a more complicate behavior (see Figure \ref{EpAlpha}), because both $E_{\rm c}$ and $\alpha$ are altered. The observations, therefore, are self-similar to the picture as described by the cartoon (Figure \ref{Cartoon}).

Two `ideal' cases can be used to the better demonstration of the evolutionary properties of the parameters: GRB 190114C (the burst has a strong thermal emission) and GRB 100707 (the burst has a dramatic thermal evolution). Figure \ref{Dis} shows the temporal evolution of the flux ratio. Within Group I, GRB 190114C has particularly significant thermal emission. It owns the highest averaged ratio of $\sim$ 30\%, and all the time-resolved ratios are greater than 0.1 (marked with a dashed line), which is much higher than other bursts with typical thermal ratios of $0.01$-$0.1$. GRB 100707 has the most obvious variation of a thermal ratio within a burst. Its ratios are very high in the early time but decrease fast with time. Correspondingly, its spectral parameters evolve significantly from the early time to the late time. $\alpha$ derived from the CPL is softer than that from the CPL+BB, with a range much higher than the late time. The CPL and CPL+BB fittings give a significantly diverse $E_{\rm c}$ in the early time but a similar value in the late time. Both $\alpha$ and $E_{\rm c}$ evolve consistently with the description shown in the cartoon (Figure \ref{Cartoon}).

\subsection{The Results in Thermal-dominated Case}

The temporal evolution of parameters for the thermal-dominated cases (Group II) with its differences are shown in Figure \ref{AlphaII}. $\Delta \alpha^{\rm I}$ (Group I) is found to be much greater than $\Delta \alpha^{\rm II}$ (Group II). This is consistent with the fact that a narrower spectrum leads to more obvious results. The parameters relations for our Group II bursts are presented in Figure \ref{FluxAlphaII}. The $F-\alpha$ relation has a marked tendency of moving to the left, which is consistent with the results of Group I.
In Figure \ref{SpectrumII}, we present the observed spectral variation between the CPL and the PL+BB models. For GRB 090902B, $E$(BB)=640.1 keV$<$$E_{\rm c}$(CPL)=2732.34 keV; $\alpha$(CPL)=-1.14 is much harder than the index (PL+BB)=-1.74. Similarly, the same results are also found in GRB 160107, $E_{\rm BB}$=32.8 keV$<$$E_{\rm c}$(CPL)=492.5 keV; $\alpha$(CPL)=-1.72 is harder than the index $\alpha$(PL+BB)=-1.88.

The results indicate that the two Groups bursts share the same results. The difference is that the magnitude of parameter variation for Group II is much higher than Group I.

\section{Conclusion} \label{sec:Conclusion}

In this paper, the {\it Fermi}-GBM bursts, of which the thermal components significantly contribute to the observed spectra as reported in previous literature, are analyzed and confirmed by our Bayesian analysis. Our samples are divided in two groups depending on whether the thermal component is dominant or not. We investigated how the thermal components affect the nonthermal spectral parameters by comparing the spectral parameters between the CPL+BB (or PL+BB) model and CPL-alone models. Our analysis supports the fact that if ignoring the strong BB component, the spectral parameters from the CPL-alone fitting model are misleading. We discussed, in detail, on two ideal cases: GRB 100707 and GRB 190114C. All of their characteristics of the spectral evolution comply with the paradigm described in the cartoon (Figure \ref{Cartoon}). 

In summary, we reach the following conclusions:
\begin{itemize}
\item Low-energy photon index $\alpha$: $\alpha$ softens when BB component is introduced, and such a characteristic is particularly evident in Group II. For Group I, $\Delta$$\alpha$=0.08$\pm$0.15. For Group II, $\Delta$$\alpha$=-0.54$\pm$0.27. For the global sample, $\Delta$$\alpha$=-0.14$\pm$0.28.
\item Peak energy $E_{\rm c}$: $E_{\rm c}$ obtained from CPL only is significantly less than that of the CPL+BB model but is typically greater than the peak of the BB temperature ($kT$ and $kT$ $\times$ 2.8), with log $\Delta$$E_{\rm c}$=2.42$\pm$0.68.
\item Parameters relations: both the $F-\alpha$ and $F-E_{\rm c}$ relations are generally present the same slope with an offset, whereas $E_{\rm c}$-$\alpha$ relation is possibly altered in all cases (the slope increases, or decreases, or remains unchanged).
\item The magnitude of change of spectral parameters ($\Delta \alpha$ and $\Delta E_{\rm c}$, $\alpha$ goes to harder and $E_{\rm c}$ goes to smaller) are positively monotonous correlated with the intensity ($F_{\rm BB}/F_{\rm tot}$) of the thermal components.
\item In GRB 190114C, a majority of $\alpha$ indices are across the synchrotron limit (-2/3) with the hard-to-soft trend when BB component is introduced.

\end{itemize}

We conclude that if the thermal component is present and strong in a GRB, the spectral parameters from the fitting adopting only the nonthermal model may deviate from parameters from the proper fitting considering the nonthermal model with an additional thermal component. We caution that the existence of the thermal component shall be duly examined during the GRB spectral analysis.

\acknowledgments

I appreciate to the referee for the constructive report. I would like to wish thank to Prof. Felix Ryde, and Dr. Yu Wang for useful discussions. I particularly thank to the support from Prof. Remo Ruffini. This research made use of the High Energy Astrophysics Science Archive Research Center (HEASARC) Online Service at the NASA/Goddard Space Flight Center (GSFC). Part of this work made use of our personal Python library. 

\vspace{5mm}
\facilities{{\it Fermi}-GBM}
\software{3ML\citep{2015arXiv150708343V}, and RMFIT} 
\bibliography{../../MyBibFiles/MyBibFile.bib}

\clearpage
\begin{deluxetable}{ccccccccc}
\tablewidth{0pt}
\tabletypesize{\scriptsize}
\tablecaption{Properties of Our Sample.}
\tablehead{
\colhead{{\it Fermi} ID}
&\colhead{$t_{90}$}
&\colhead{Fluence}
&\colhead{Dector}
&\colhead{Source}
&\colhead{Background}
&\colhead{$S>$20/Total}
&\colhead{Best Model}
&\colhead{Reference}\\
&\colhead{(s)}
&\colhead{($10^{-5}$ erg cm$^{-2}$)}
&&&&\colhead{(Bin)}
&&\colhead{(For Best Model)}
}
\colnumbers
\startdata
081221681&29.70$\pm$0.41&3.00$\pm$0.01&n1(n2)b0&-1.-45.&[-20--10,60-80]&14/18&PL+BB&\cite{2013ApJ...768..187B}\\
081224887&16.45$\pm$1.16&3.76$\pm$0.02&(n6)n7n9b1&0.-20.&[-20--10,40-60]&5/8&Band+BB&\cite{2016MNRAS.456.2157I}\\
090719063&11.39$\pm$0.47&4.68$\pm$0.02&n7(n8)b1&0.-15.&[-20--10,40-60]&11/11&Band+BB&\cite{2016MNRAS.456.2157I}\\
090902462&19.33$\pm$0.29&22.12$\pm$0.03&n0(n1)n9b0&-1.-25.&[-20--10,60-80]&48/57&PL+BB&\cite{2010ApJ...709L.172R}\\
100507577&44.03$\pm$5.22&1.00$\pm$0.01& (n9)nanbb1&-1.-30.&[-20--10,60-80]&0/4&Band+BB&\cite{2013MNRAS.432.3237G}\\ 
100707032&81.79    $\pm$1.21&8.77$\pm$0.02&n7(n8)b1&0.-80.& [-50--10,80-100]&11/13&Band+BB&\cite{2016MNRAS.456.2157I}\\
100724029&114.69$\pm$3.24&21.75$\pm$0.06&n0(n1)n2b0&-1.-140.&[-20--10,150-170]&33/38&Band+BB&\cite{2011ApJ...727L..33G}\\
101219686&51.01$\pm$1.78&0.40$\pm$0.005&n3n6(n7)b1&-1.-30.&[-20--10,60-80]&0/2&Band+BB&\cite{2015ApJ...800L..34L}\\
110721200&21.82$\pm$0.57&3.70$\pm$0.004&(n6)n7n9b1&0.-25.&[-20--10,40-60]&8/9&Band+BB&\cite{2013MNRAS.433.2739I}\\
110920546&160.77$\pm$5.22&17.23$\pm$0.02&(n0)n1n3b0&0.-15.&[-18--10,180-190]&8/10&PL+2BB&\cite{2015MNRAS.450.1651I}\\
120323162&4.35$\pm$0.72&0.14$\pm$0.002&n9na(nb)b1&-1.-5.&[-20--10,60-80]&0/3&Band+BB&\cite{2013ApJ...770...32G}\\
160107931&113.92$\pm$17.76&2.73$\pm$0.01&(n1)n2n5b0&-1.-110.&[-100--80,150-170]&9/32&PL+BB&\cite{2018PASJ...70....6K}\\
190114873&116.35$\pm$2.56&44.33$\pm$0.05&n3(n4)n7b0&0.-25.&[-20--10,180-200]&48/51&Band+BB&\cite{2019arXiv190107505W}\\
\enddata

Notes. {\it Fermi} ID (Column 1), the $t_{90}$ (Column 2), the fluence in 10 keV-40 MeV (Column 3), the detector (Column 4), the source (Column 5) and background (Column 6) intervals, and $S>$20/total time bins using the BBlocks method across the source intervals (Column 7), and best model reported in previous literature (Column 8), and its reference (Column 9). Five bursts are known to redshift. GRB 081221681: 2.26; GRB 090902462: 1.822; GRB 101219686: 0.55; GRB 110721200: 0.382/3.512; and GRB 190114873: 0.425.
\end{deluxetable}

\clearpage
\begin{figure*}
\includegraphics[angle=0,scale=0.450]{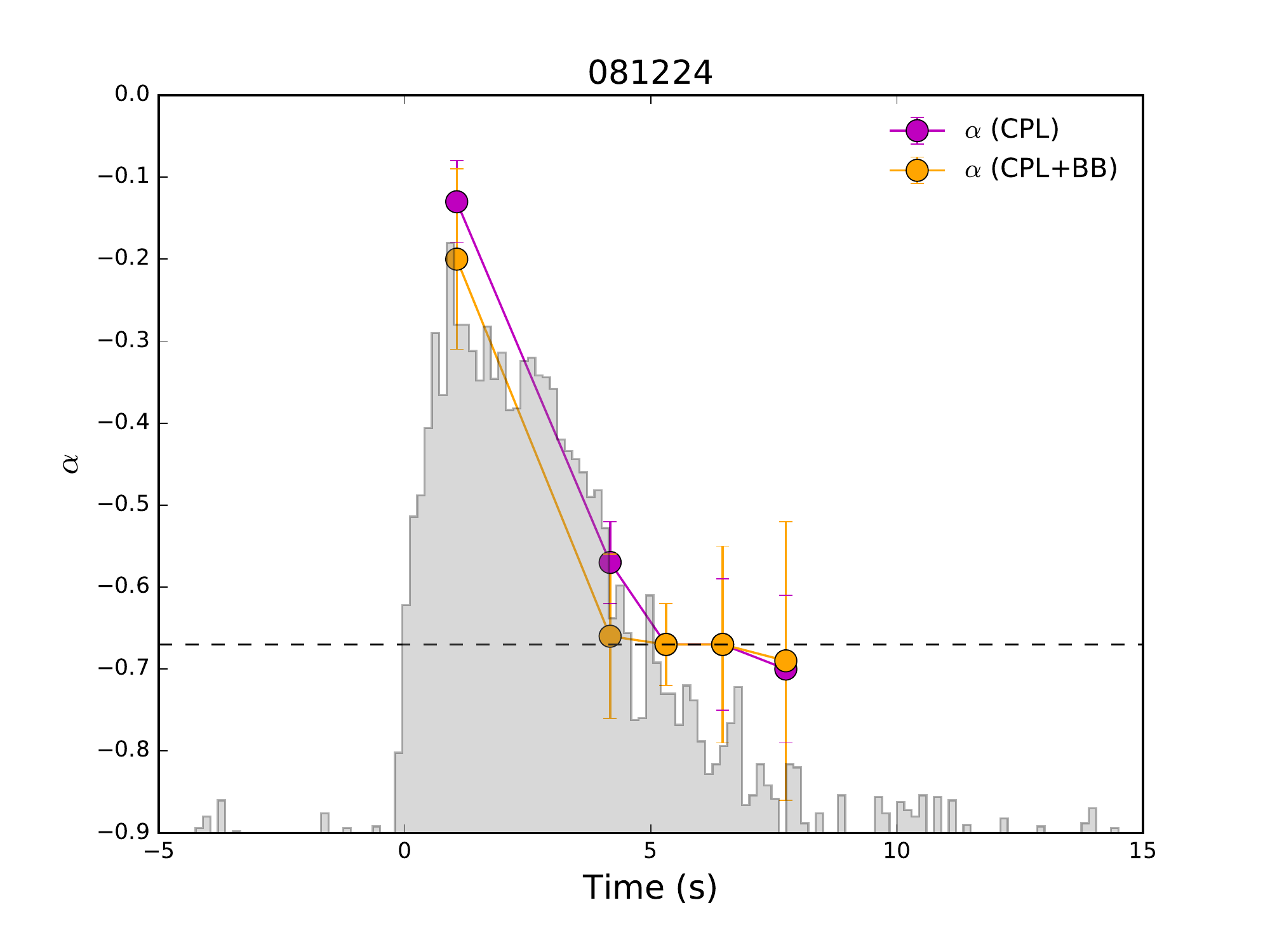}
\includegraphics[angle=0,scale=0.450]{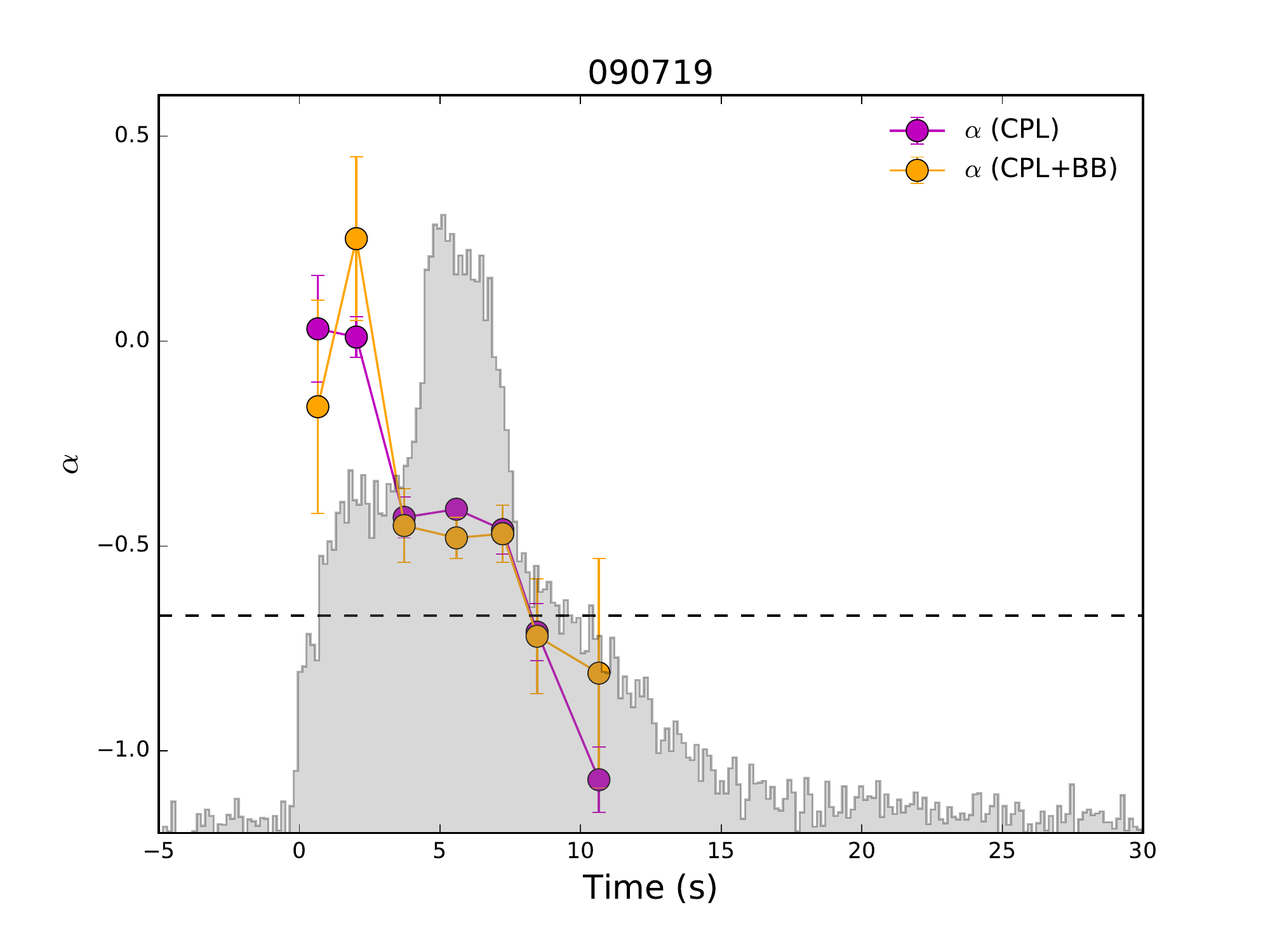}
\includegraphics[angle=0,scale=0.450]{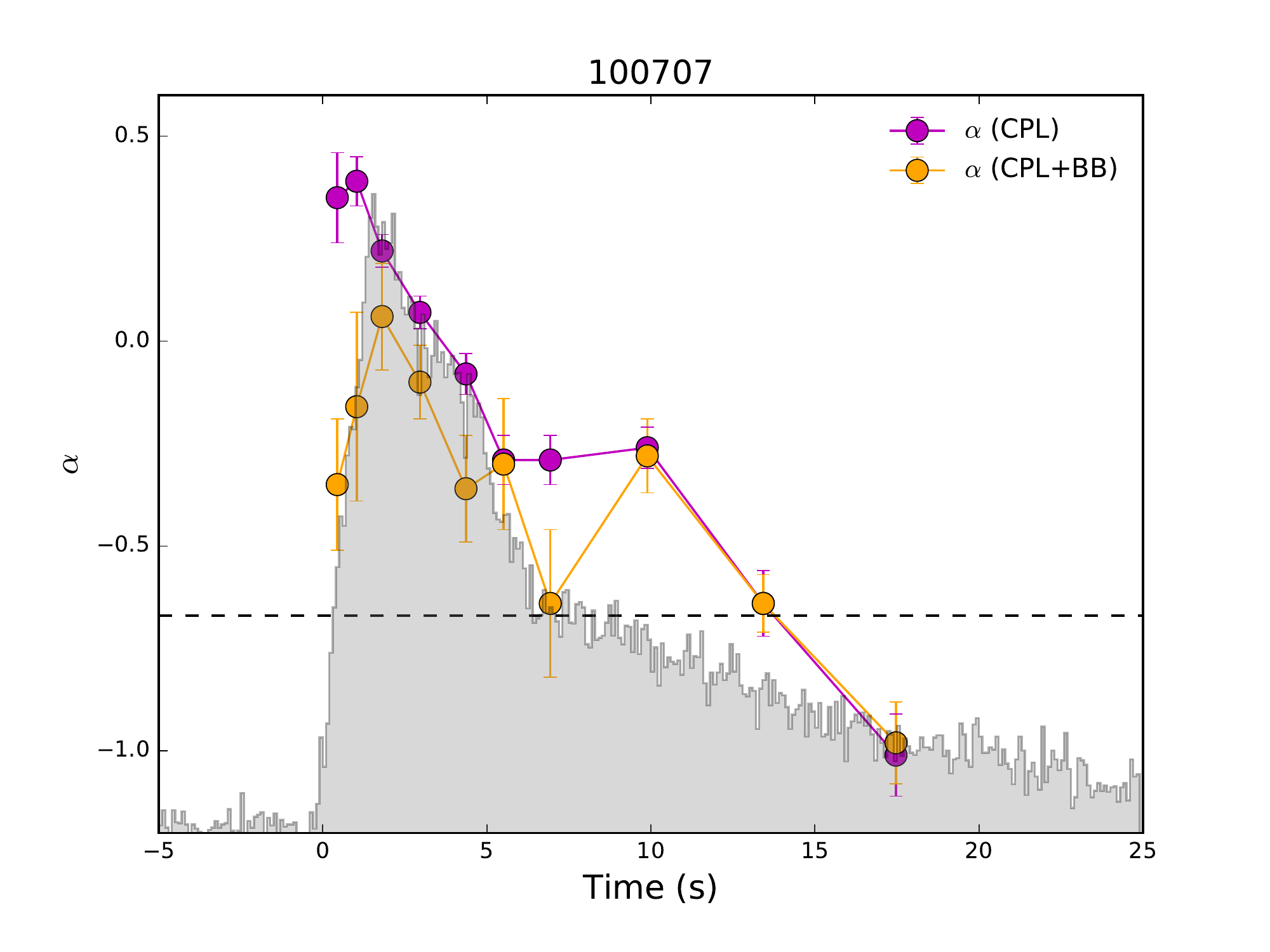}
\includegraphics[angle=0,scale=0.450]{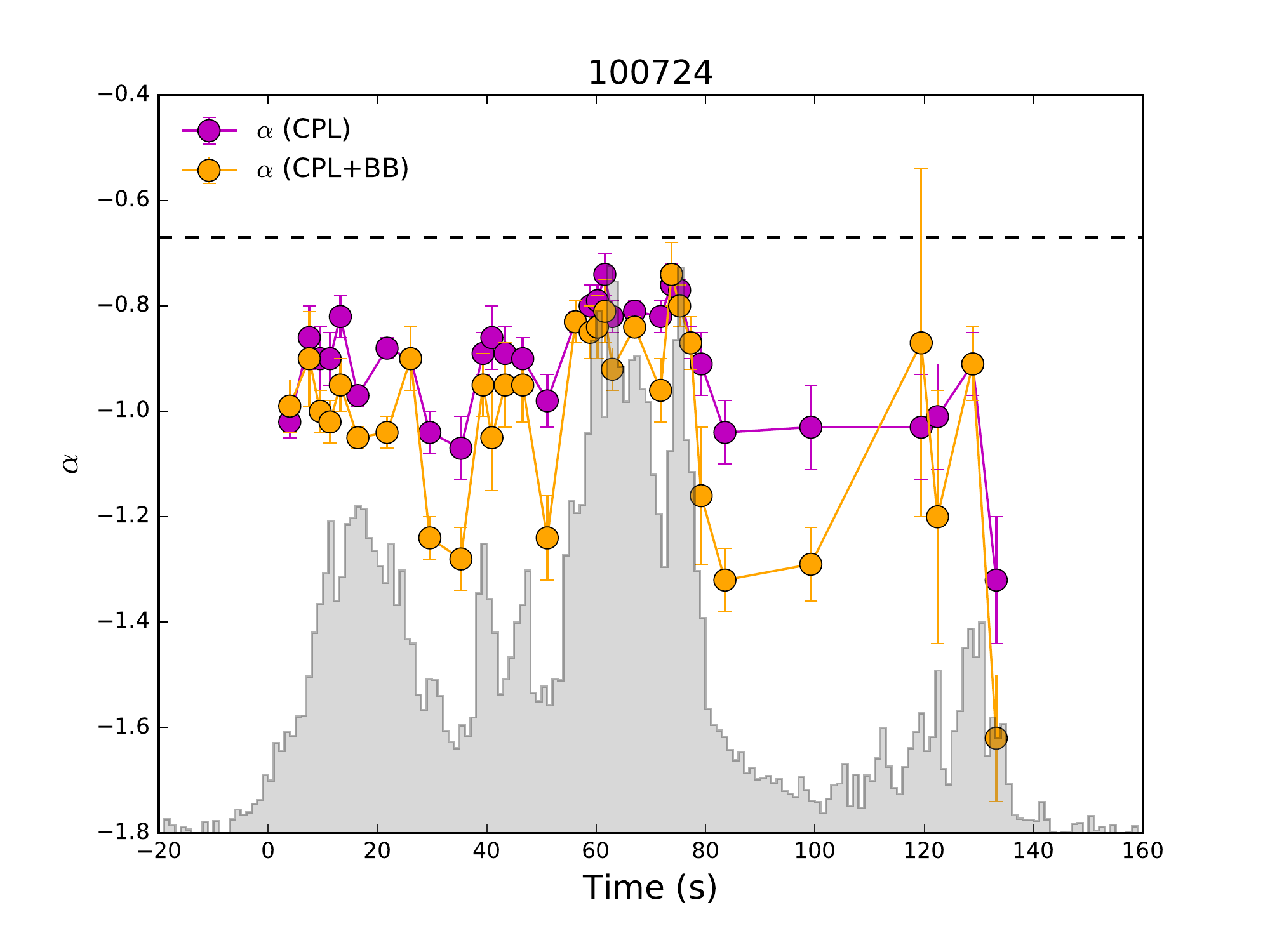}
\includegraphics[angle=0,scale=0.450]{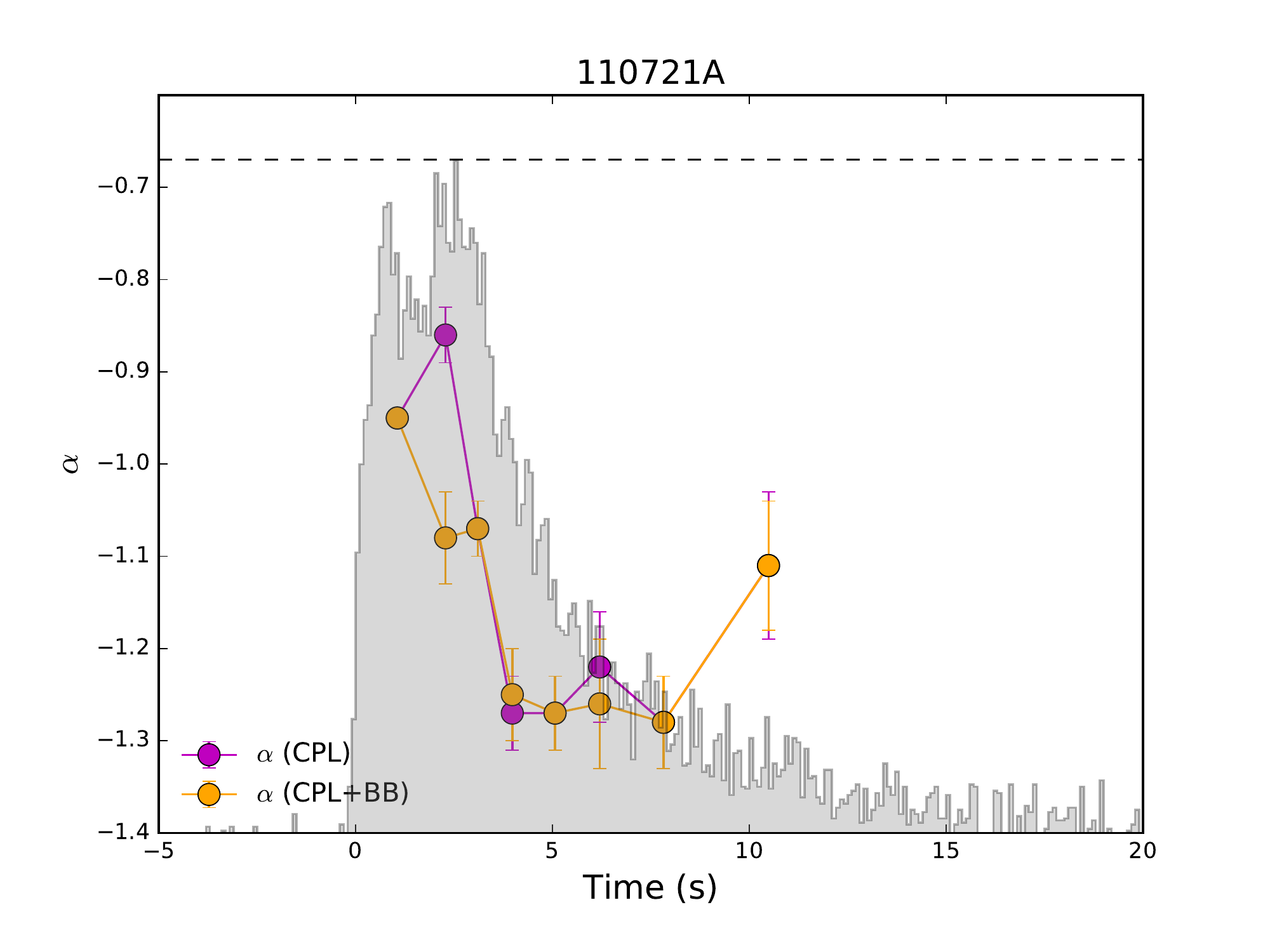}
\includegraphics[angle=0,scale=0.450]{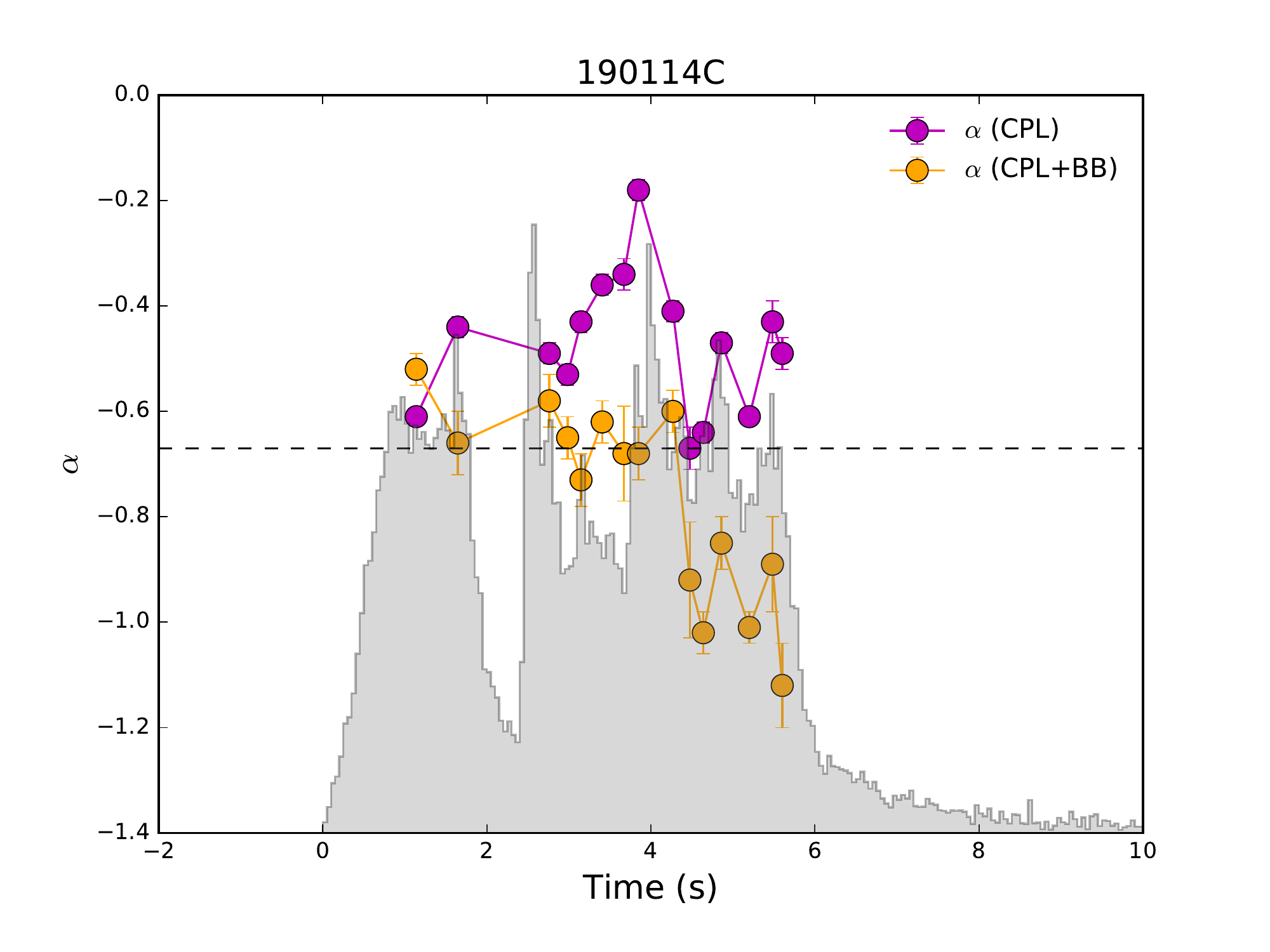}
\caption{Temporal evolution of $\alpha$. Data points with orange and pink indicate the CPL+BB model and the CPL alone model, respectively. Light curves are overlaid in grey. 
The horizontal dashed line represents the limiting value of $\alpha$=-2/3.}
\end{figure*}\label{Alpha}

\clearpage
\begin{figure*}
\includegraphics[angle=0,scale=0.450]{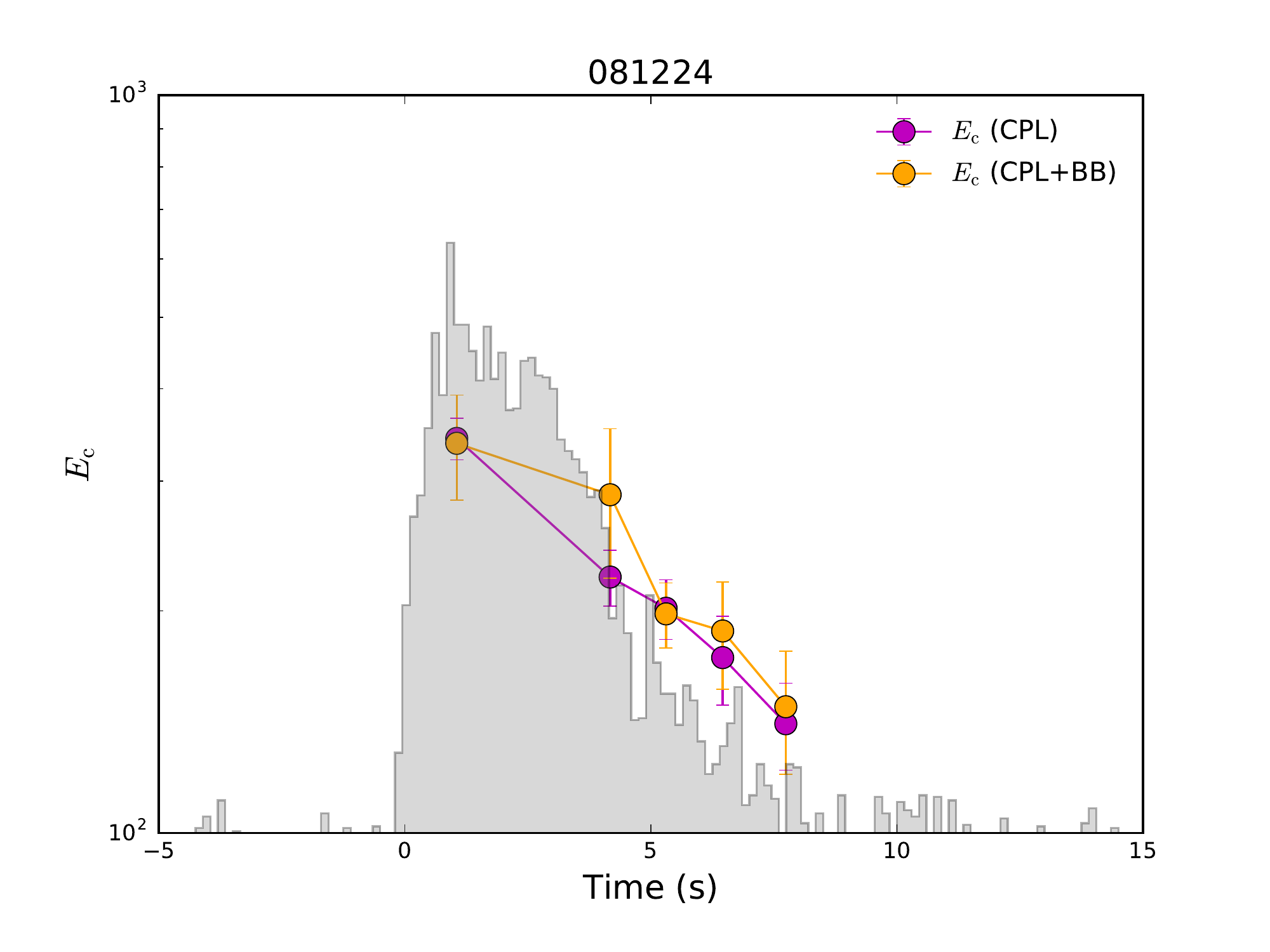}
\includegraphics[angle=0,scale=0.450]{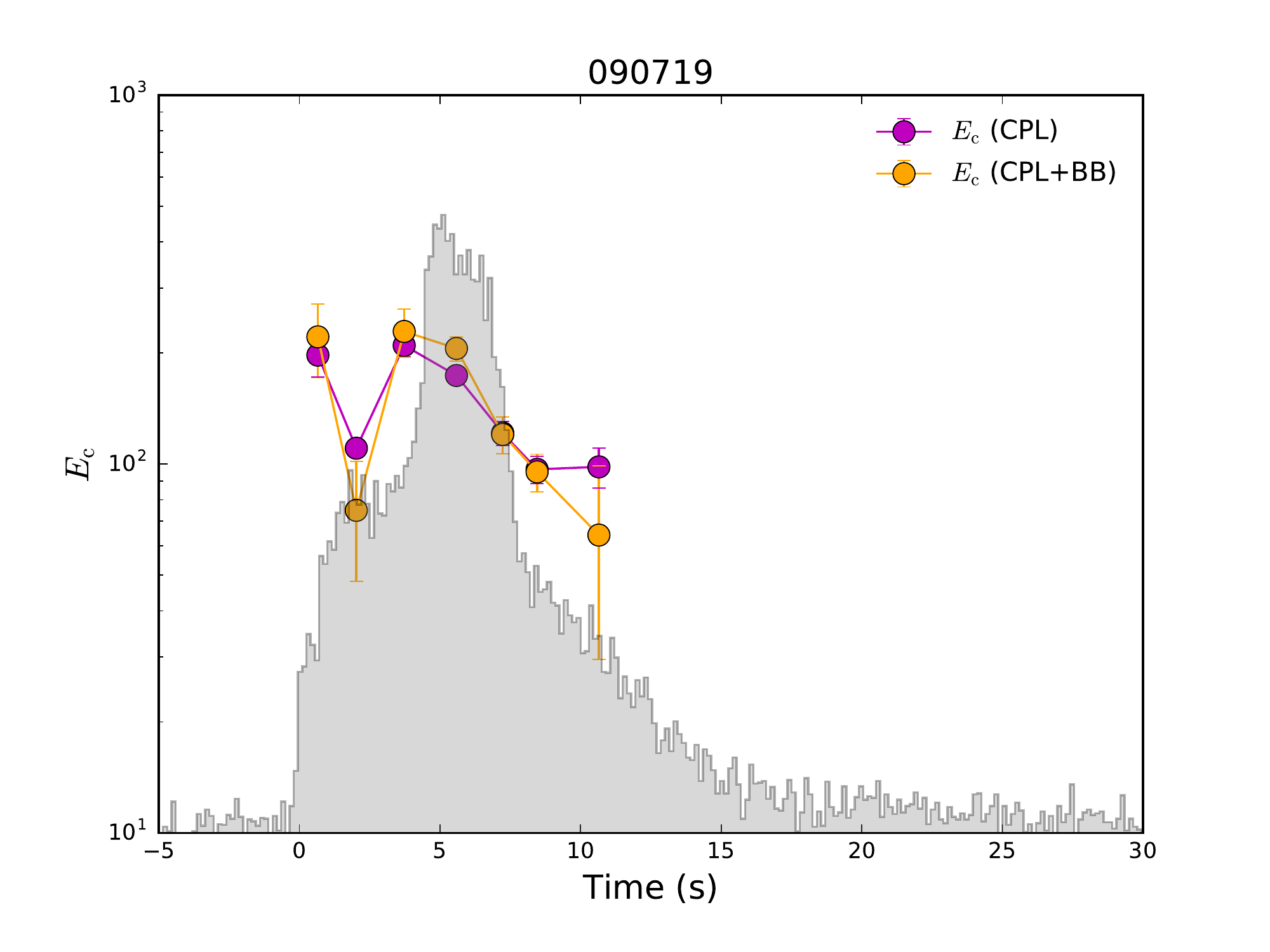}
\includegraphics[angle=0,scale=0.450]{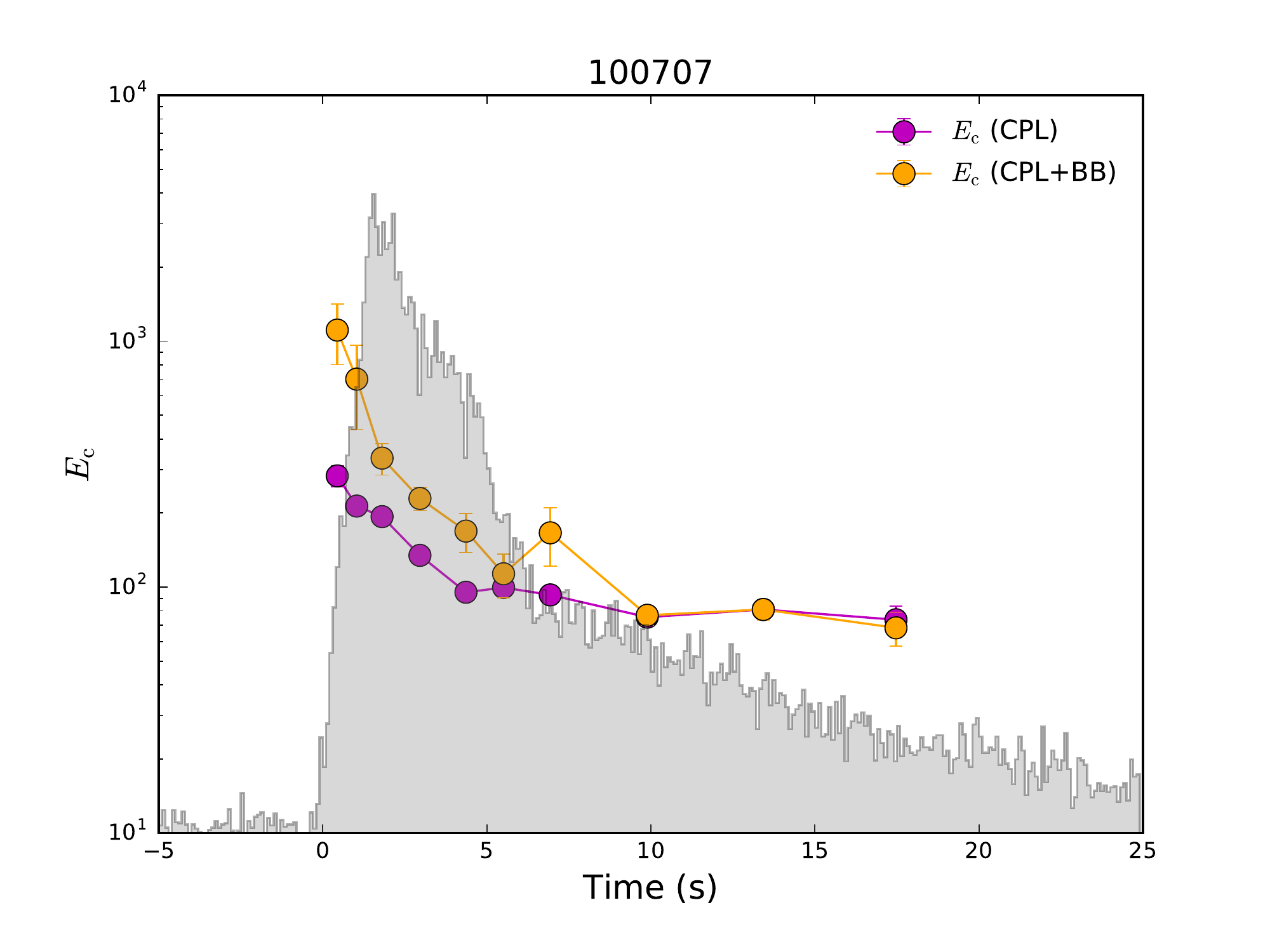}
\includegraphics[angle=0,scale=0.450]{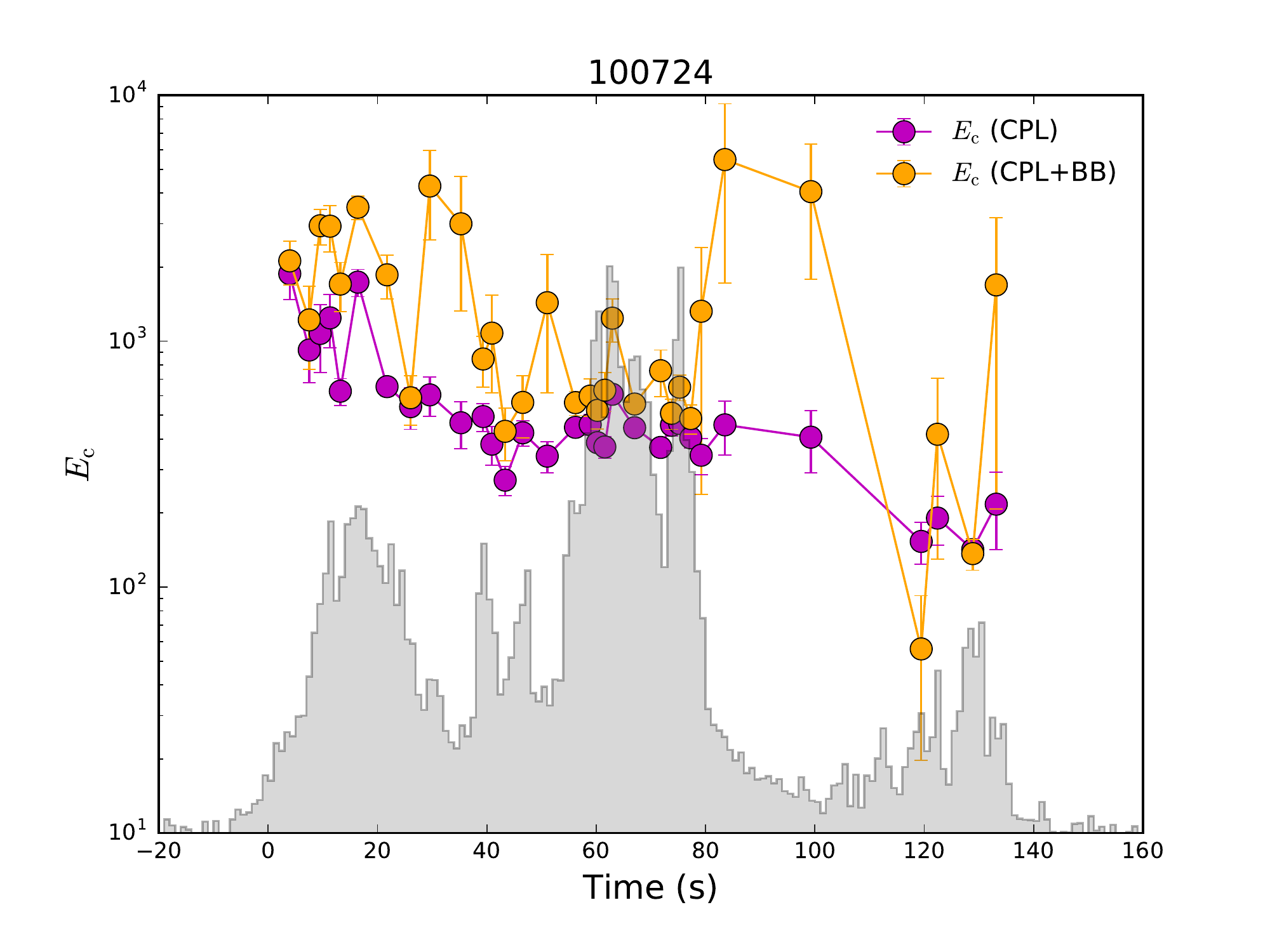}
\includegraphics[angle=0,scale=0.450]{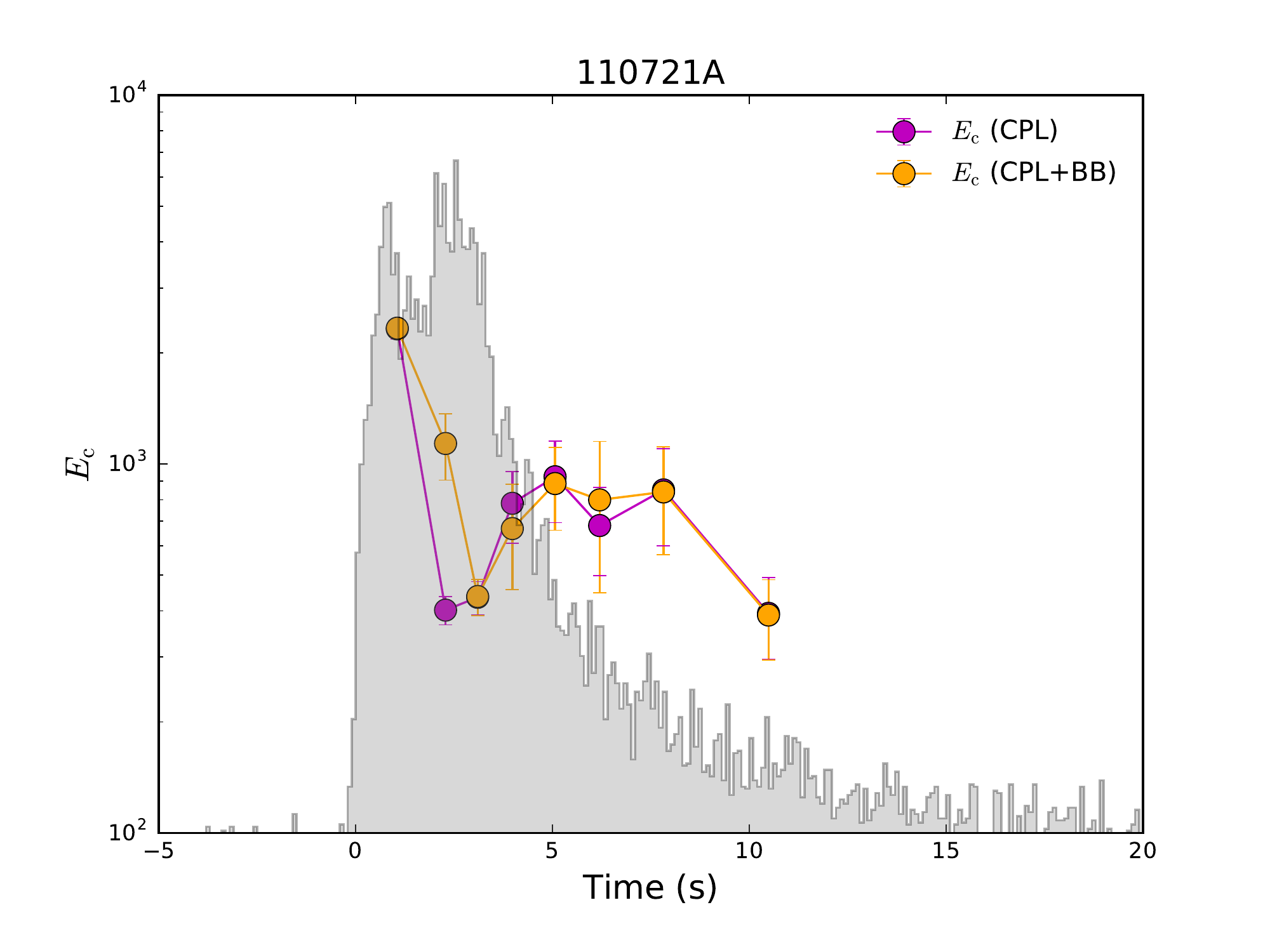}
\includegraphics[angle=0,scale=0.450]{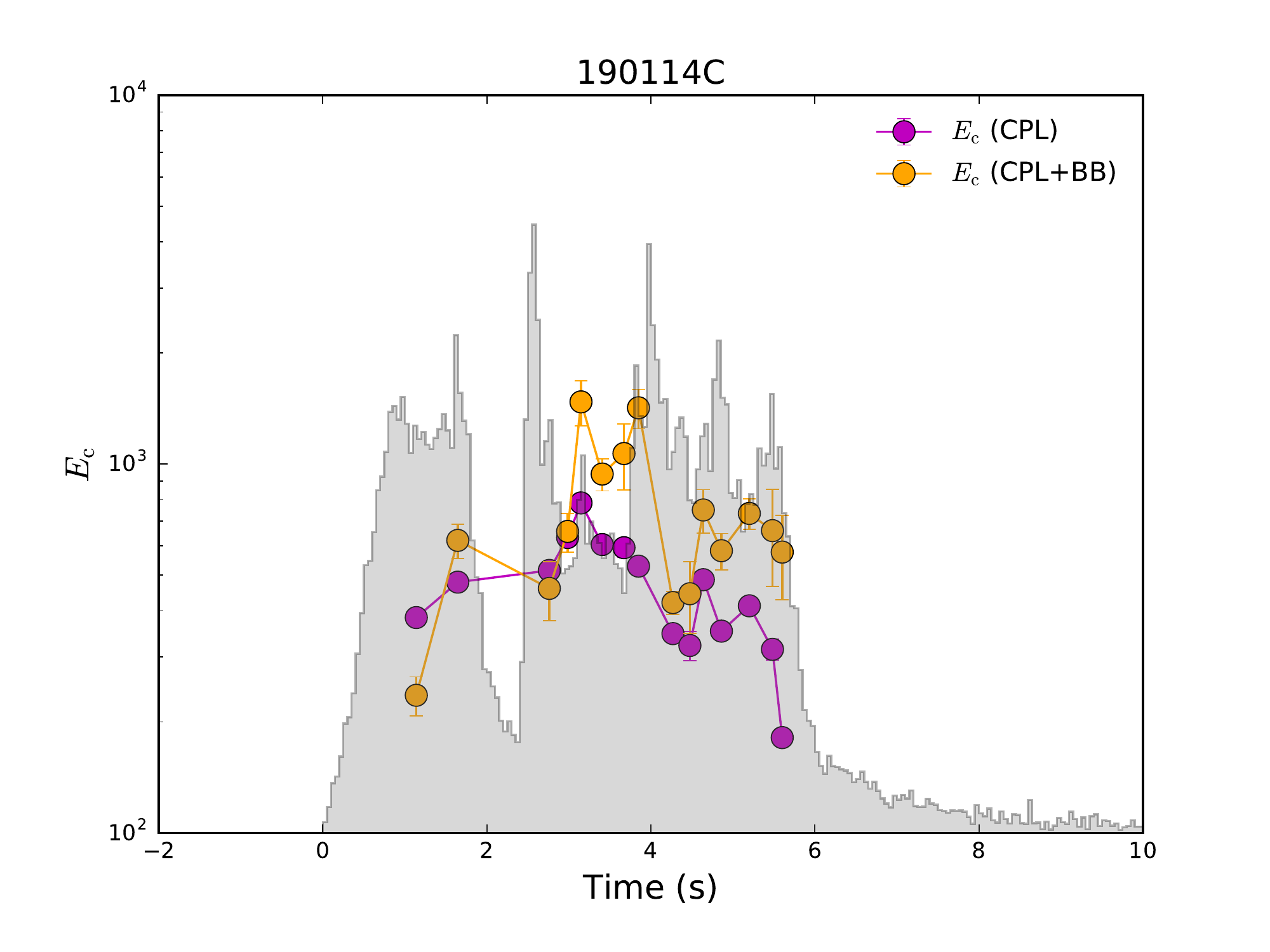}
\caption{Same as Figure \ref{Alpha}, but for the temporal evolution of $E_{\rm c}$.}
\end{figure*}\label{Ec}

\clearpage
\begin{figure*}
\includegraphics[angle=0,scale=0.450]{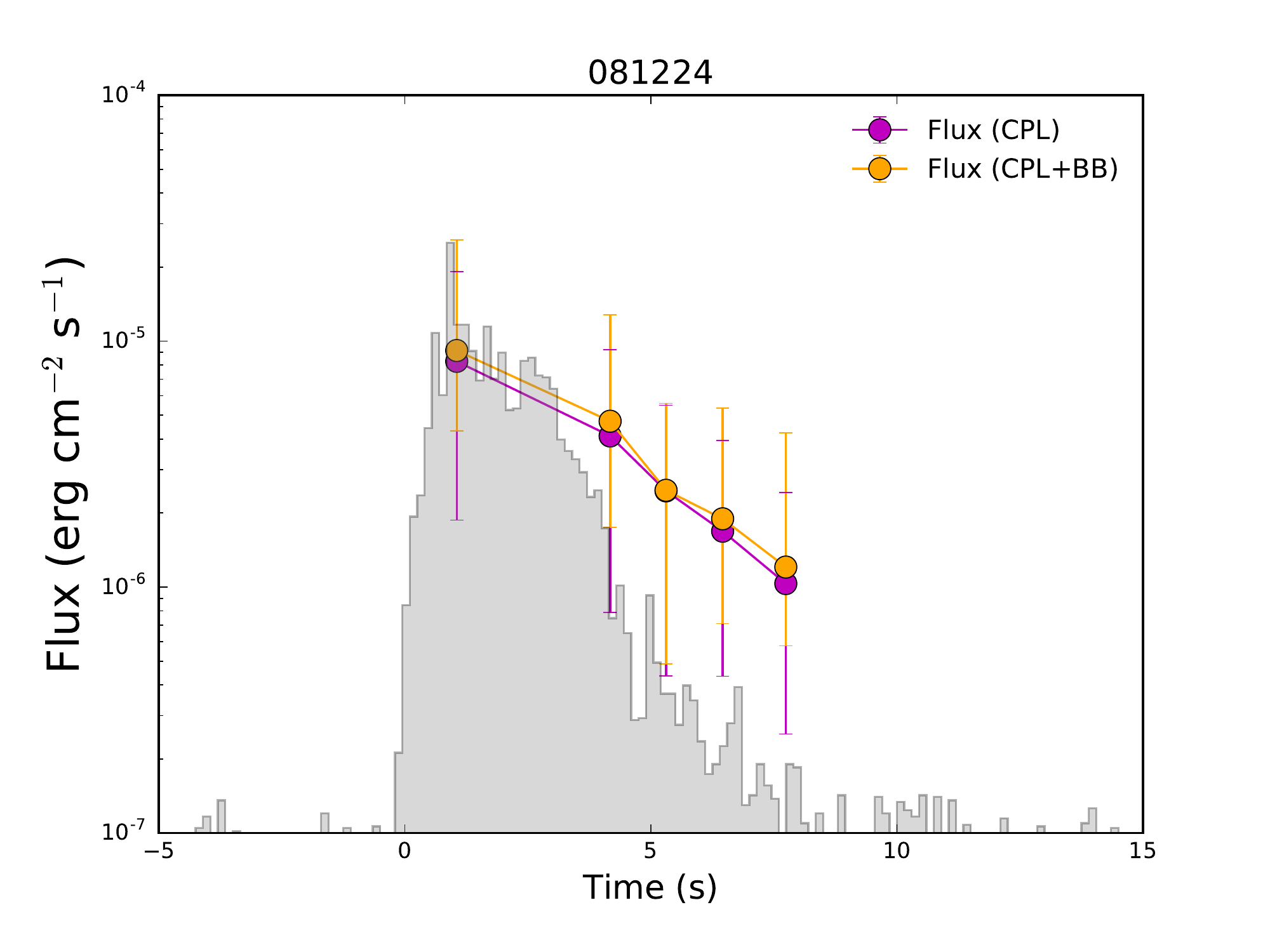}
\includegraphics[angle=0,scale=0.450]{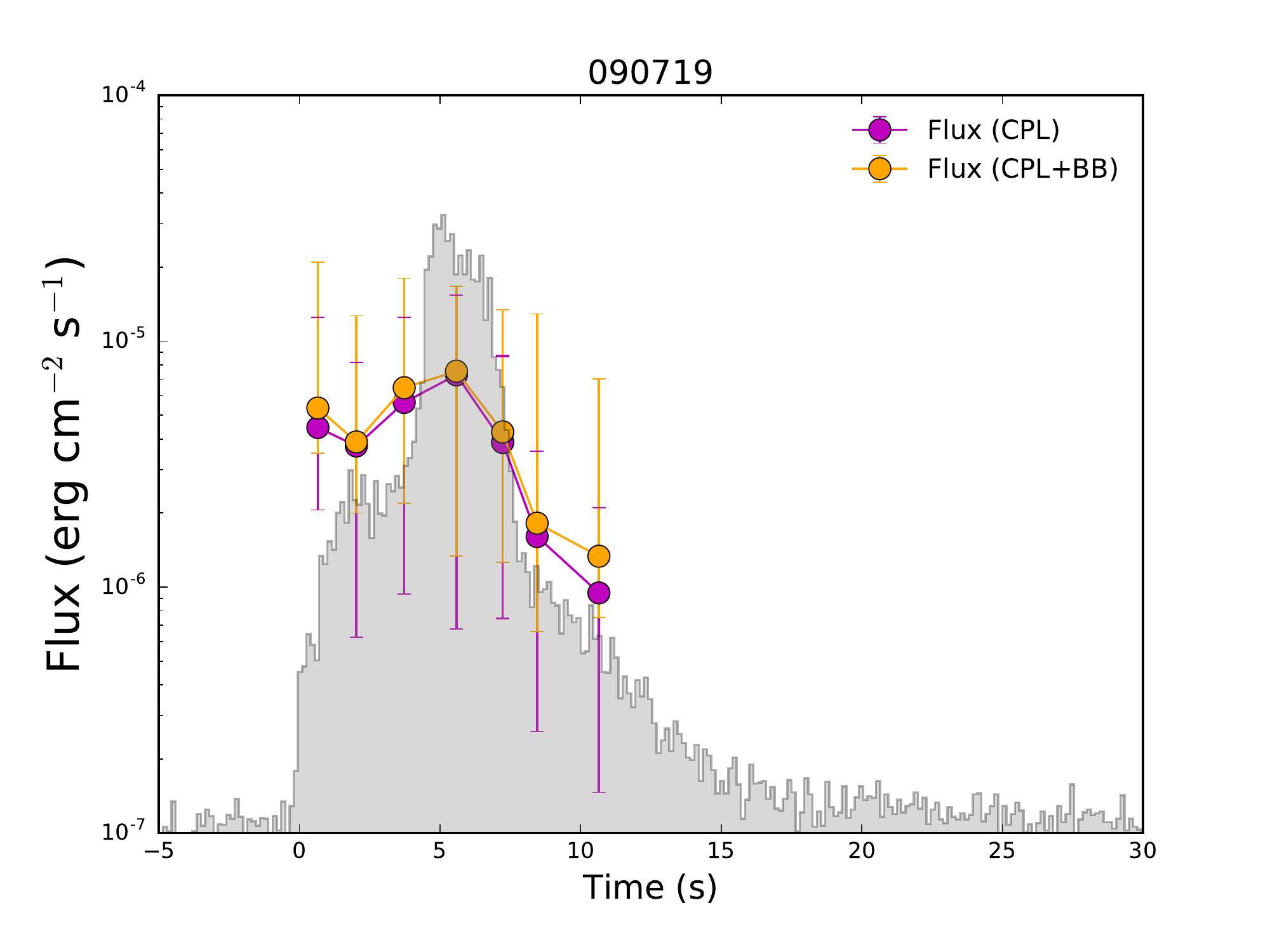}
\includegraphics[angle=0,scale=0.450]{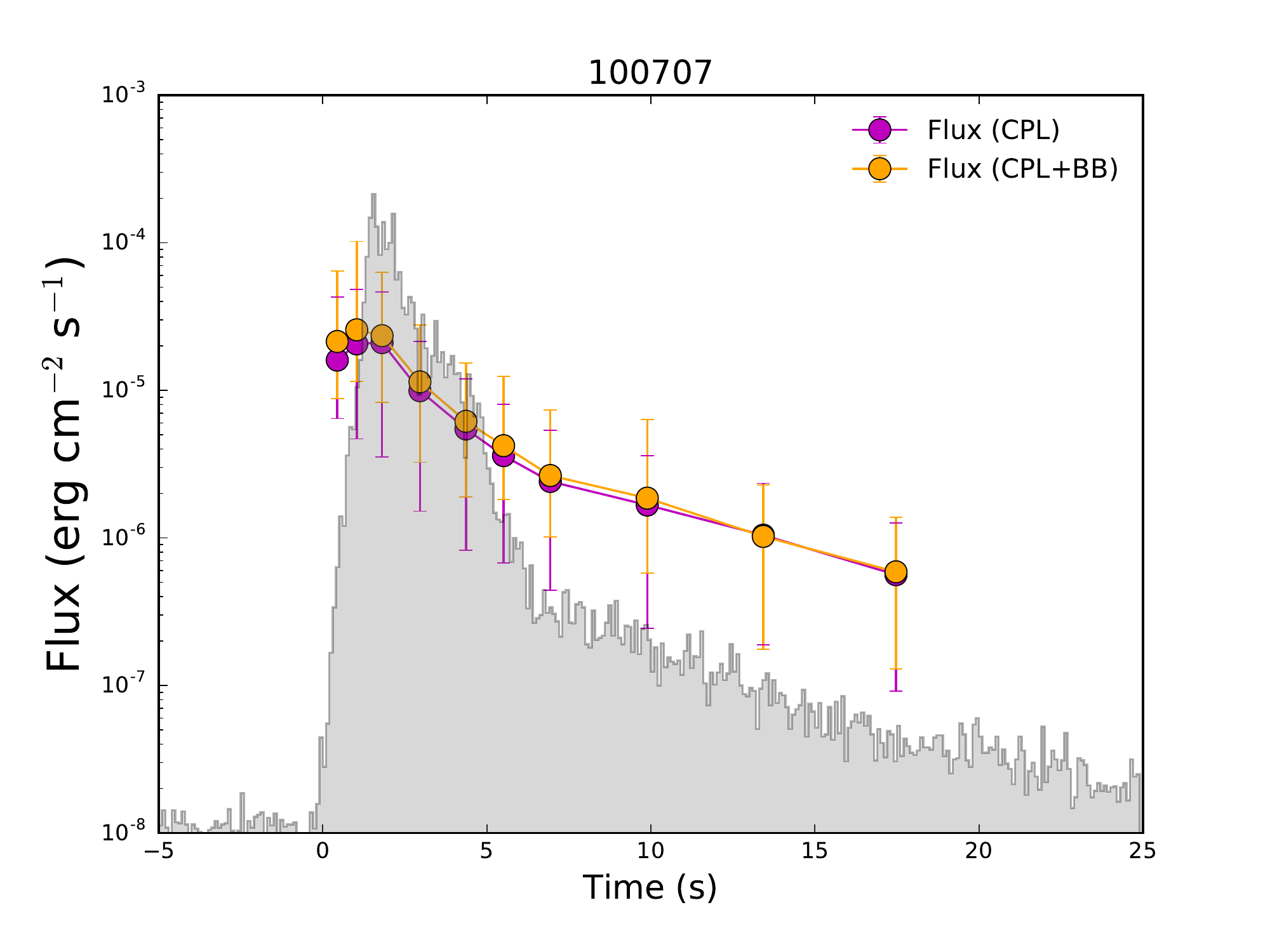}
\includegraphics[angle=0,scale=0.450]{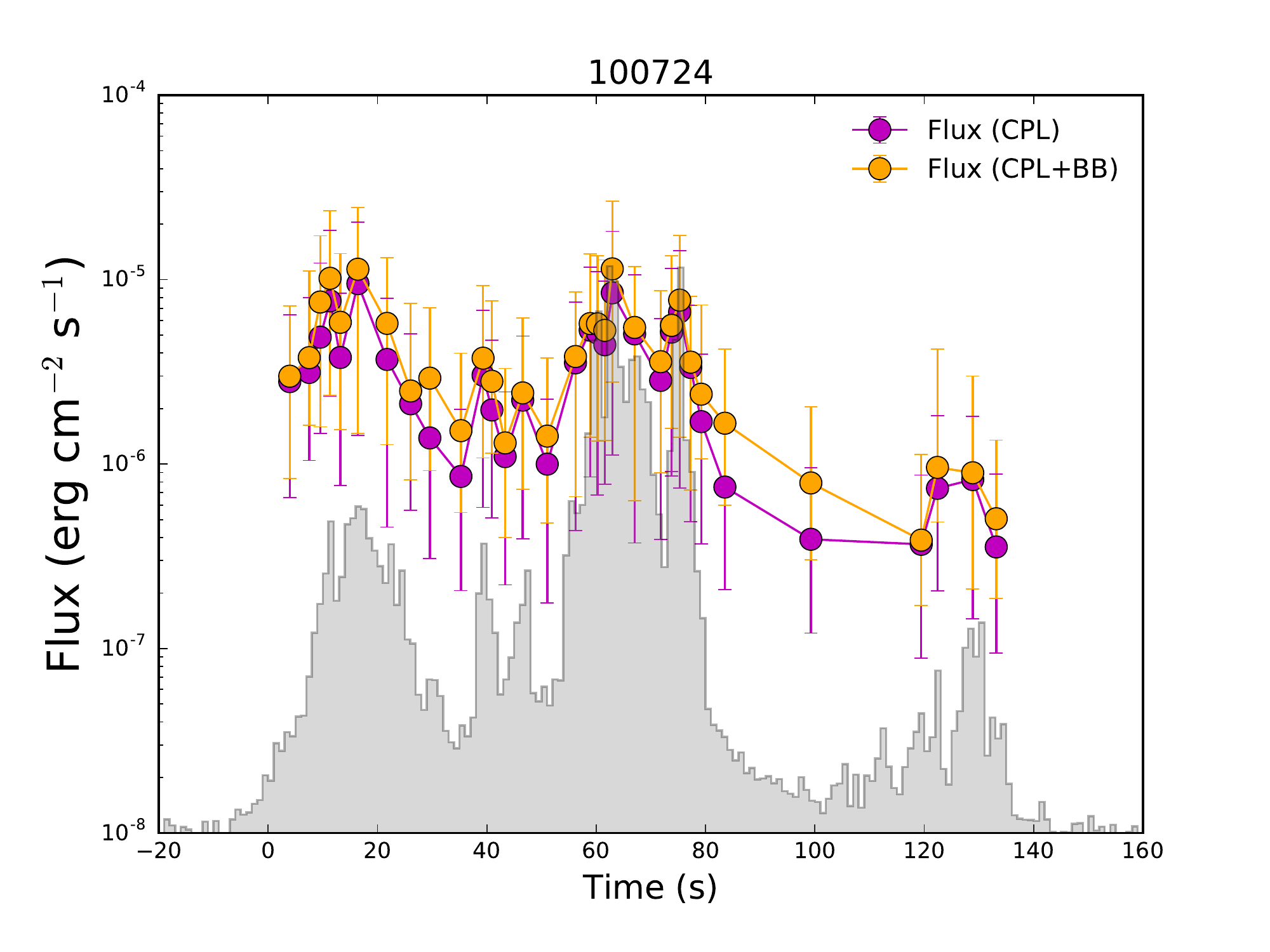}
\includegraphics[angle=0,scale=0.450]{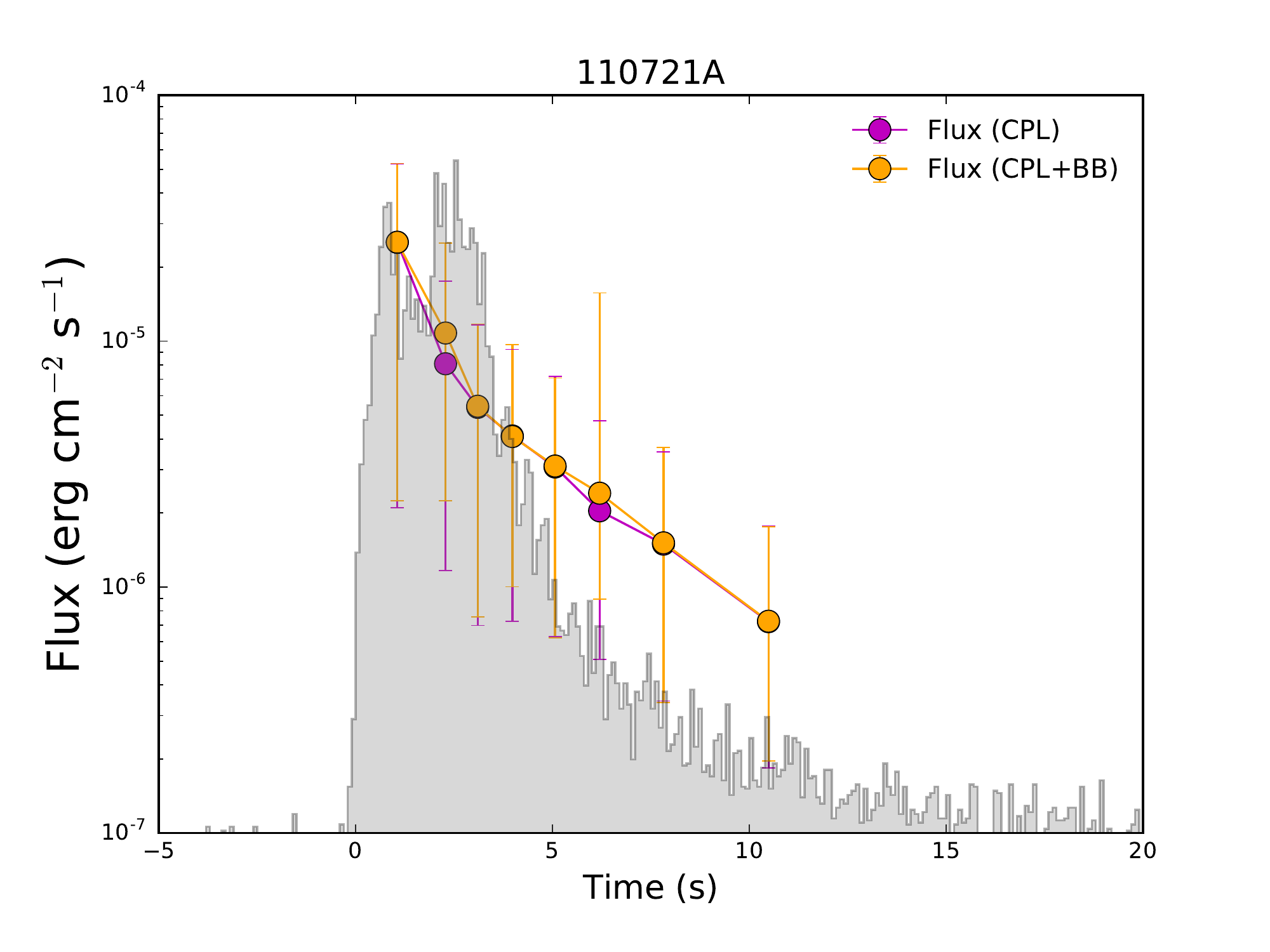}
\includegraphics[angle=0,scale=0.450]{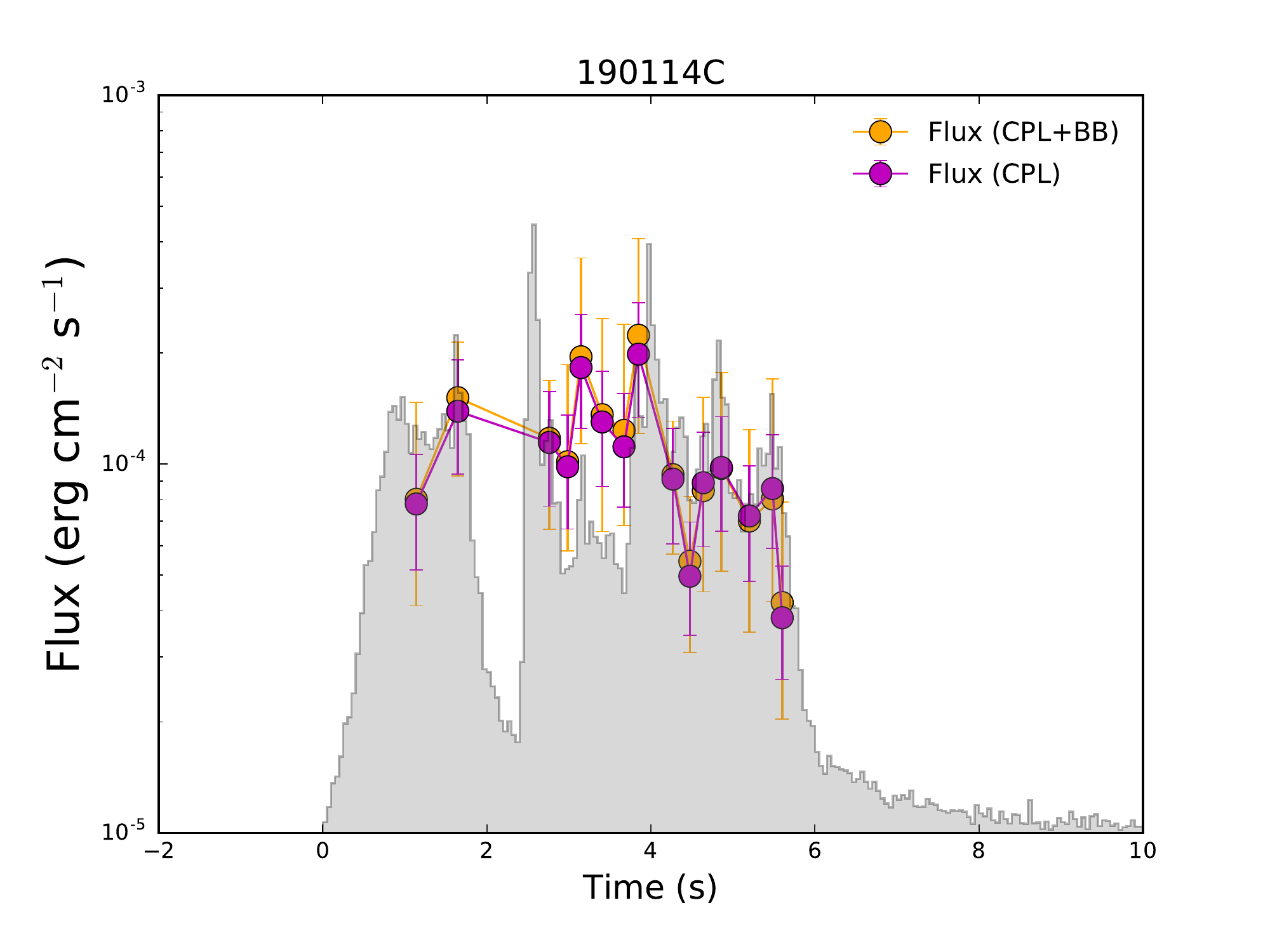}
\caption{Same as Figure \ref{Alpha}, but for the temporal evolution of energy flux.}
\end{figure*}\label{Flux}

\begin{figure*}
\includegraphics[angle=0,scale=0.45]{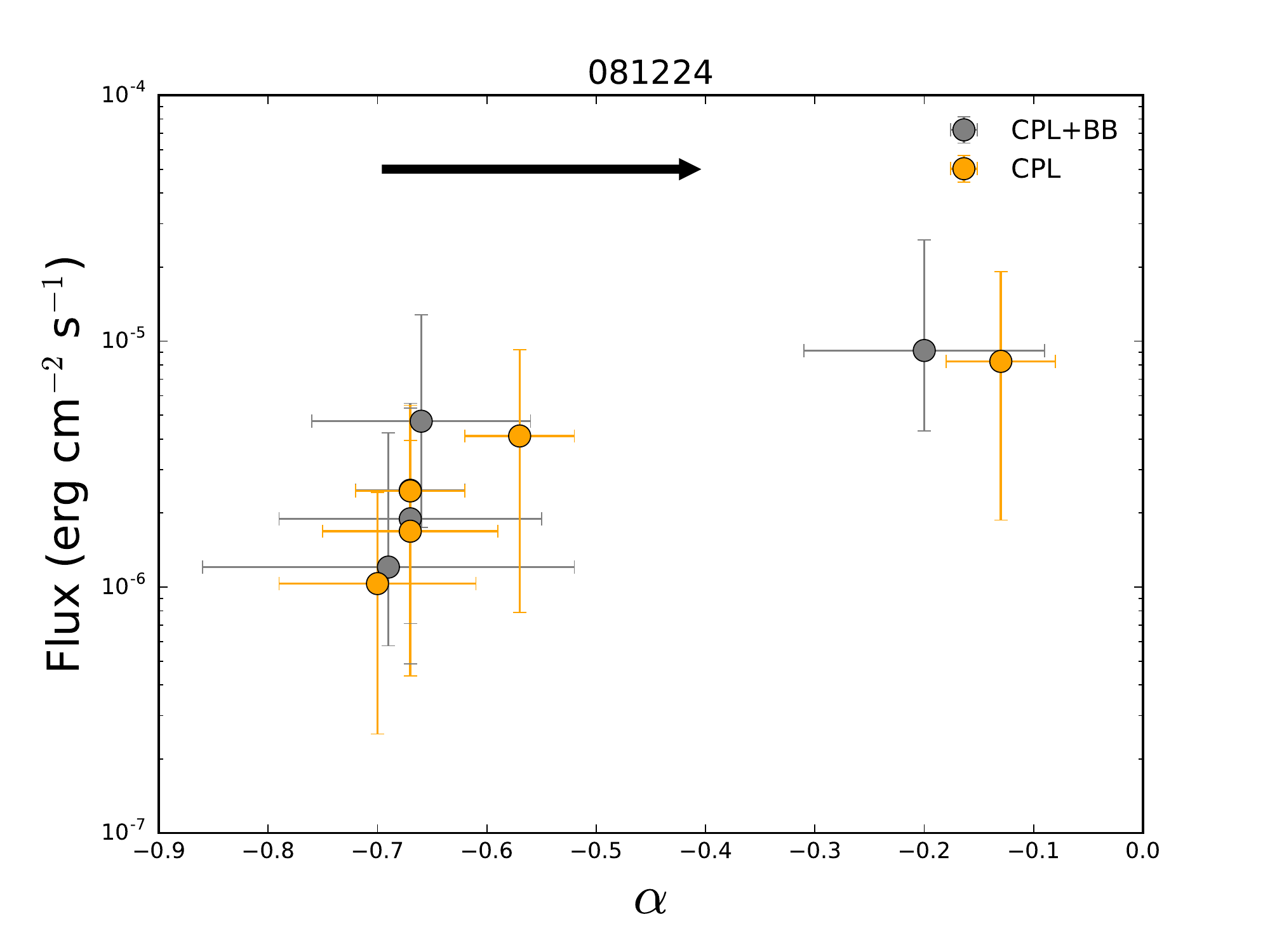}
\includegraphics[angle=0,scale=0.45]{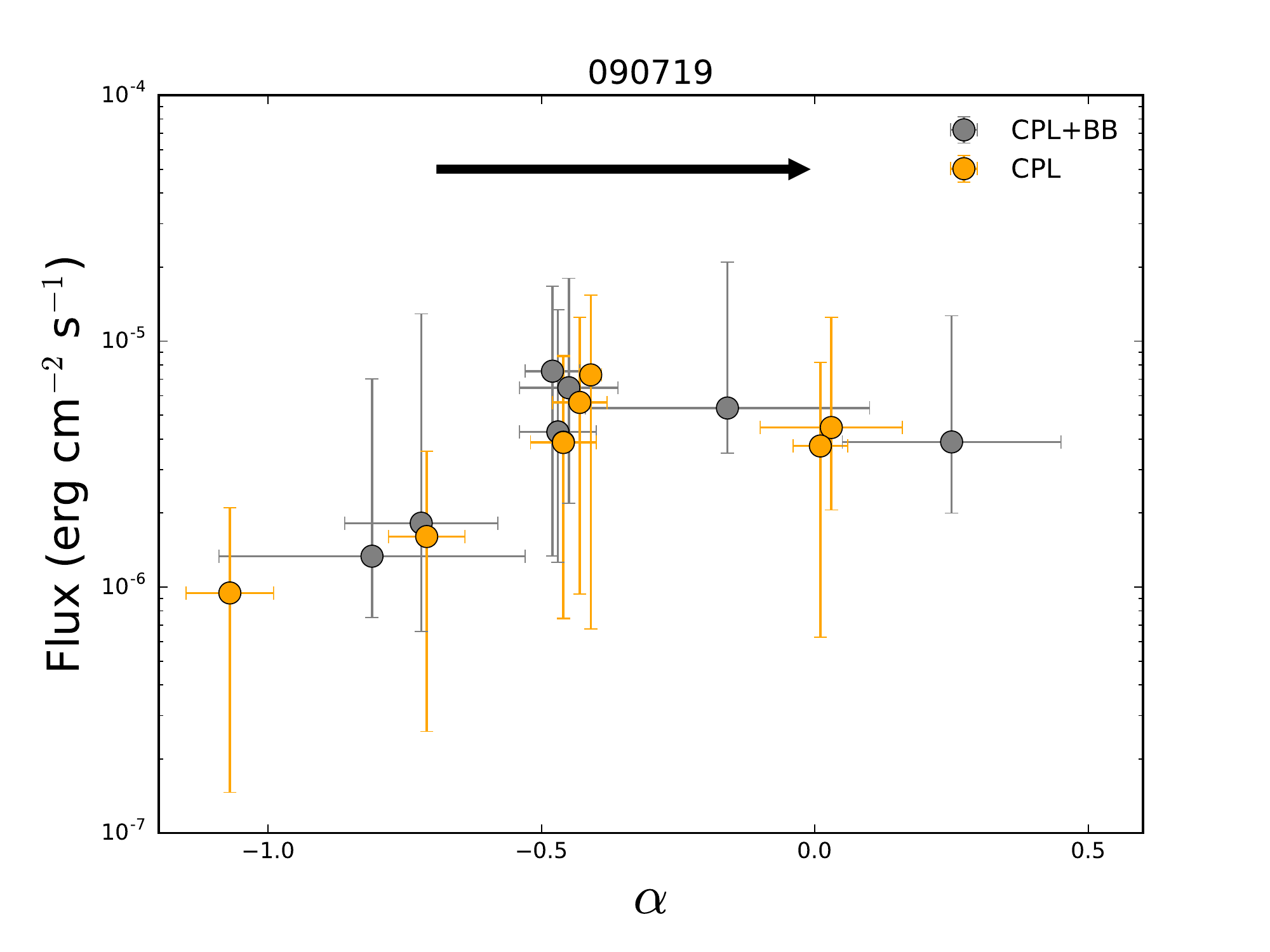}
\includegraphics[angle=0,scale=0.45]{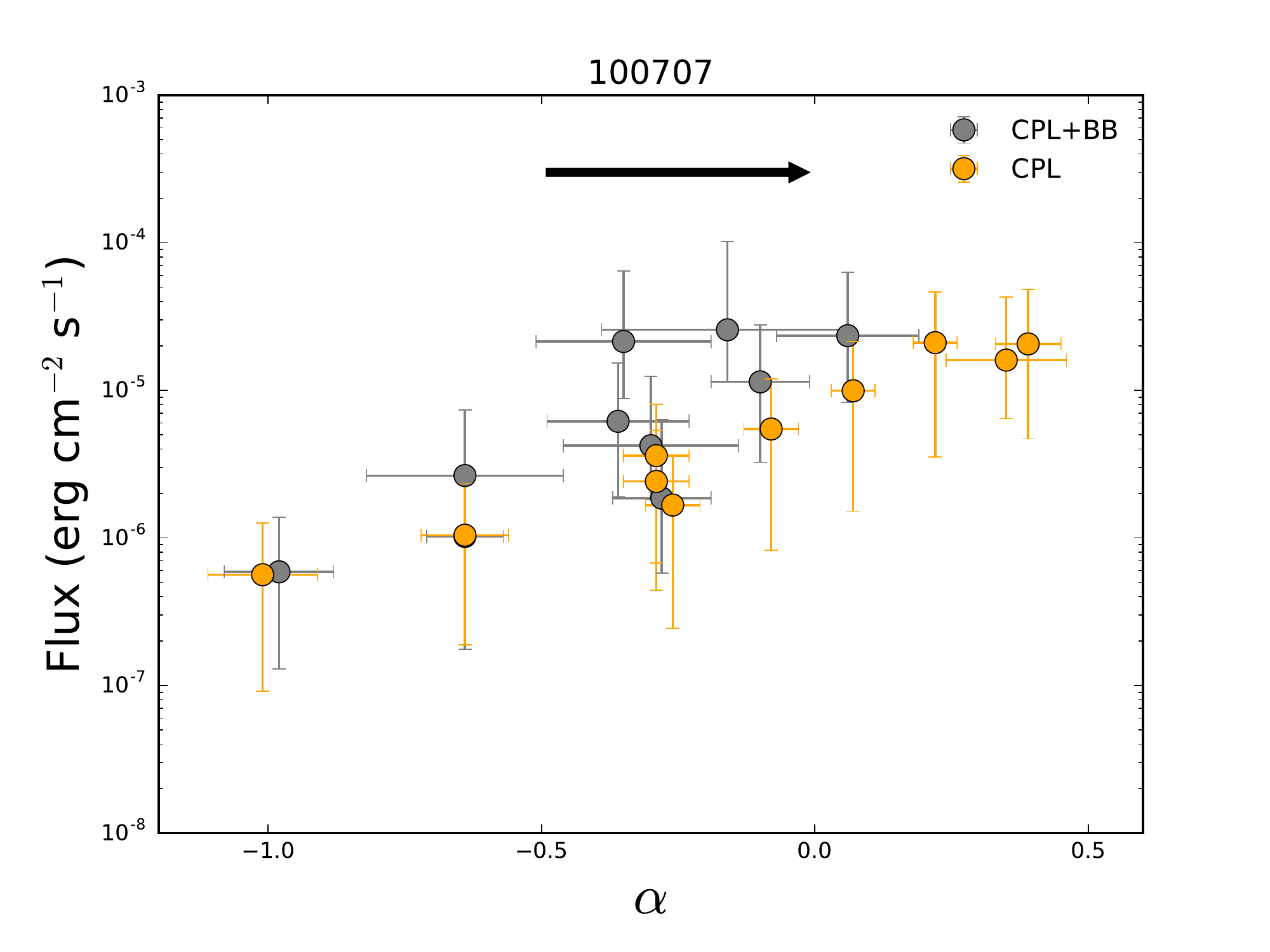}
\includegraphics[angle=0,scale=0.45]{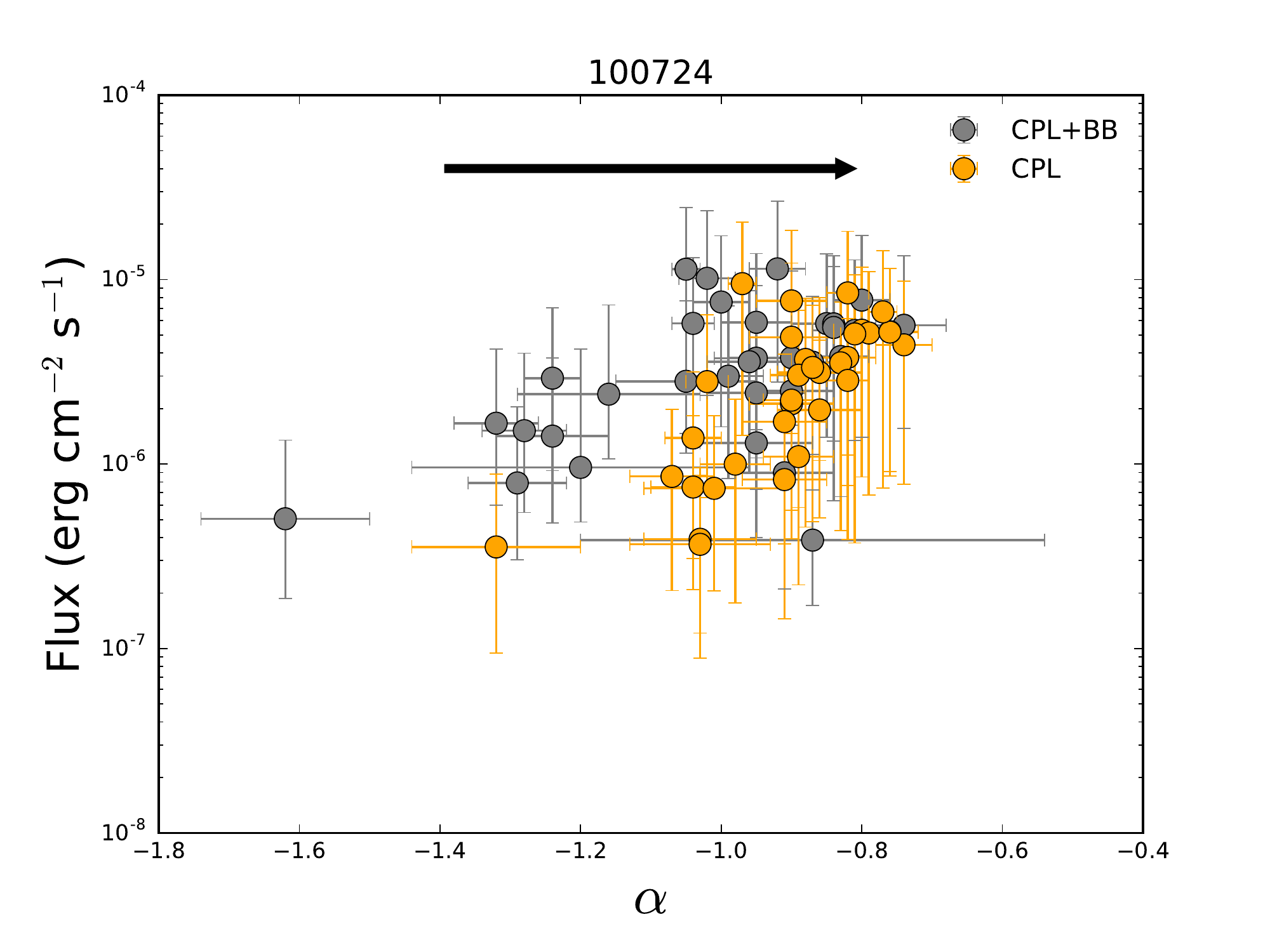}
\includegraphics[angle=0,scale=0.45]{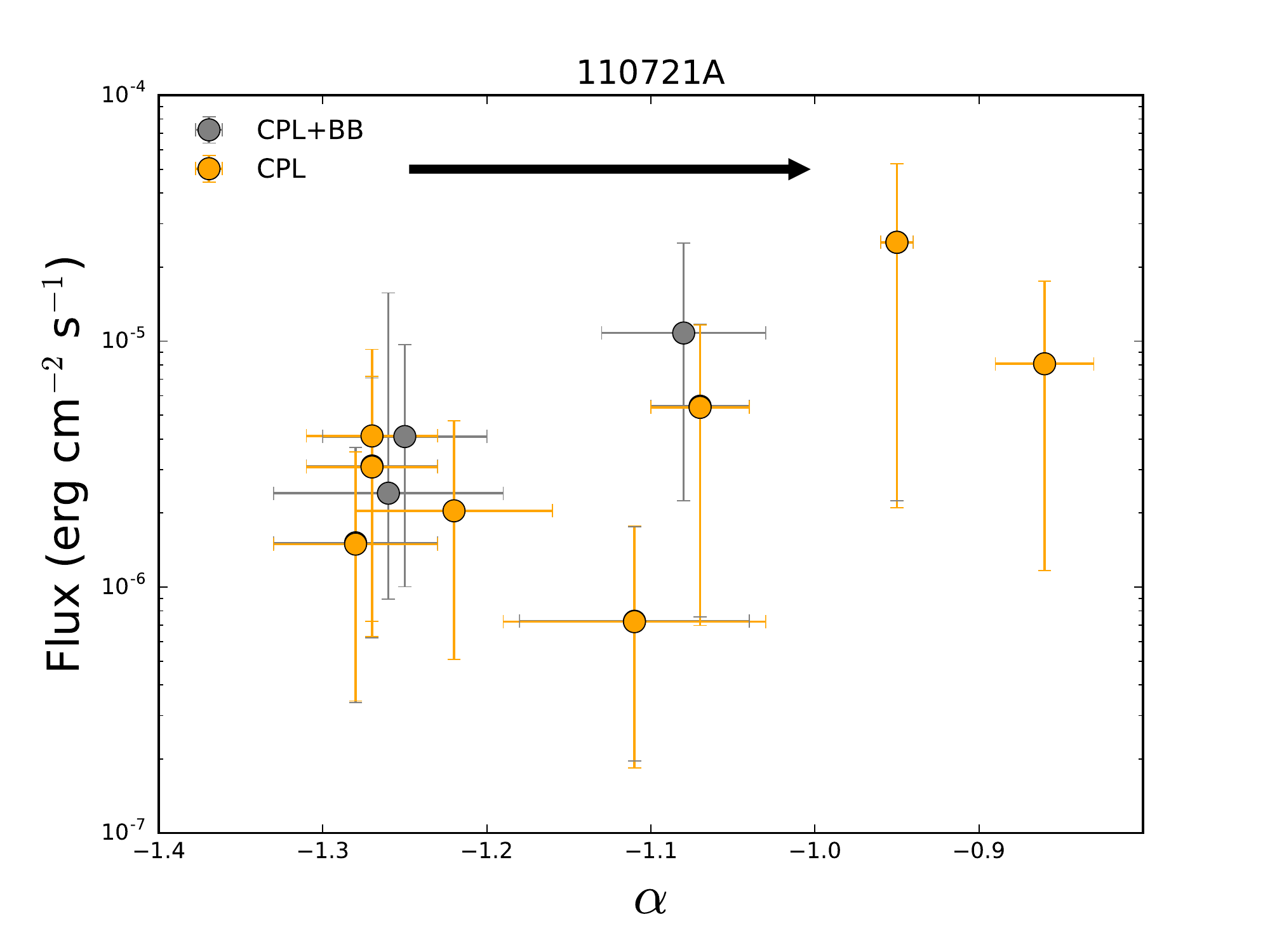}
\includegraphics[angle=0,scale=0.45]{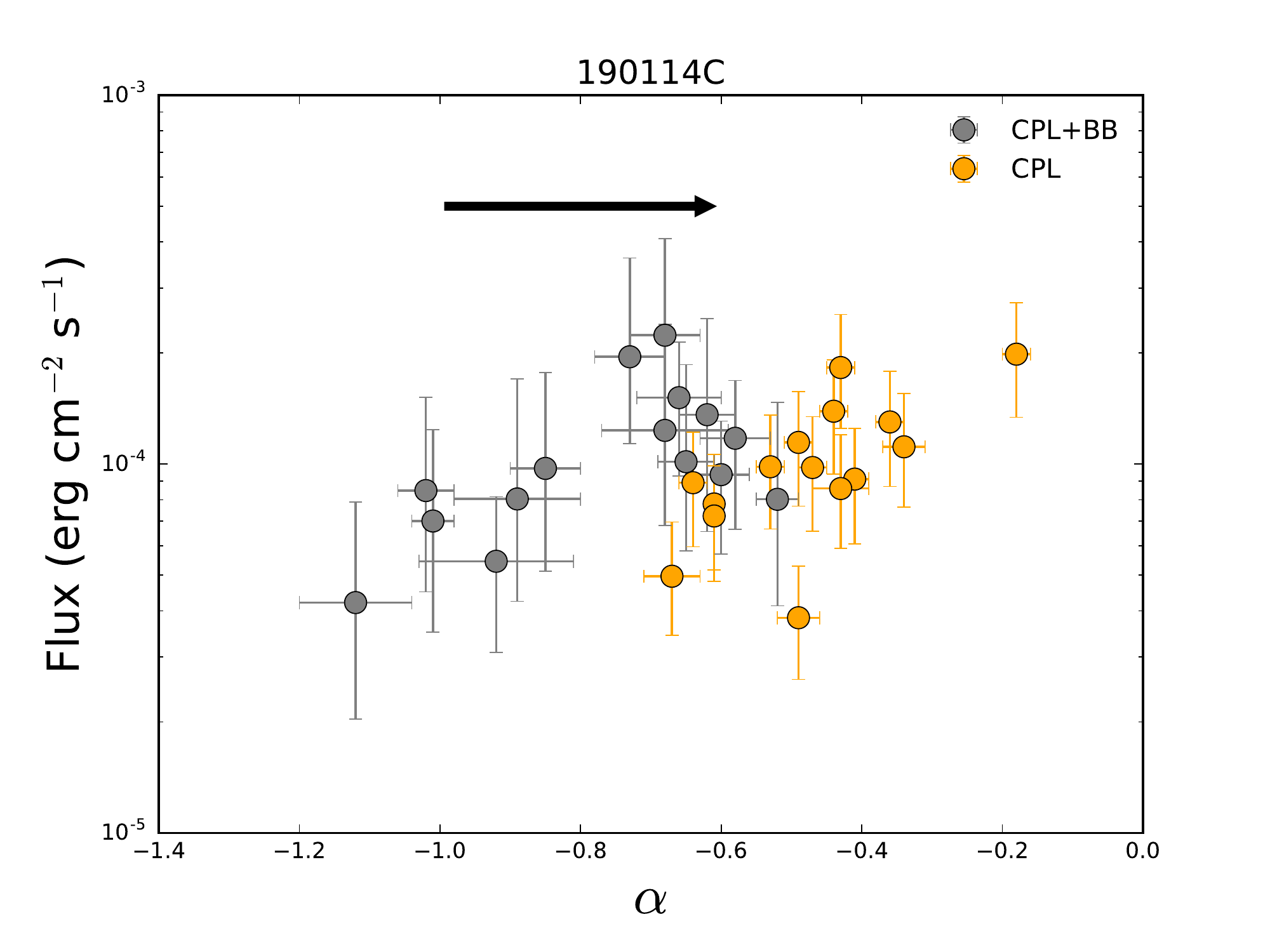}
\caption{Relation of $F-\alpha$. Data points with grey and orange indicate the CPL+BB model and the CPL alone model, respectively. The trend from CPL+BB to CPL alone is marked by an arrow for each burst.}
\end{figure*}\label{FluxAlpha}

\begin{figure*}
\includegraphics[angle=0,scale=0.45]{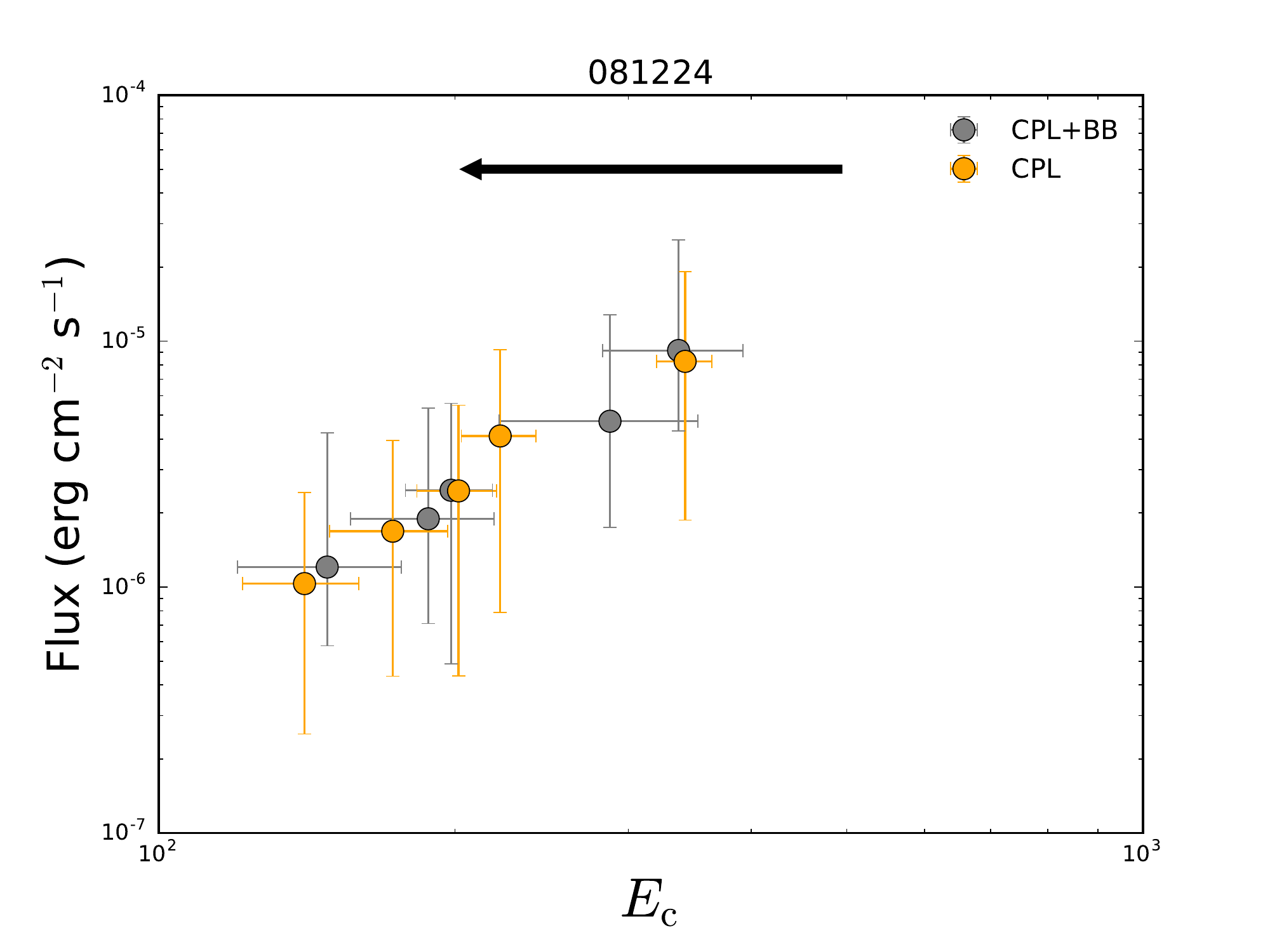}
\includegraphics[angle=0,scale=0.45]{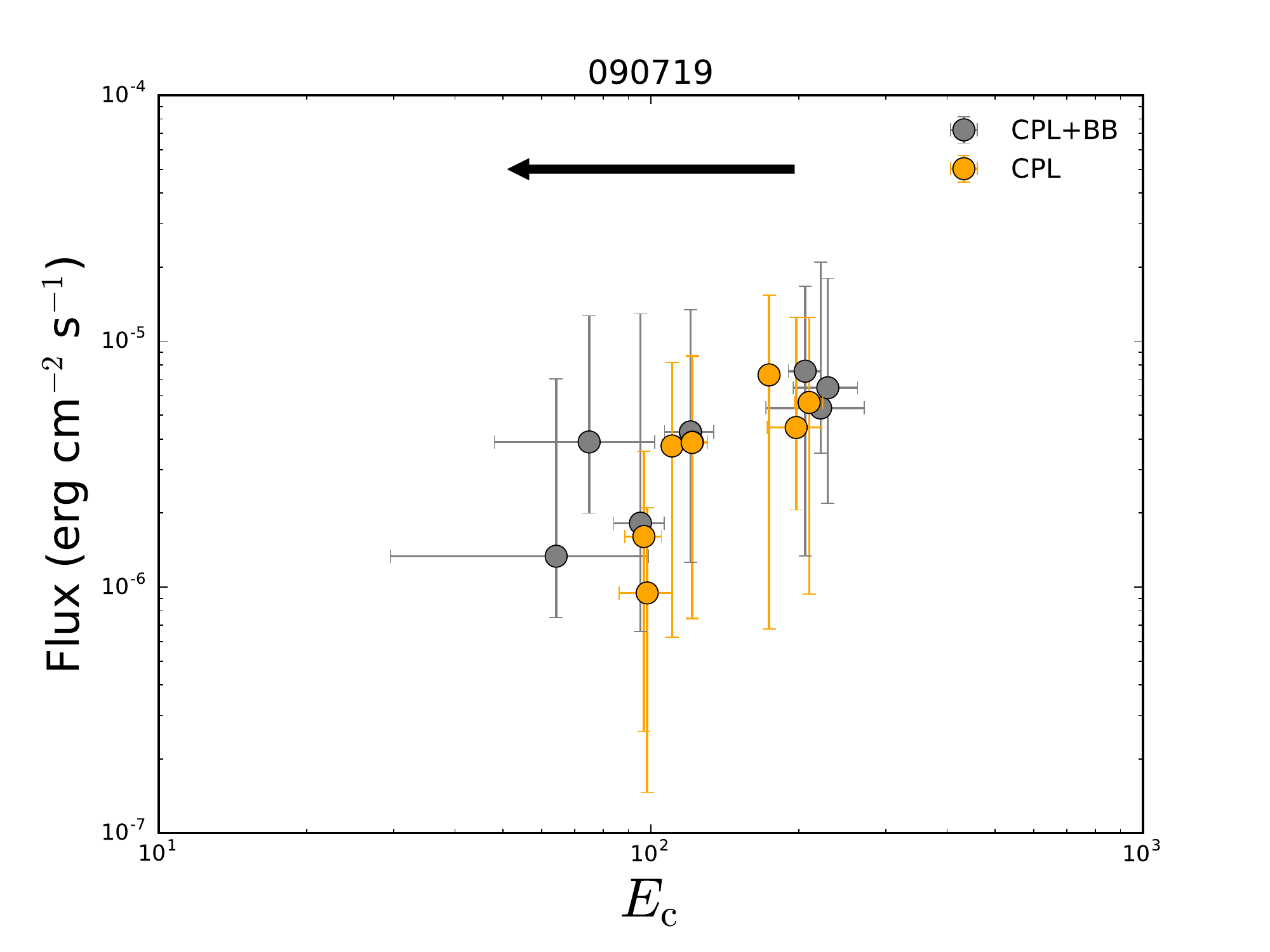}
\includegraphics[angle=0,scale=0.45]{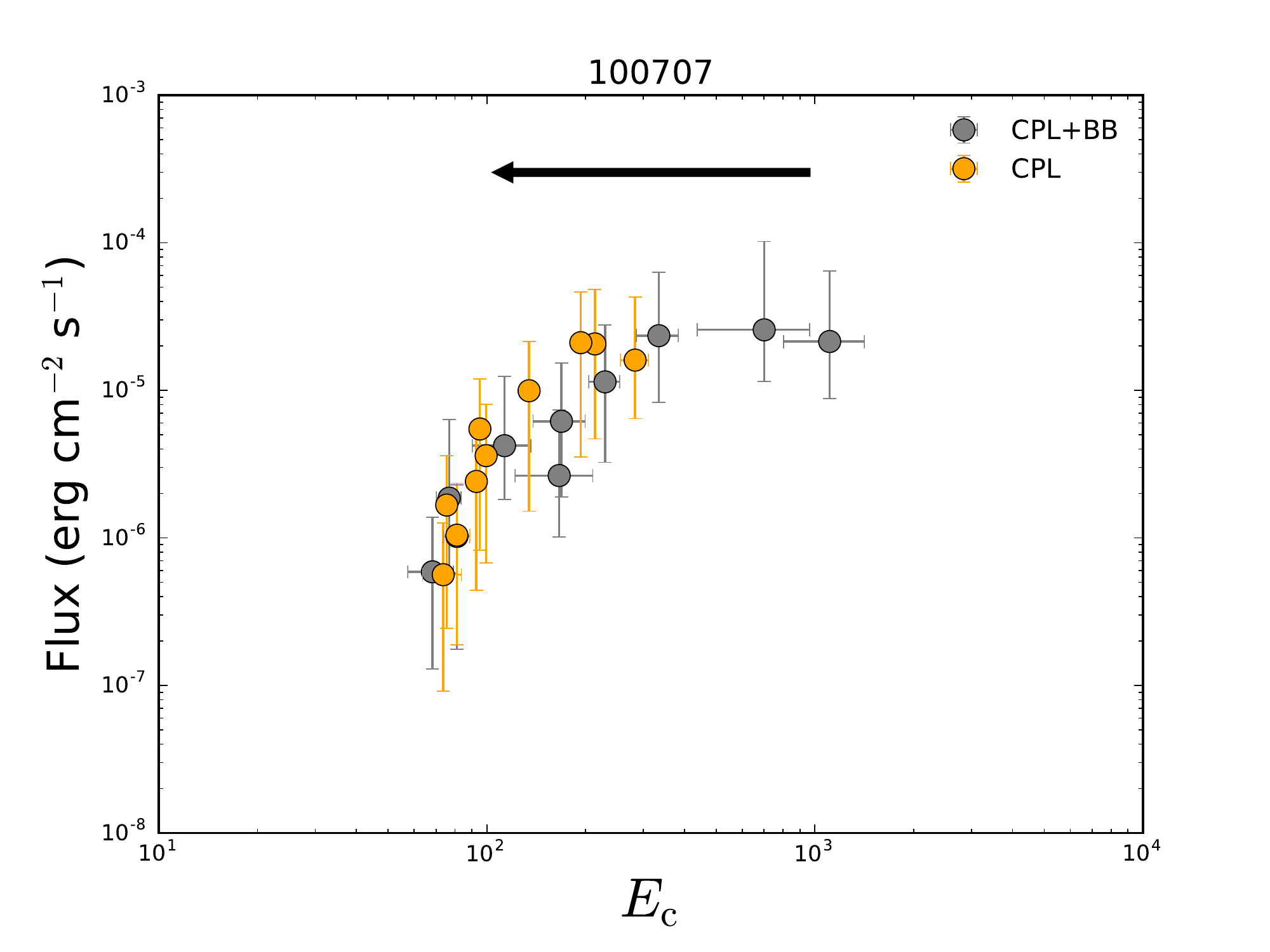}
\includegraphics[angle=0,scale=0.45]{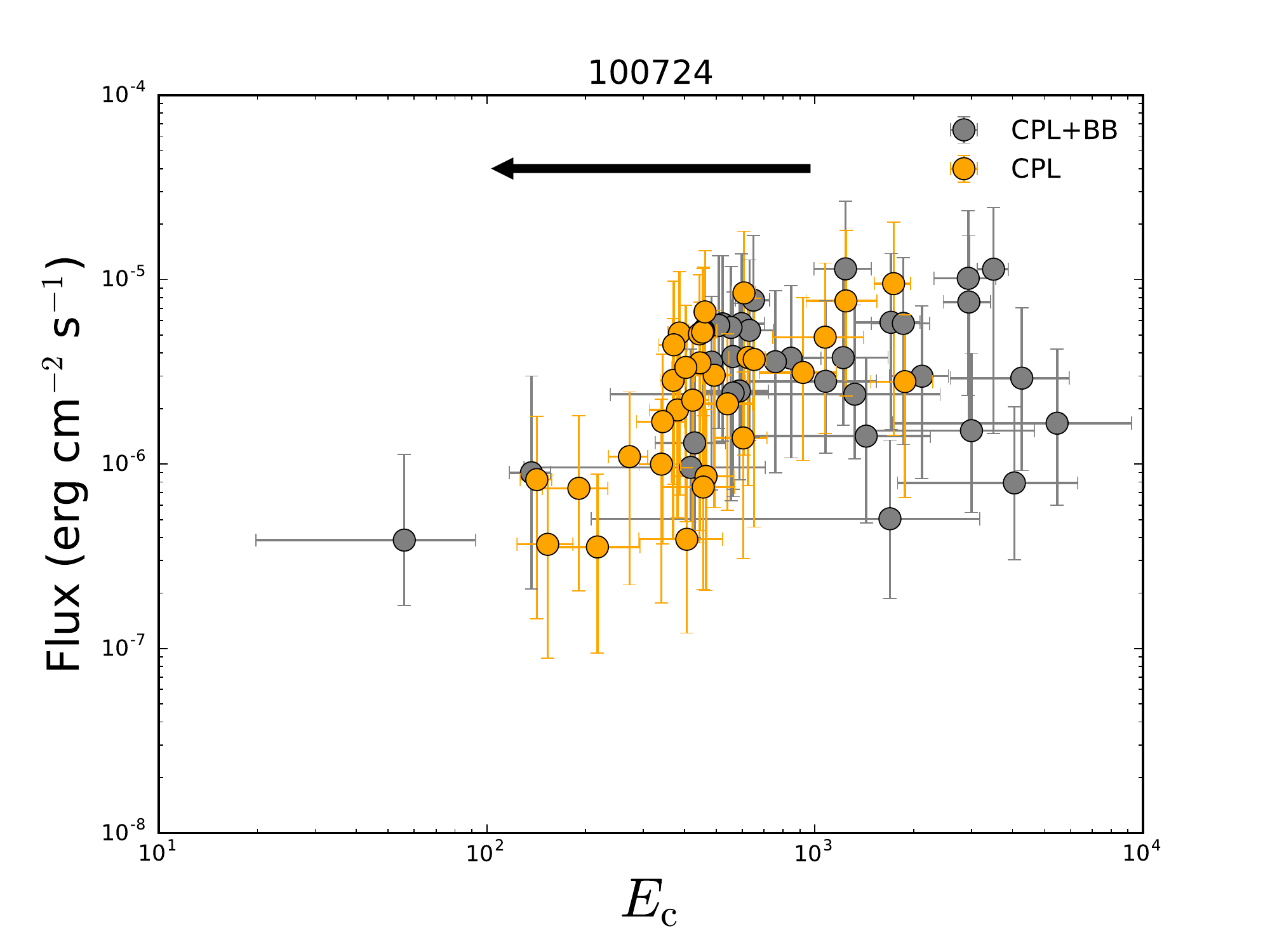}
\includegraphics[angle=0,scale=0.45]{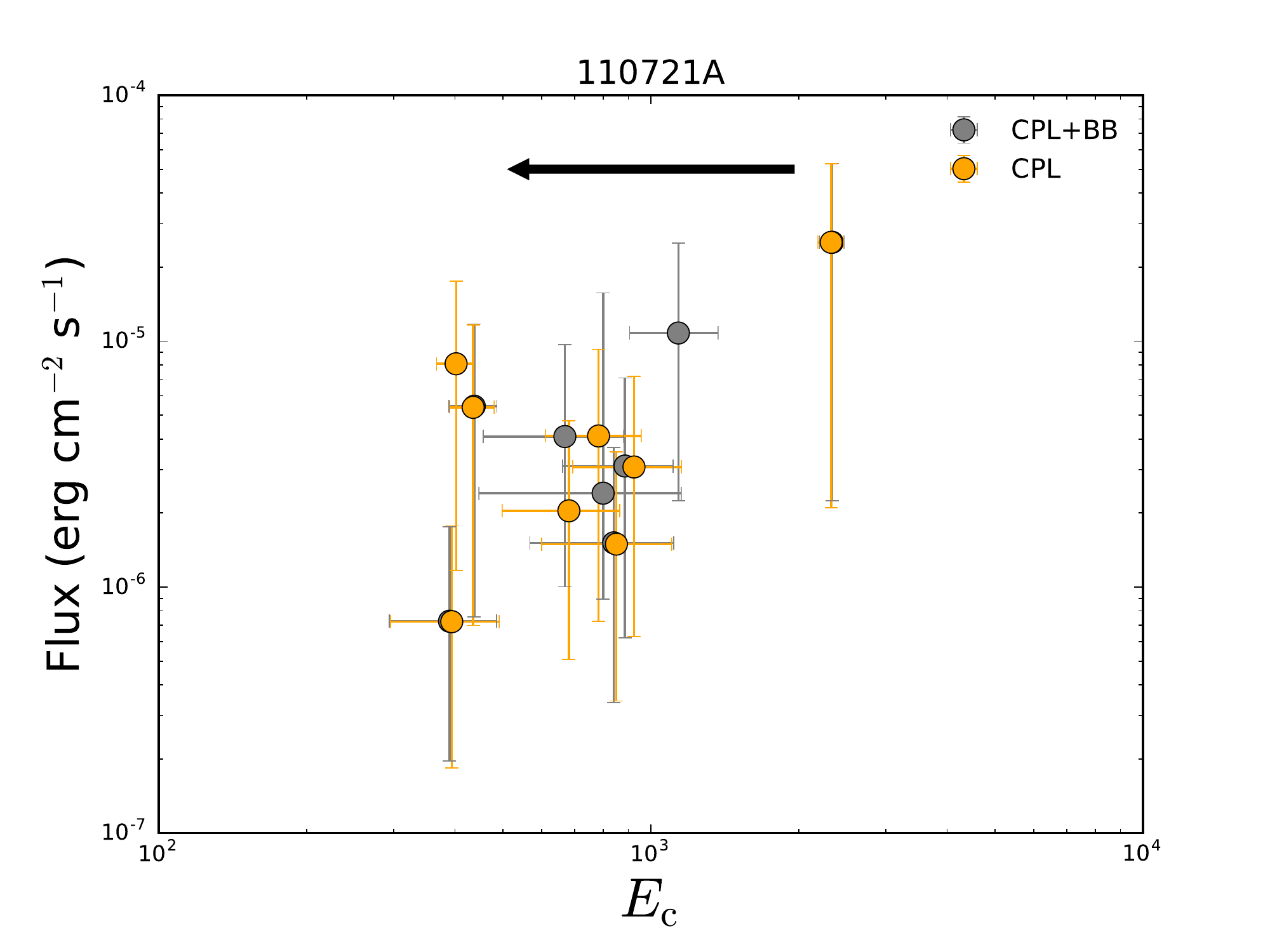}
\includegraphics[angle=0,scale=0.45]{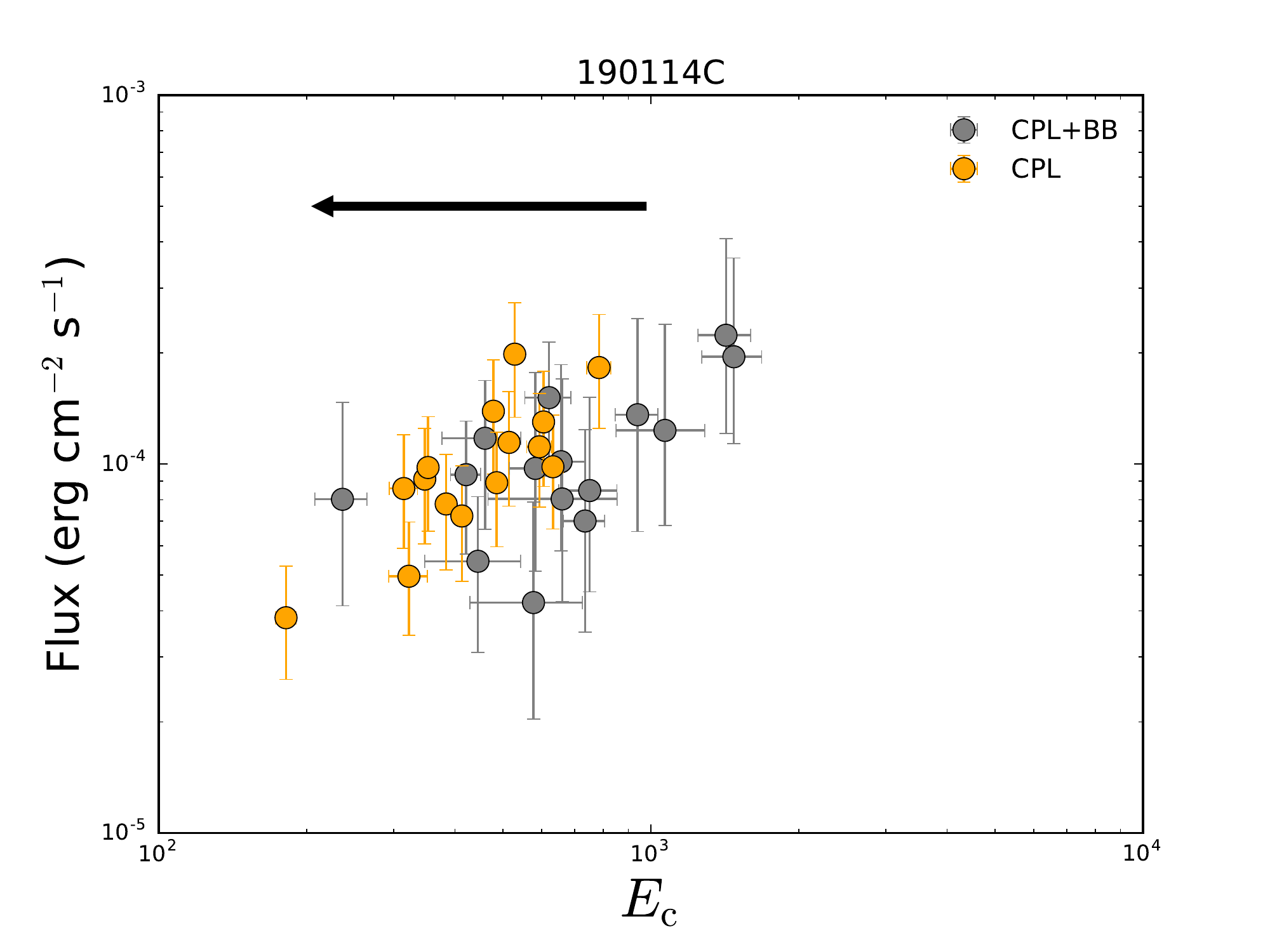}
\caption{Same as Figure \ref{FluxAlpha}, but for $F-E_{\rm c}$ relation.}
\end{figure*}\label{FluxEp}

\begin{figure*}
\includegraphics[angle=0,scale=0.45]{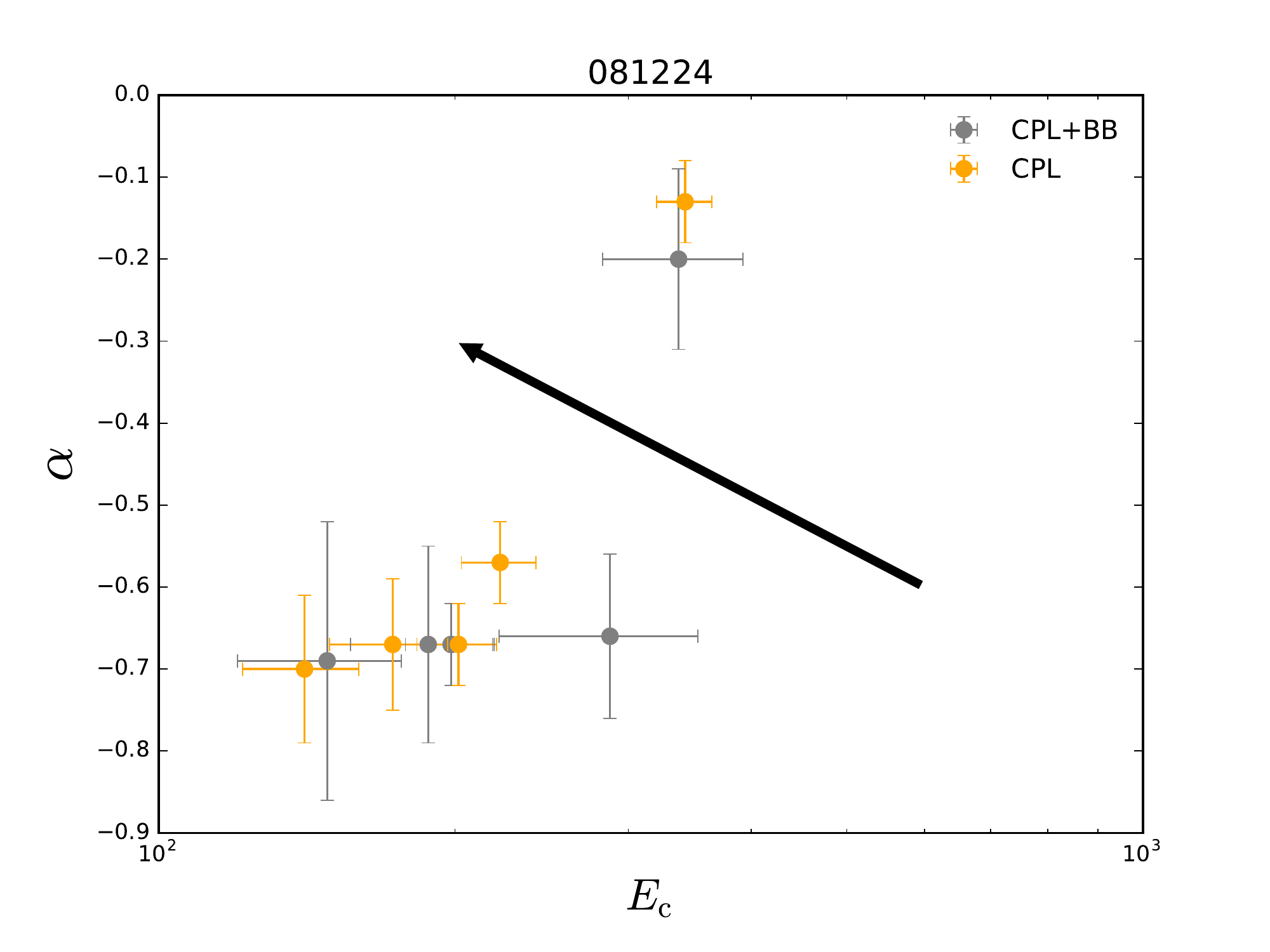}
\includegraphics[angle=0,scale=0.45]{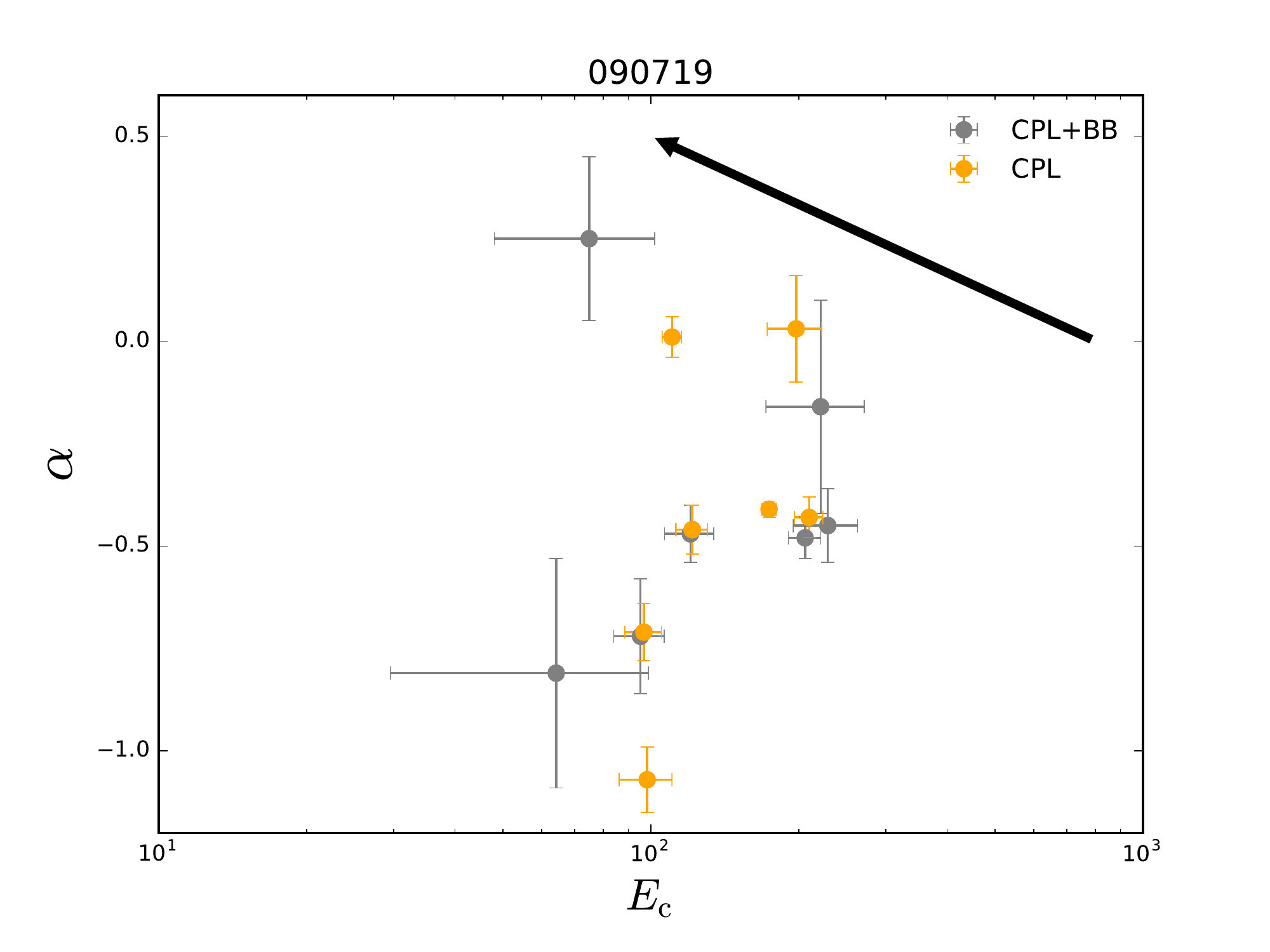}
\includegraphics[angle=0,scale=0.45]{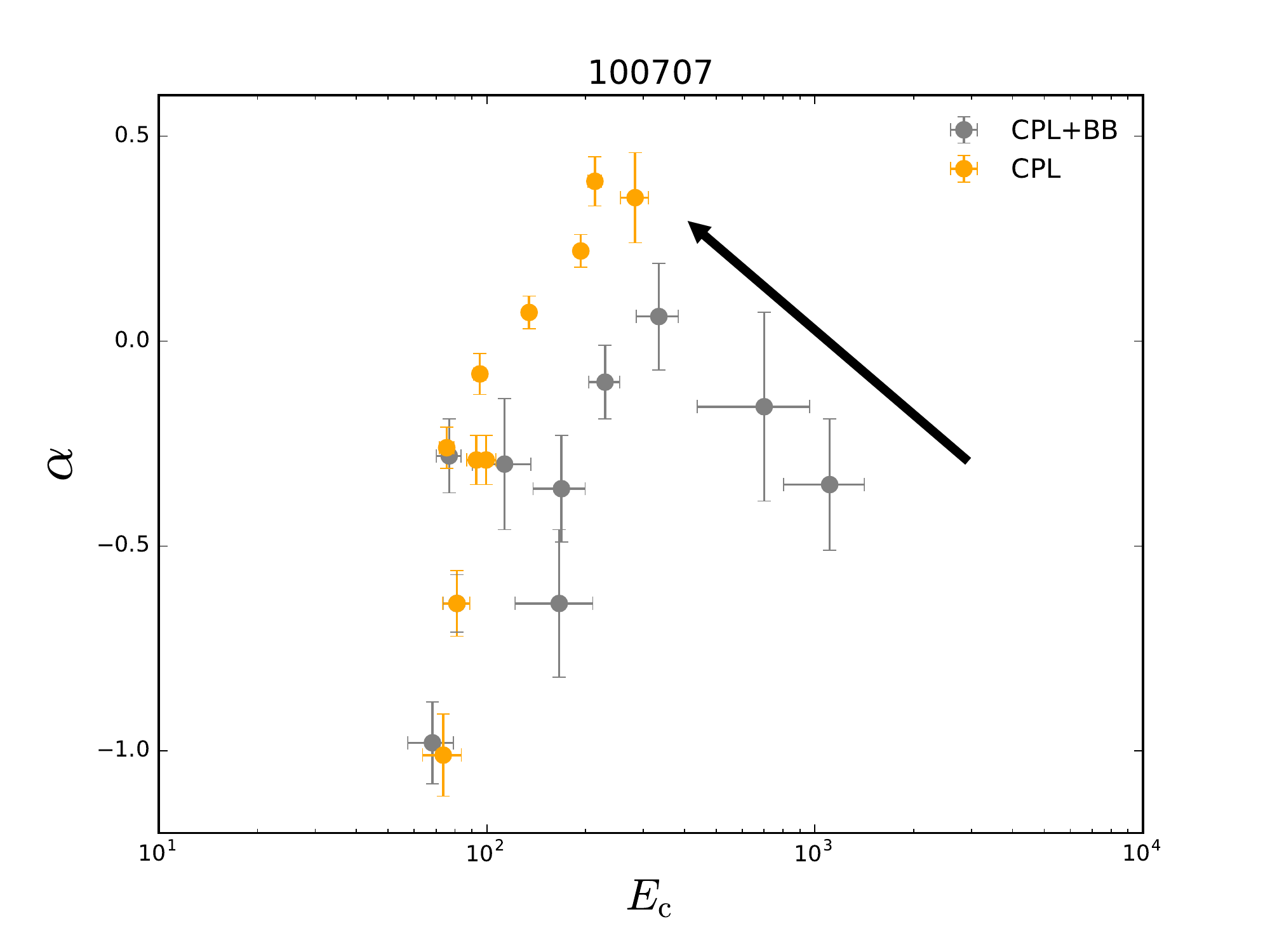}
\includegraphics[angle=0,scale=0.45]{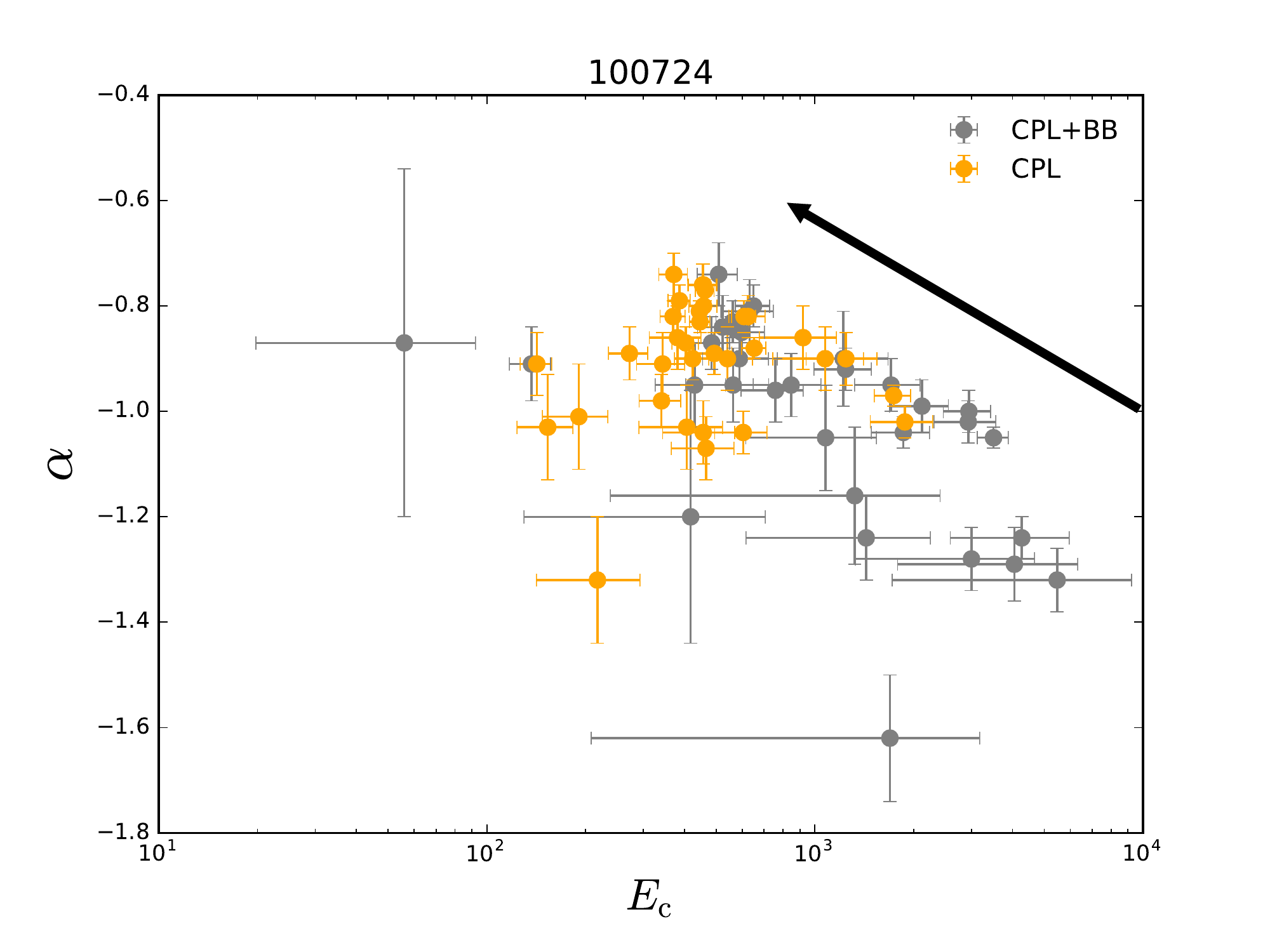}
\includegraphics[angle=0,scale=0.45]{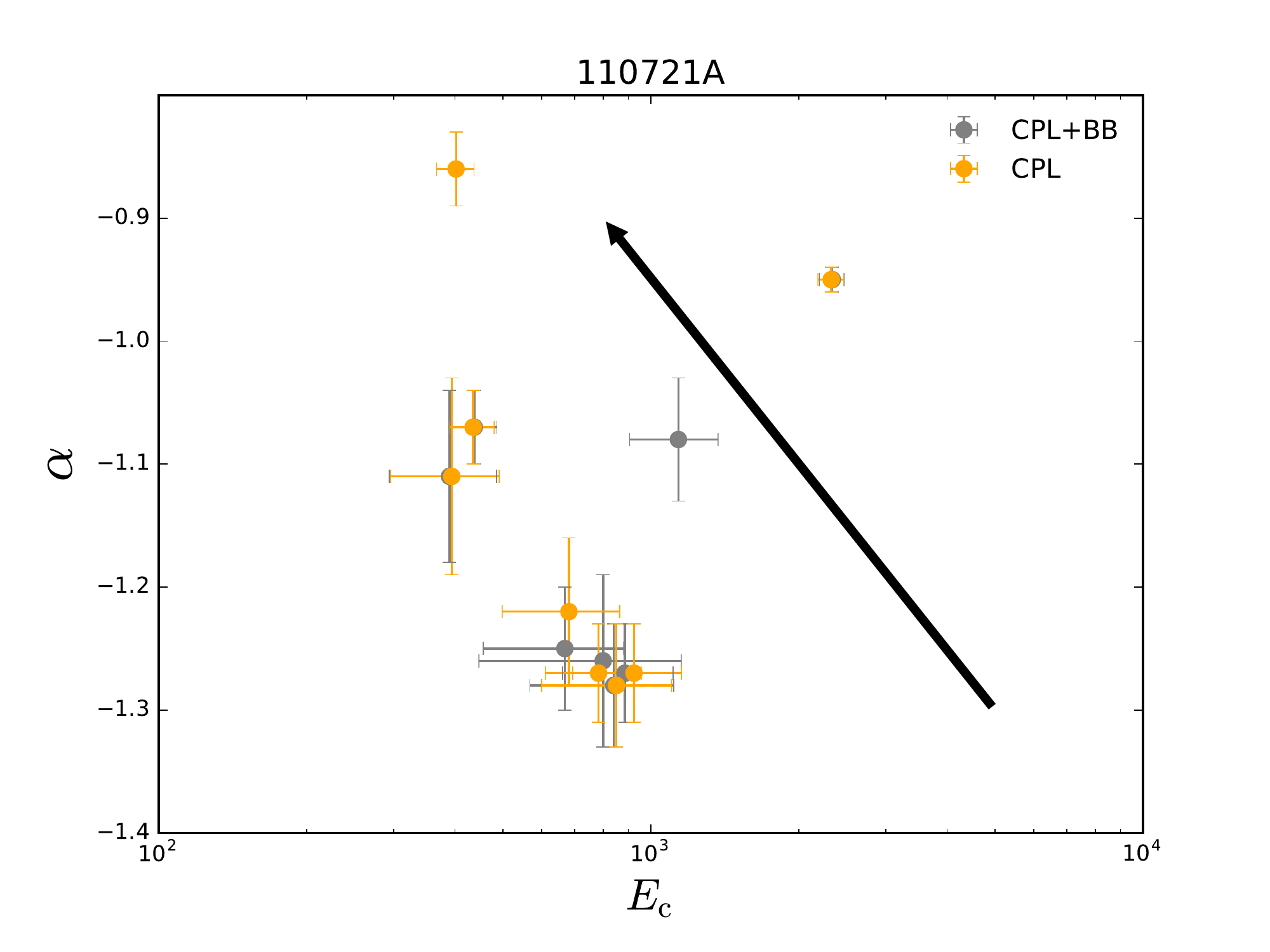}
\includegraphics[angle=0,scale=0.45]{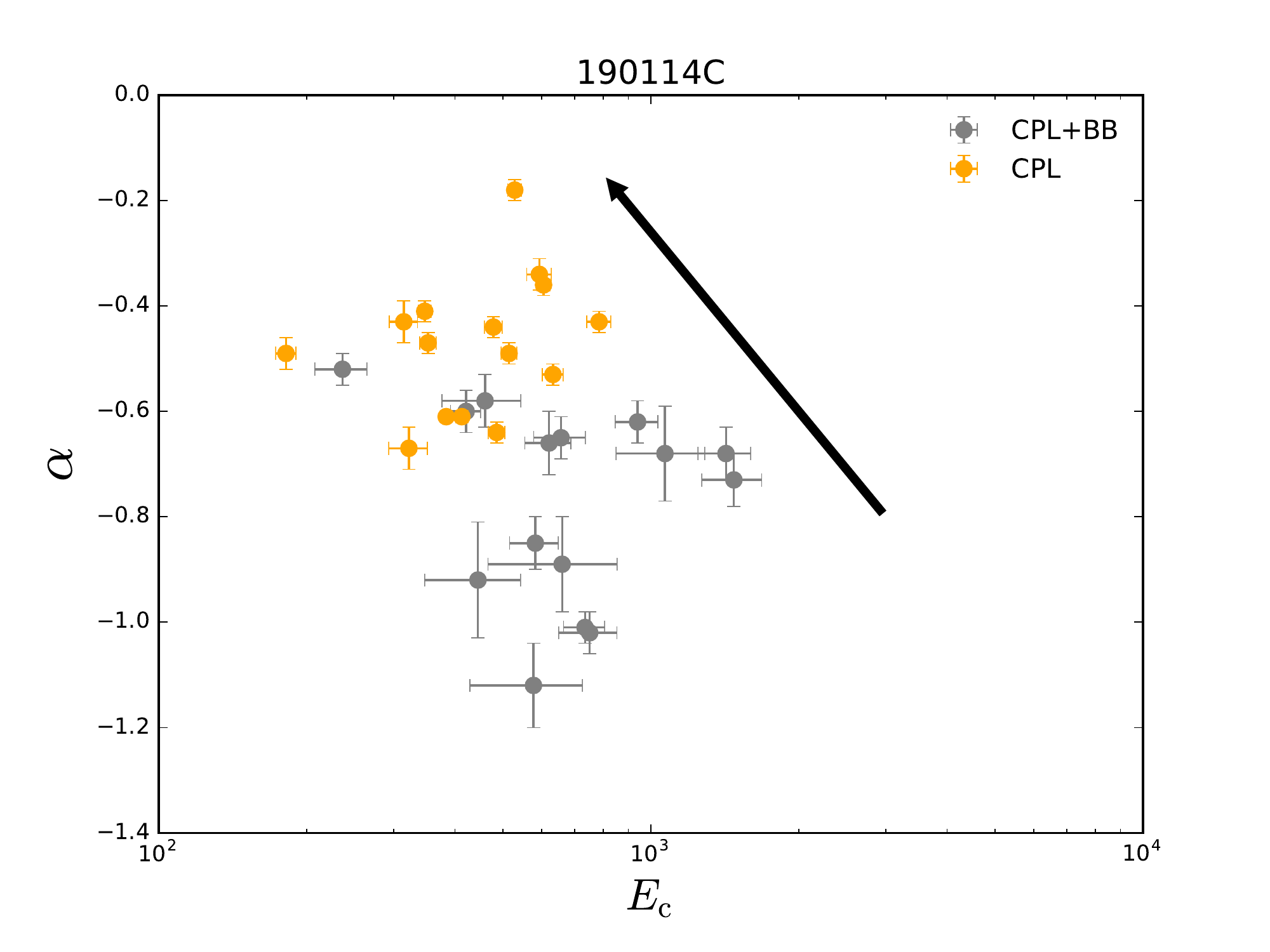}
\caption{Same as Figure \ref{FluxAlpha}, but for $E_{\rm c}$-$\alpha$ relation.}
\end{figure*}\label{EpAlpha}

\clearpage
\begin{figure*}
\includegraphics[angle=0,scale=0.45]{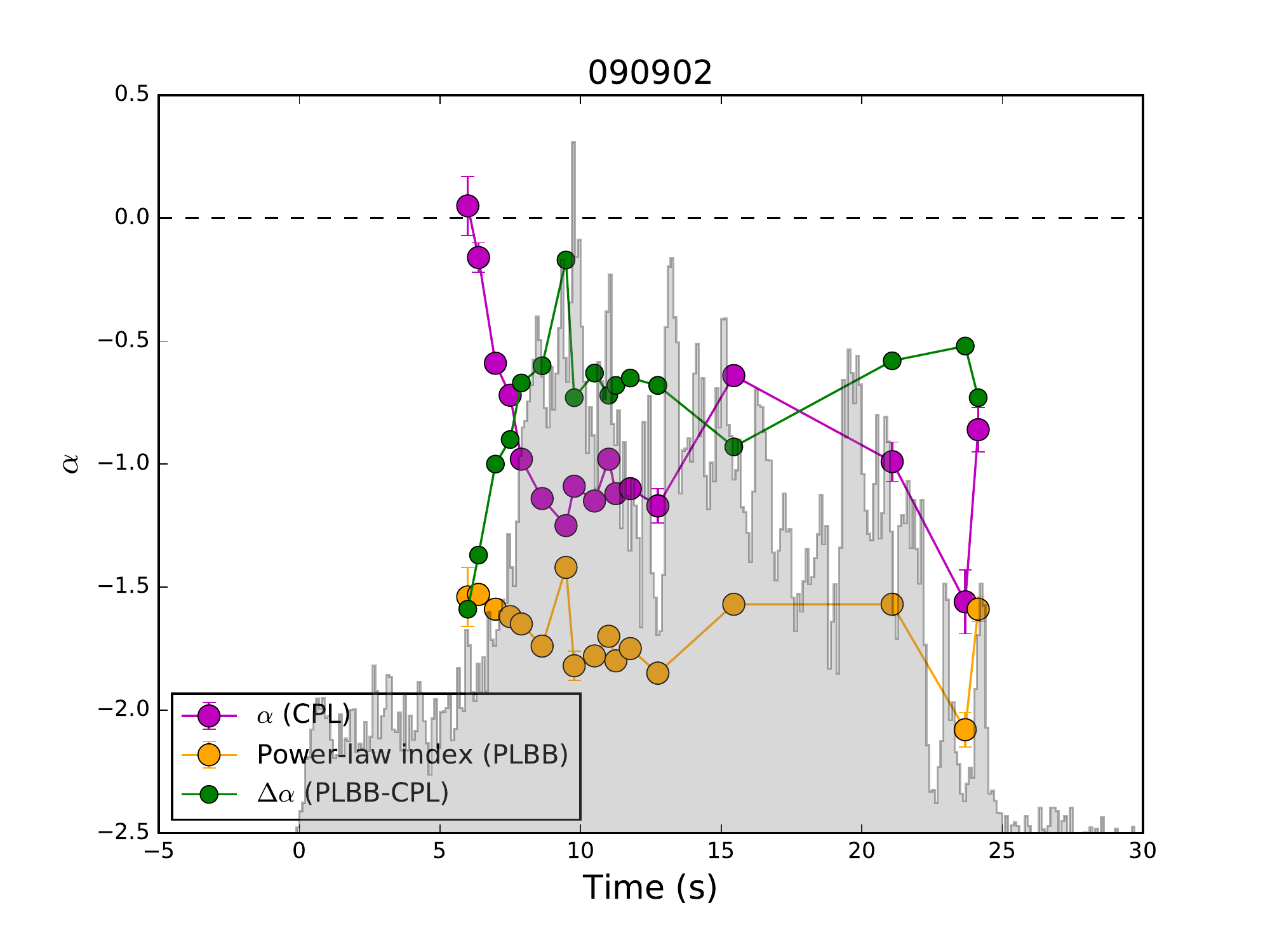}
\includegraphics[angle=0,scale=0.45]{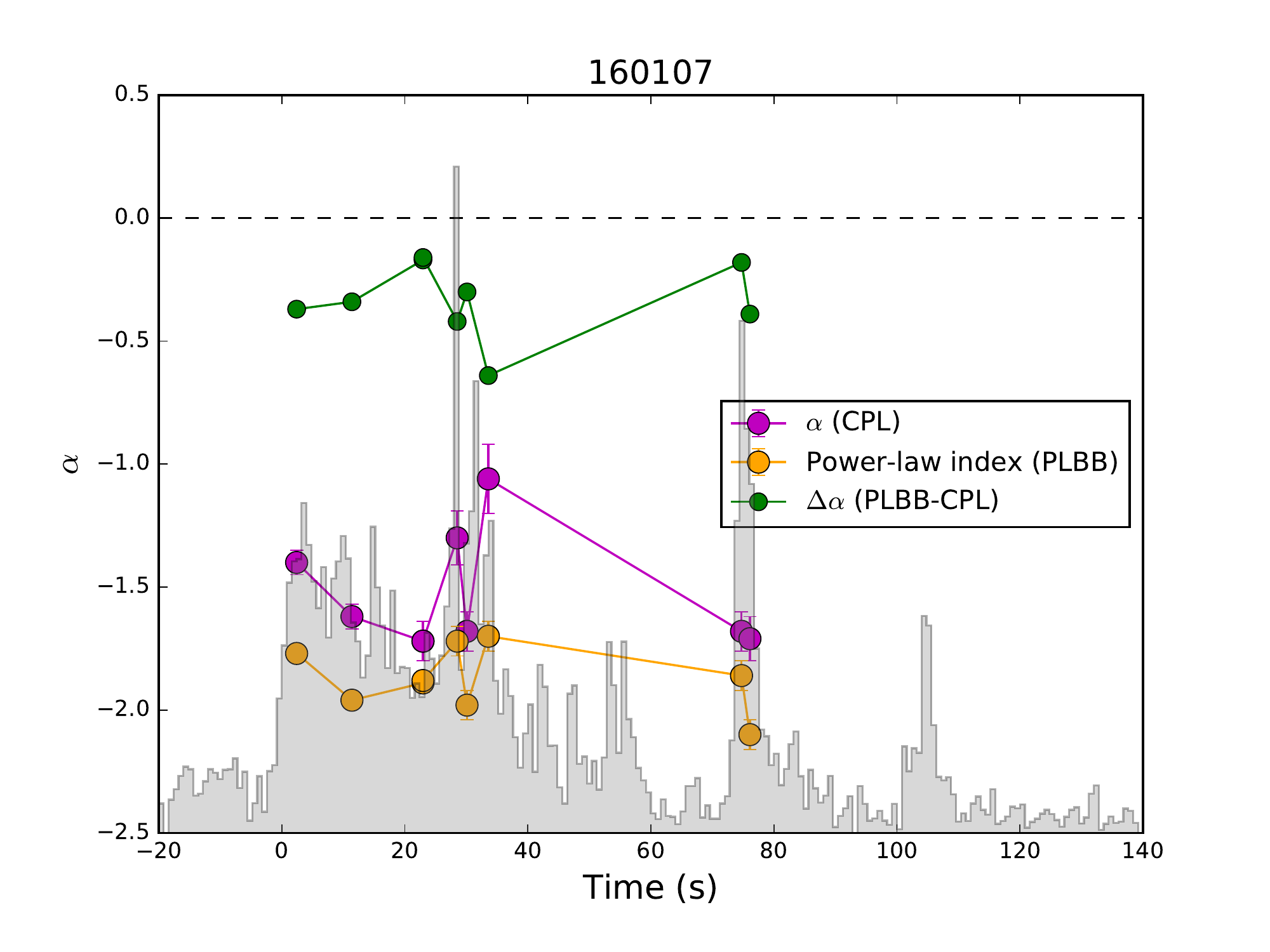}
\caption{Same as Figure \ref{Alpha}, but for Group II bursts.}
\end{figure*}\label{AlphaII}

\begin{figure*}
\includegraphics[angle=0,scale=0.45]{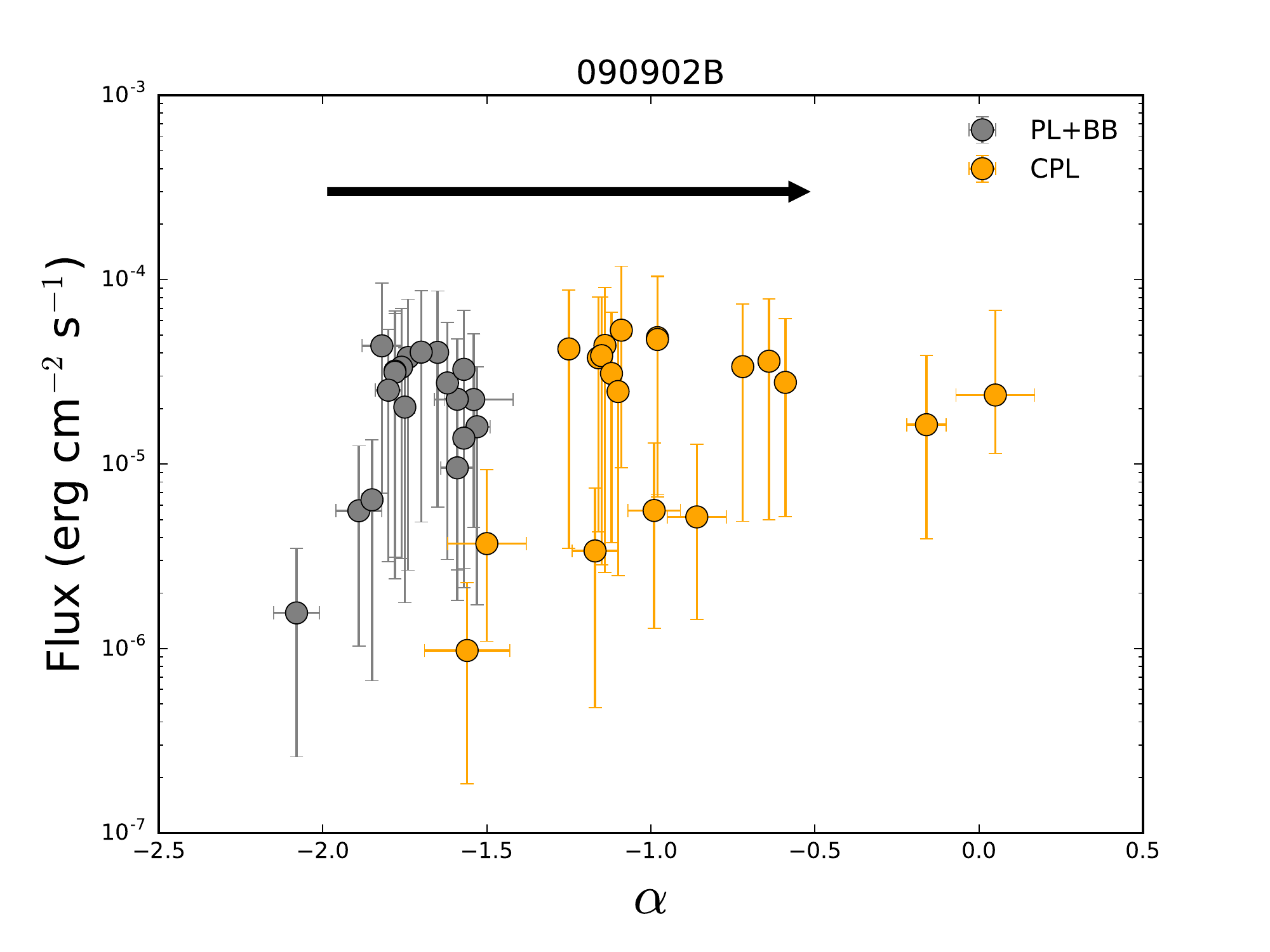}
\includegraphics[angle=0,scale=0.45]{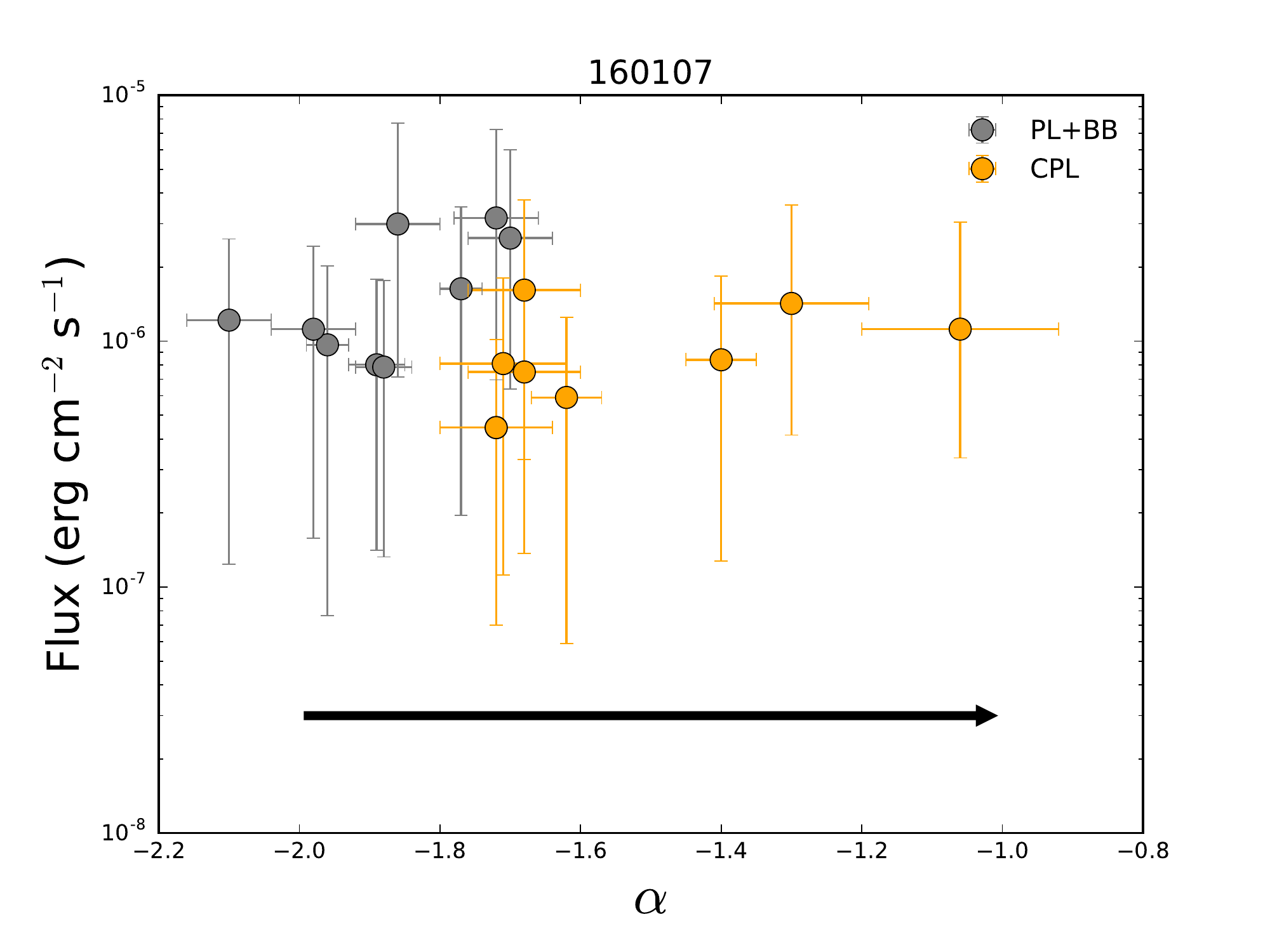}
\caption{Same as Figure \ref{FluxAlpha}, but for Group II bursts.}
\end{figure*}\label{FluxAlphaII}

\begin{figure*}
\includegraphics[angle=0,scale=0.30]{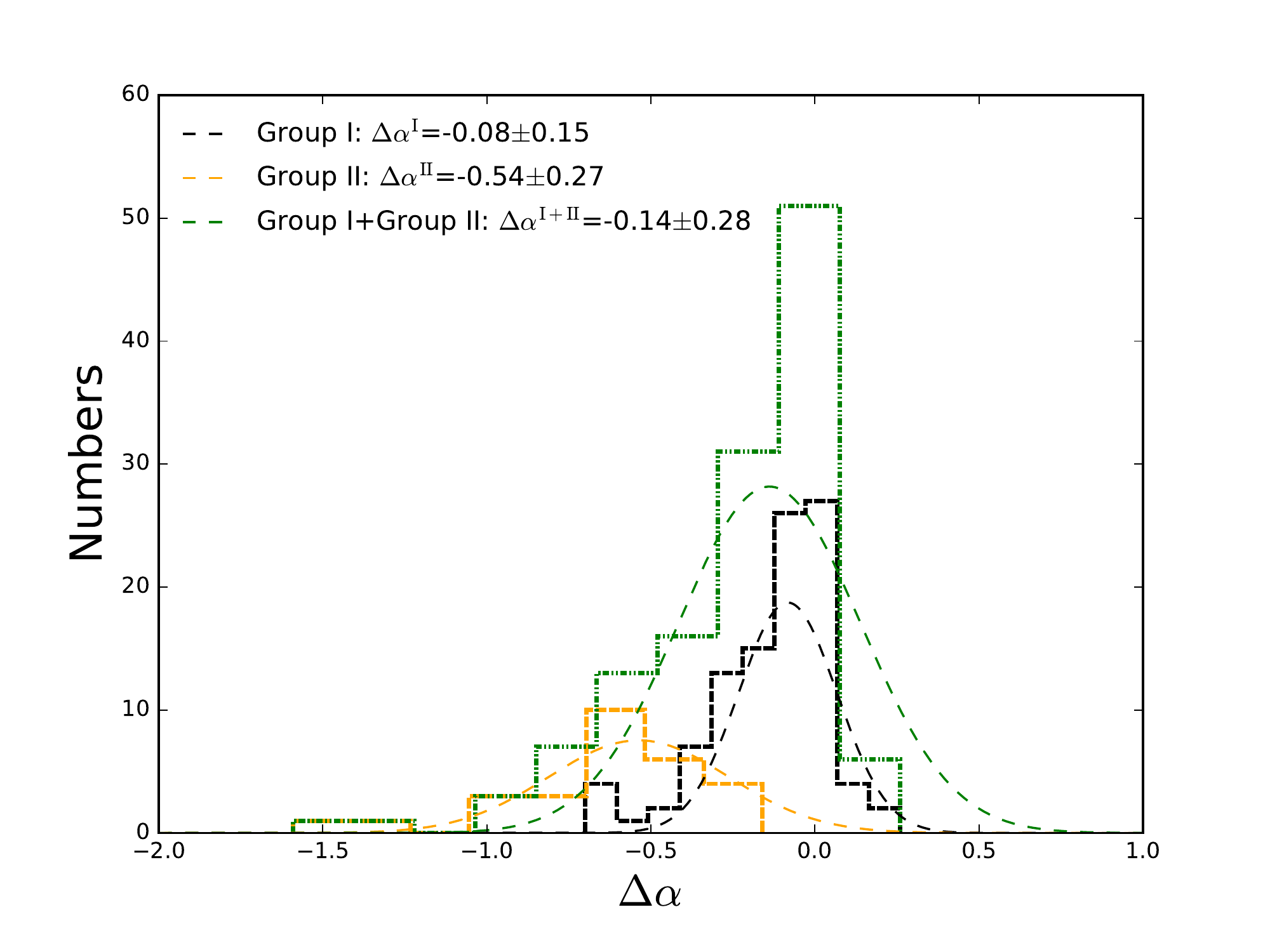}
\includegraphics[angle=0,scale=0.30]{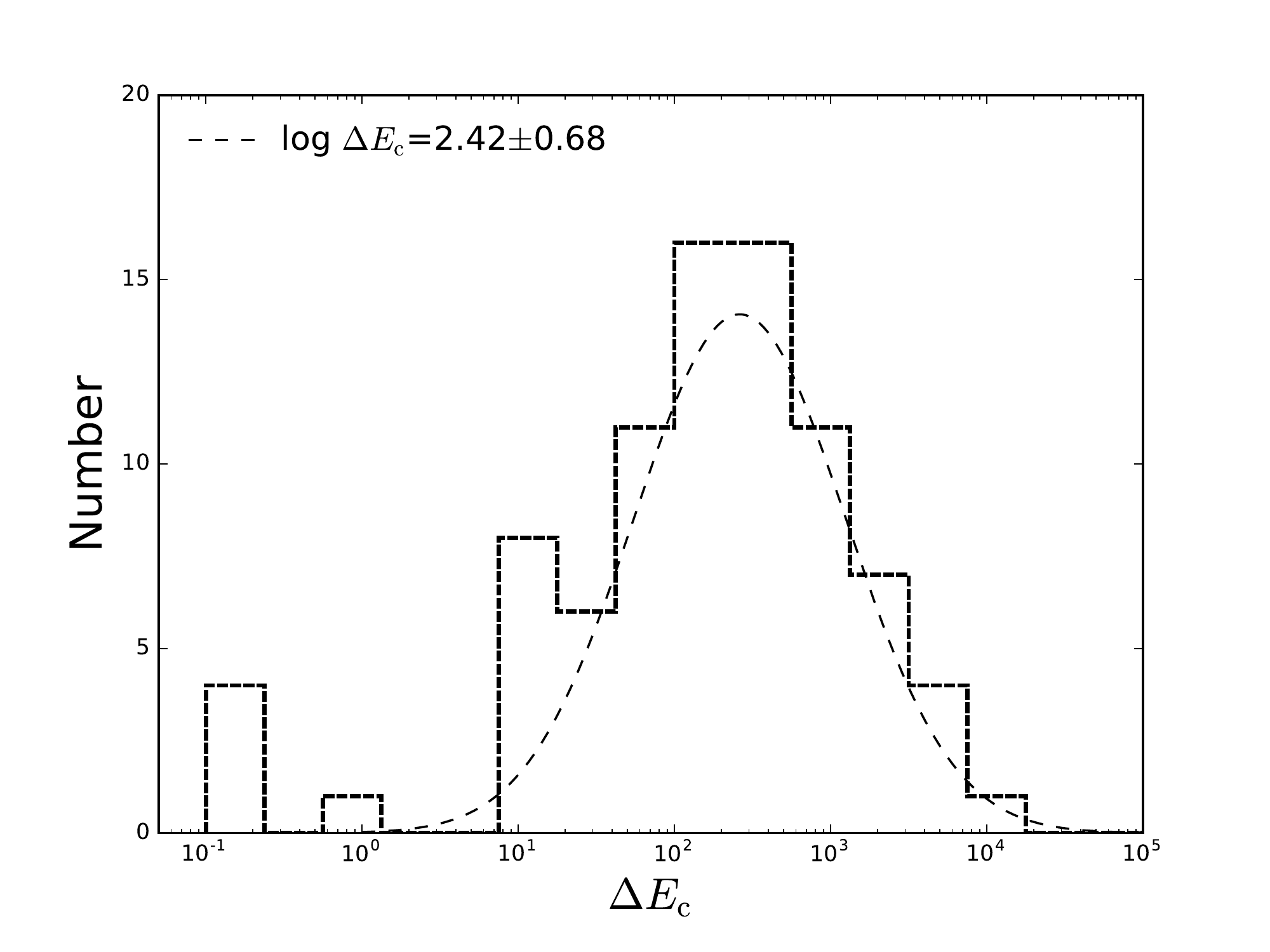}
\includegraphics[angle=0,scale=0.30]{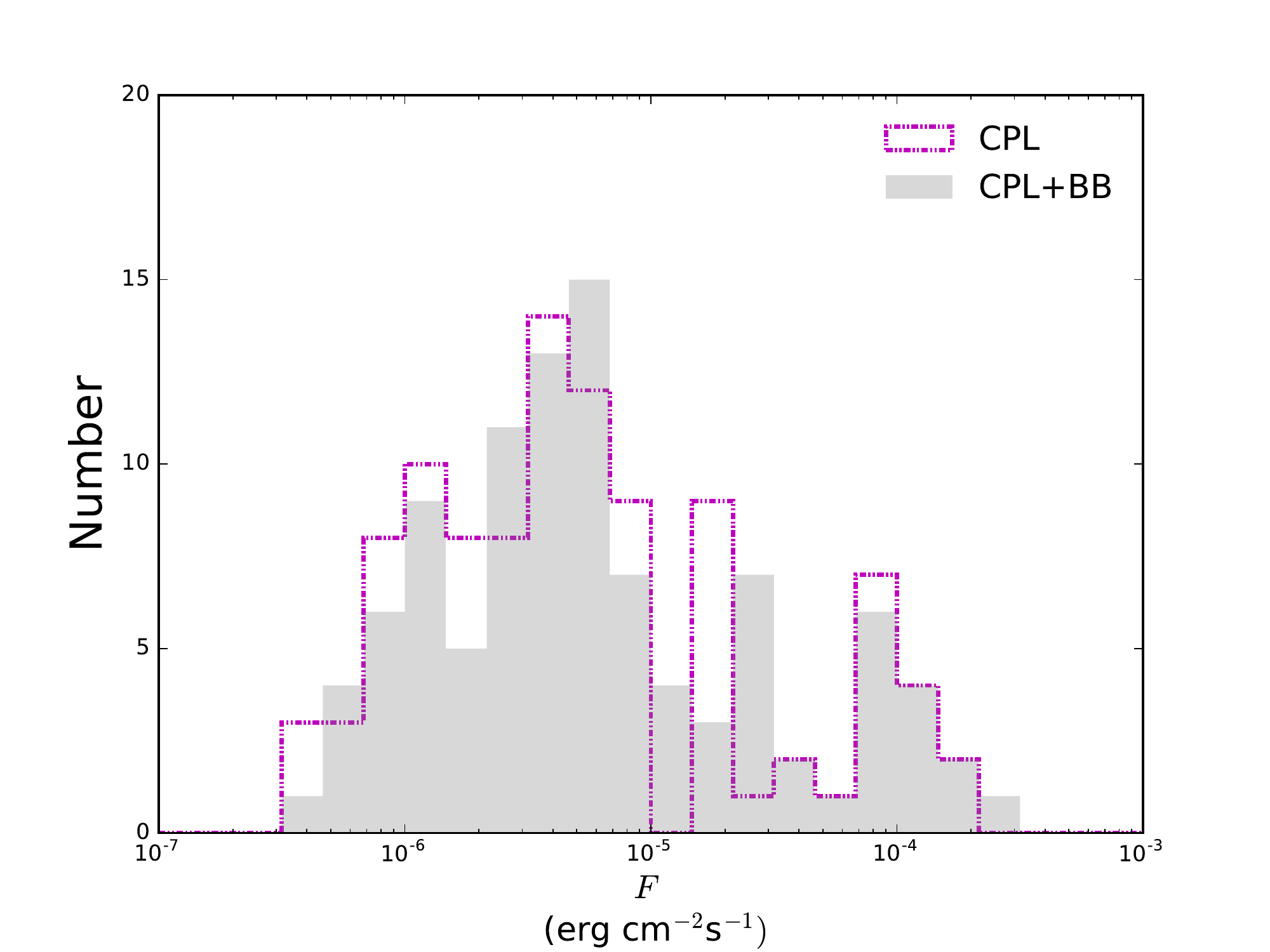}
\includegraphics[angle=0,scale=0.30]{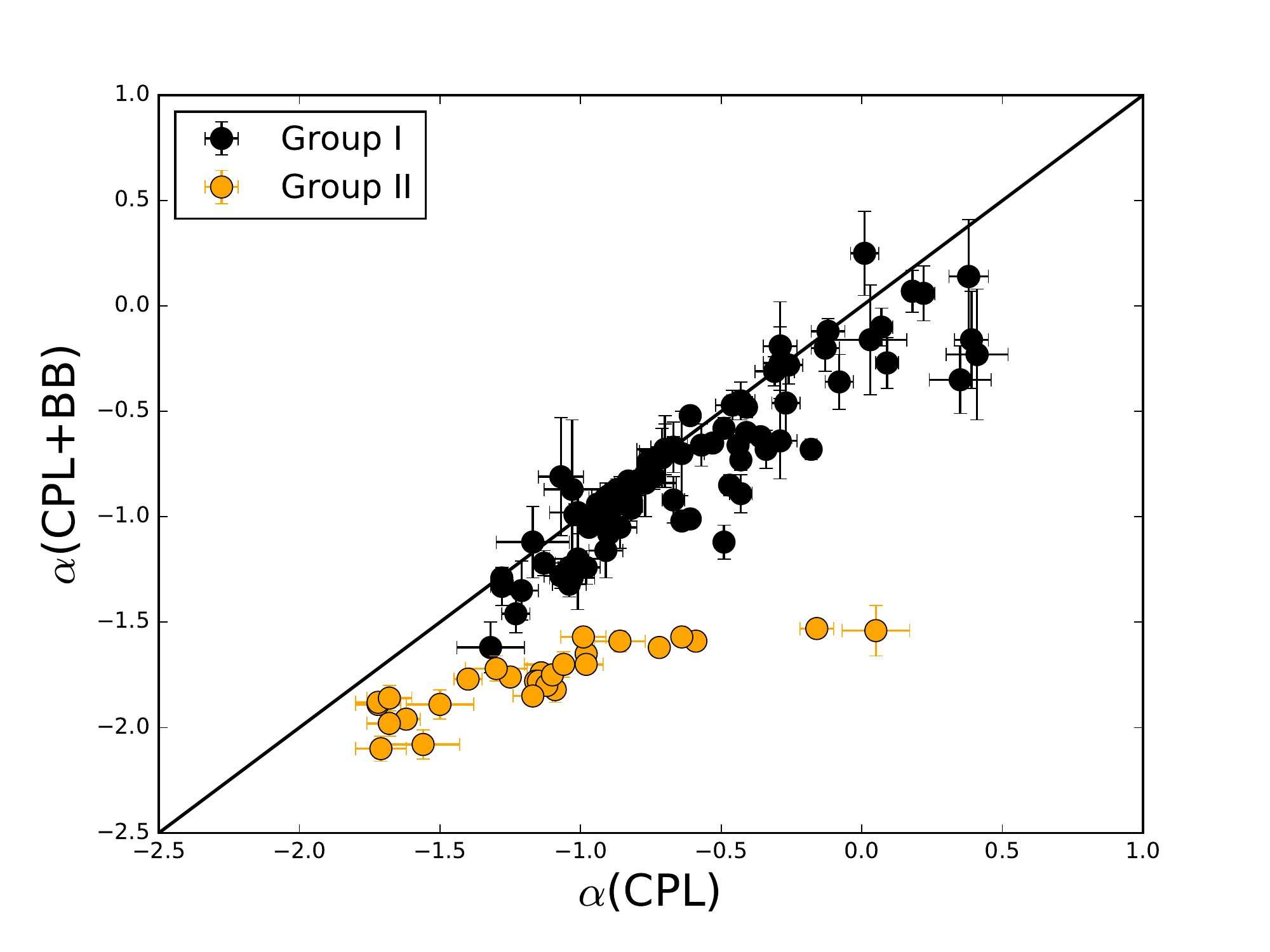}
\includegraphics[angle=0,scale=0.30]{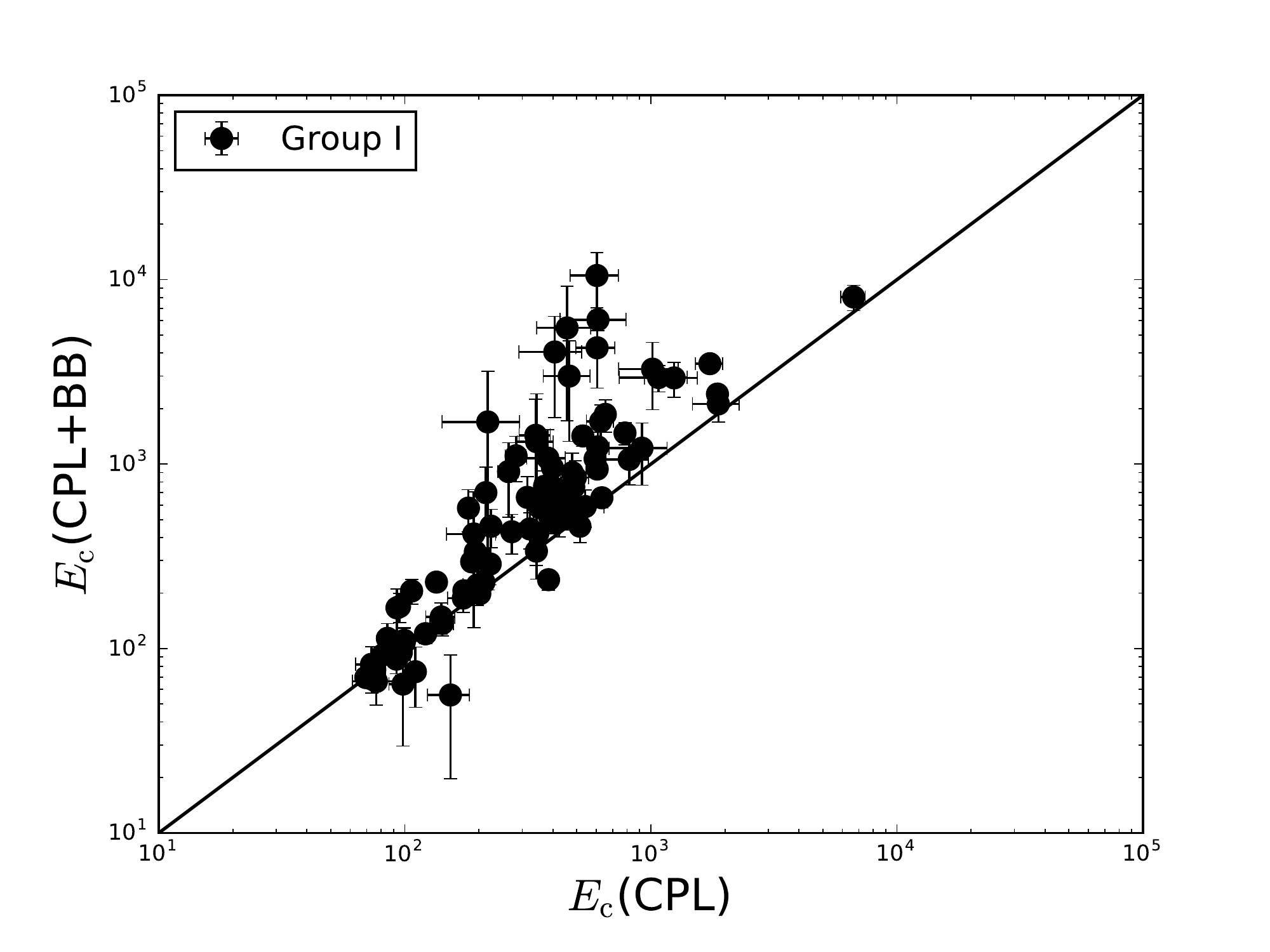}
\includegraphics[angle=0,scale=0.30]{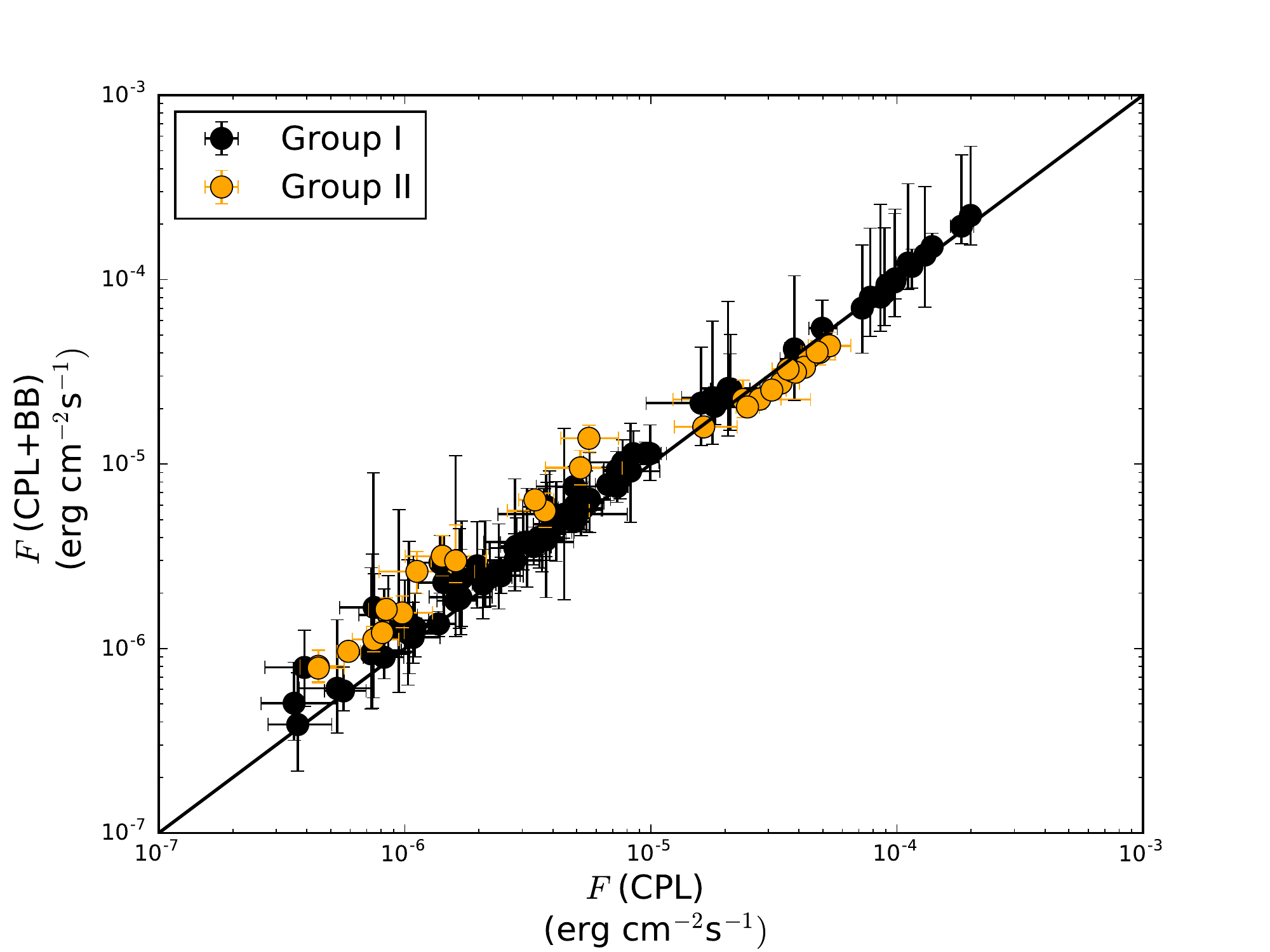}
\includegraphics[angle=0,scale=0.30]{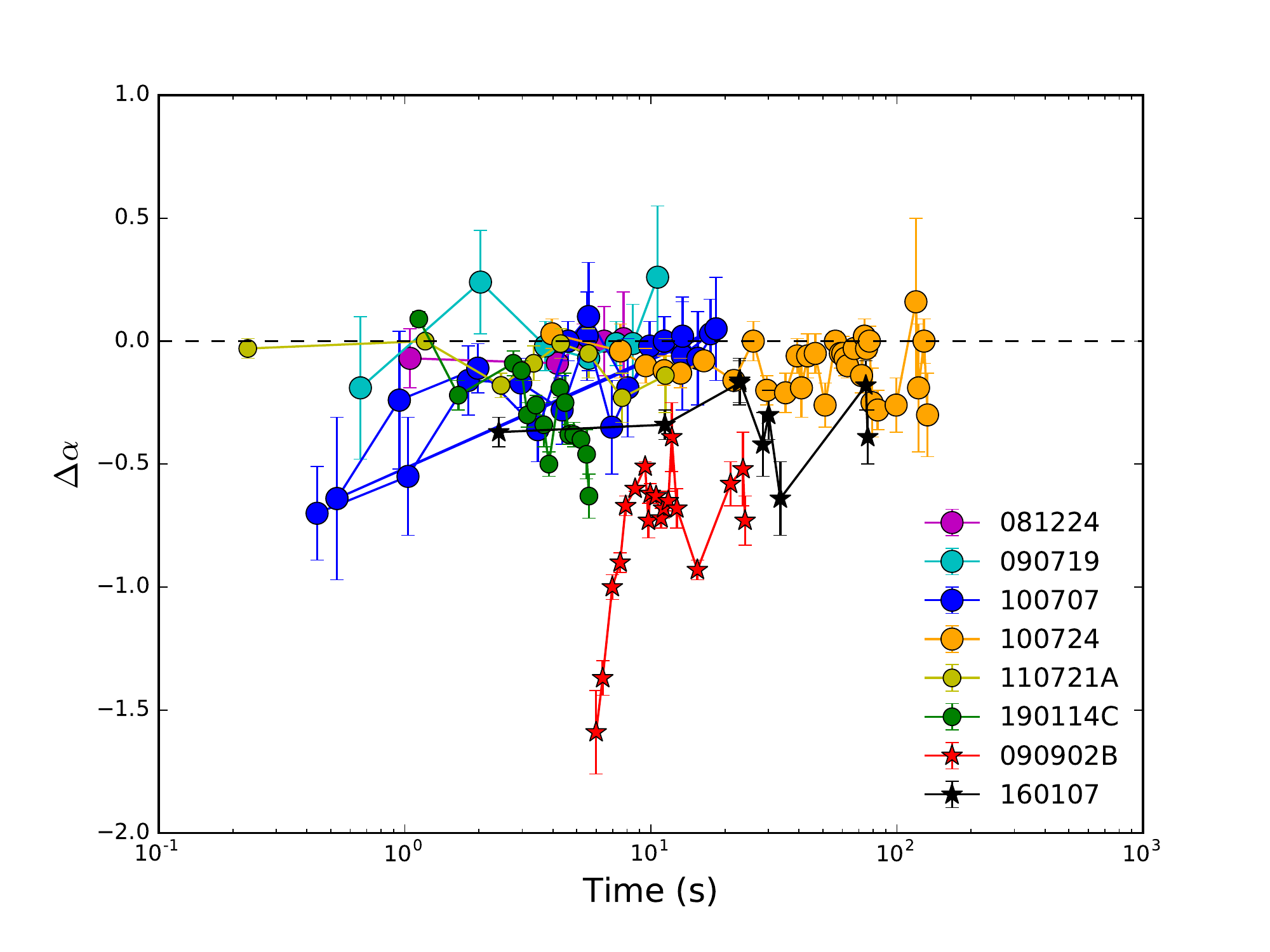}
\includegraphics[angle=0,scale=0.30]{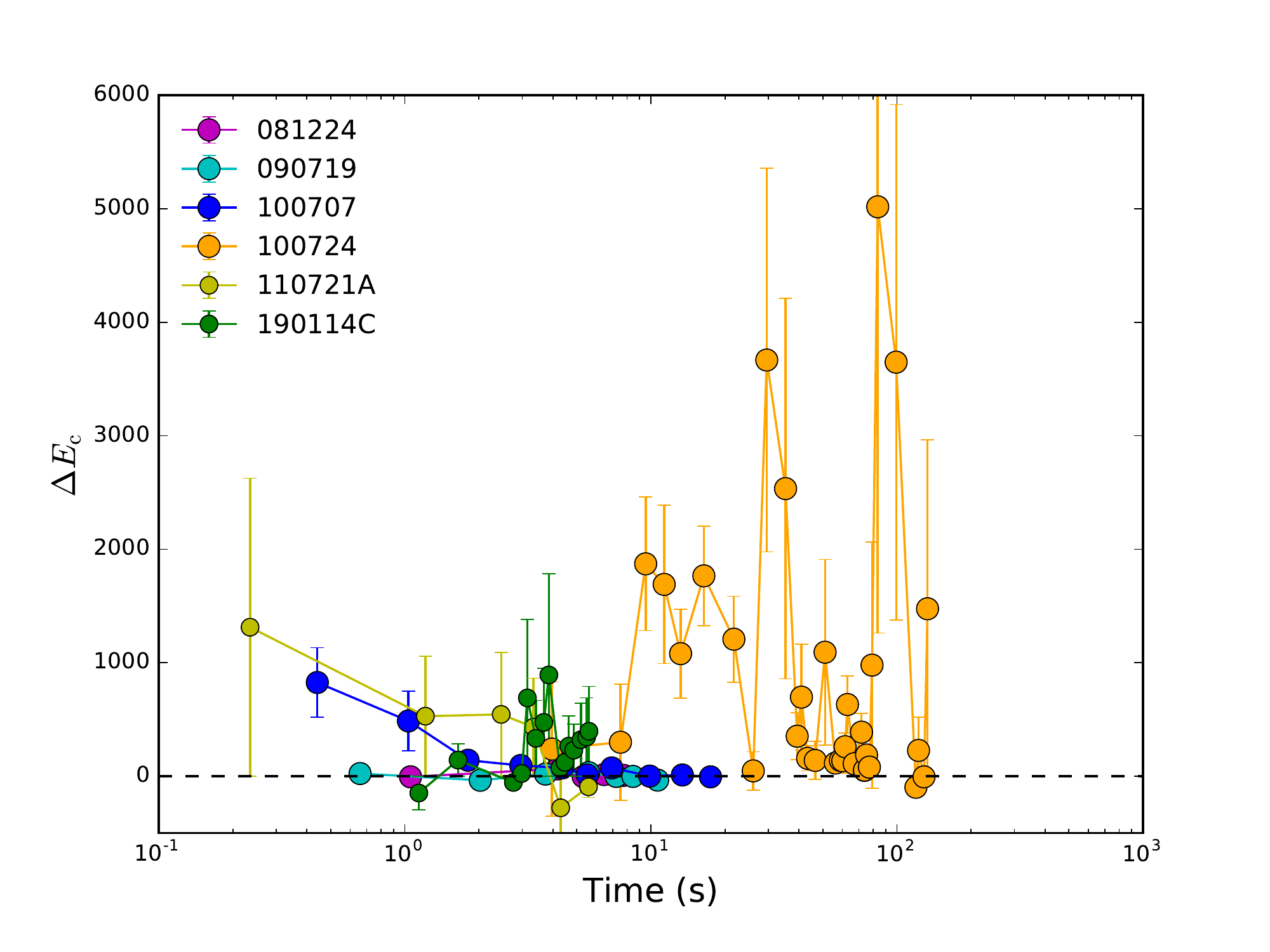}
\includegraphics[angle=0,scale=0.30]{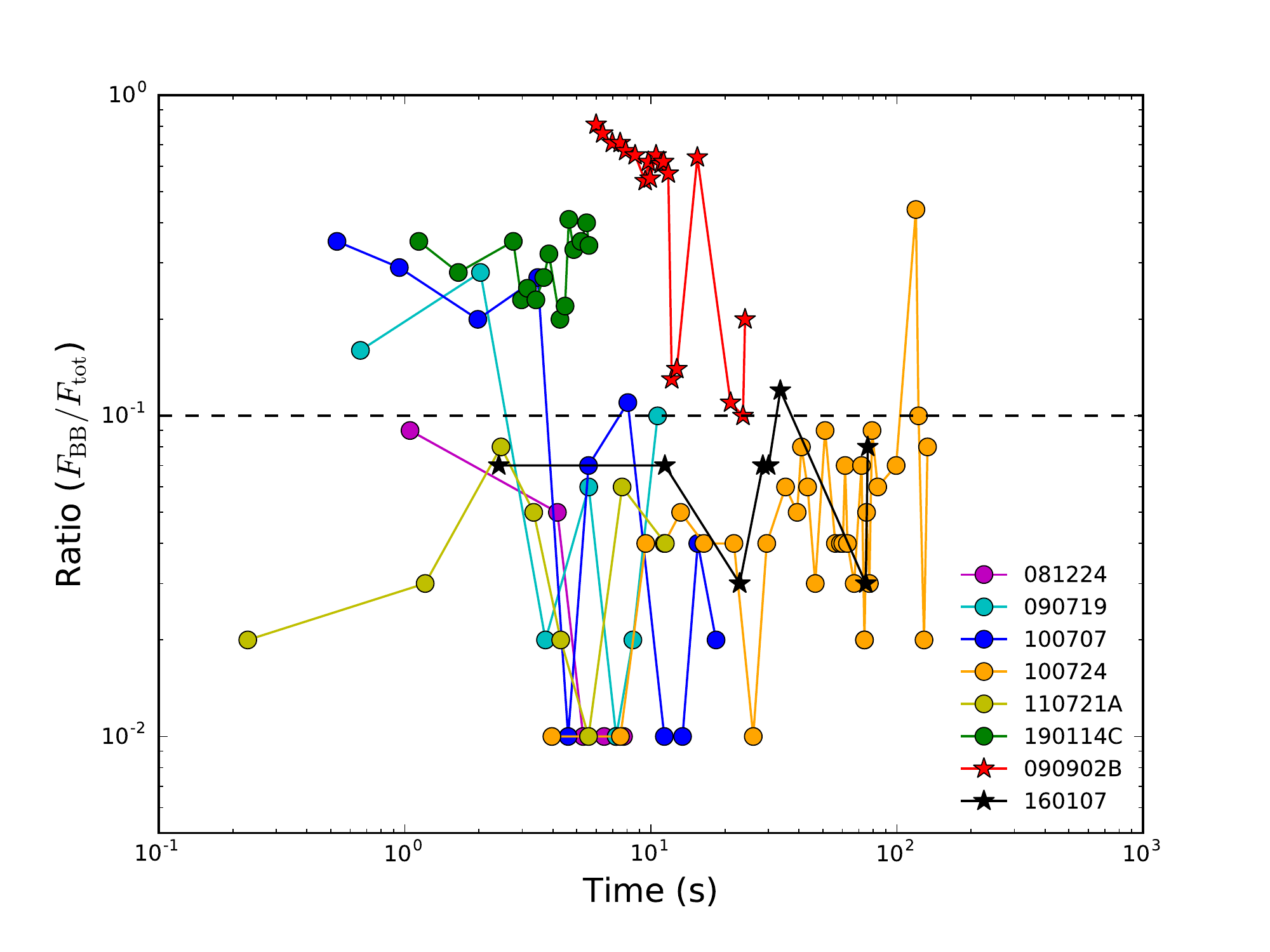}
\caption{Top panels: distributions of $\Delta \alpha$, $\Delta E_{\rm c}$ and flux for our sample. Middle panels: global parameter relations between CPL+BB (or PL+BB) model and CPL-alone model, and the solid lines represent the equal lines. Bottom panels: temporal evolution of $\Delta \alpha$, $\Delta E_{\rm c}$ and  flux ratio ($F_{\rm BB}/F_{\rm tot}$), and the horizontal-dashed lines represent the value of $\Delta \alpha$, $\Delta E_{\rm c}$=0 and ratio=0.1. Different colors represent different bursts.}
\end{figure*}\label{Dis}

\begin{figure*}
\centering
\includegraphics[angle=0,scale=0.80]{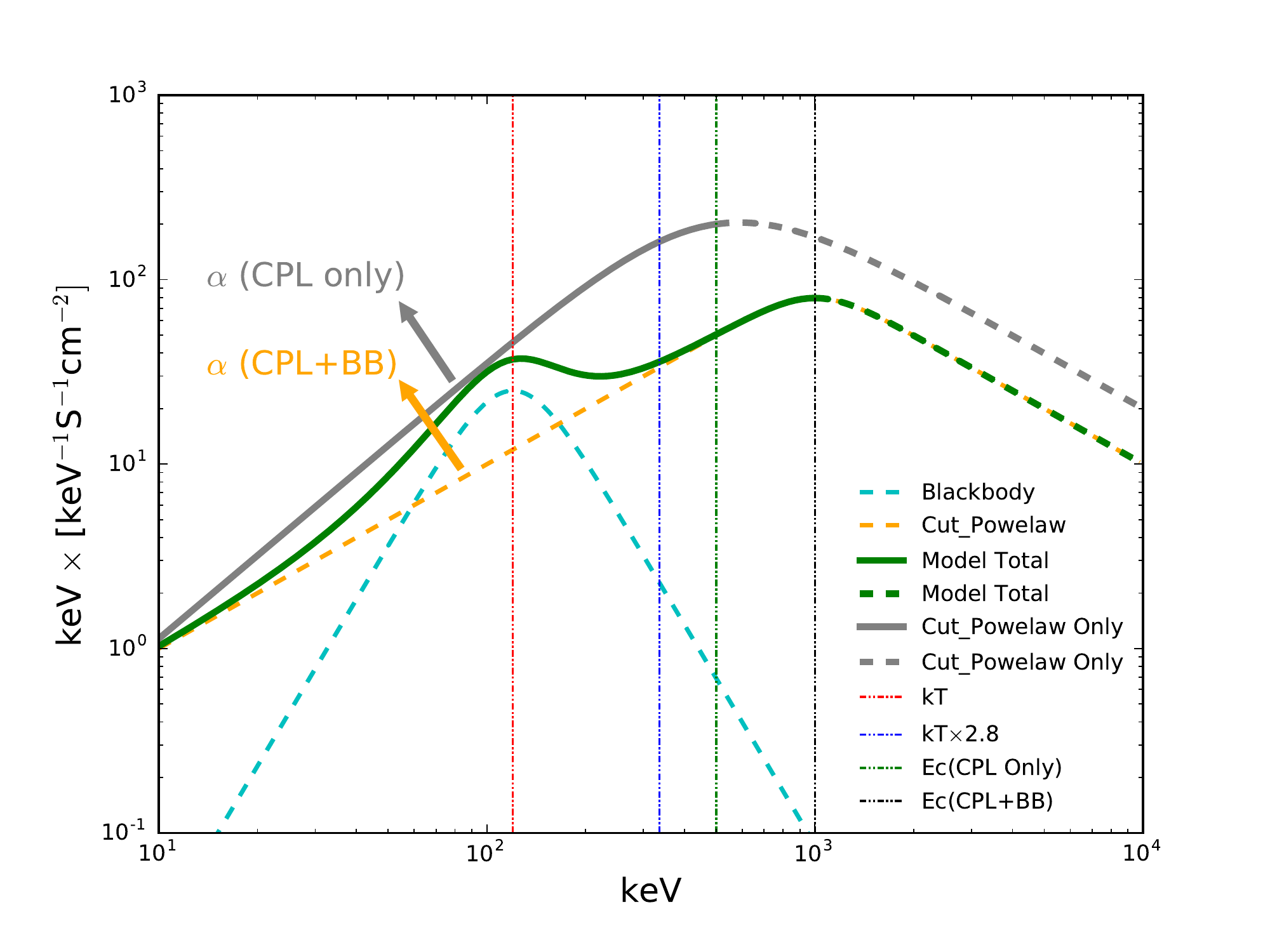}
\caption{Cartoon show that the spectral parameters ($\alpha$ and $E_{\rm c}$) are altered, if without considering the contribution of the thermal component, which is based on an assumption of typically observed parameters and the same normalization: $E_{\rm c}$=1000 keV, $\alpha$=-1, $kT$=150 keV. The CPL+BB model (green solid line) consists of the CPL model (orange dashed line) and the BB model (cyan dashed line); while the CPL only model is denoted by grey sold line. Two arrows indicate the $\alpha$ indices obtained from the CPL+BB model (orange) and the CPL-alone model (grey), respectively. Four perpendiculars dashed lines represent the BB temperature (red), the peak energy of the BB (blue), the peak energy of the CPL model (green), and the peak energy of the CPL+BB model (black), respectively.}
\end{figure*}\label{Cartoon}

\begin{figure*}
\includegraphics[angle=0,scale=0.45]{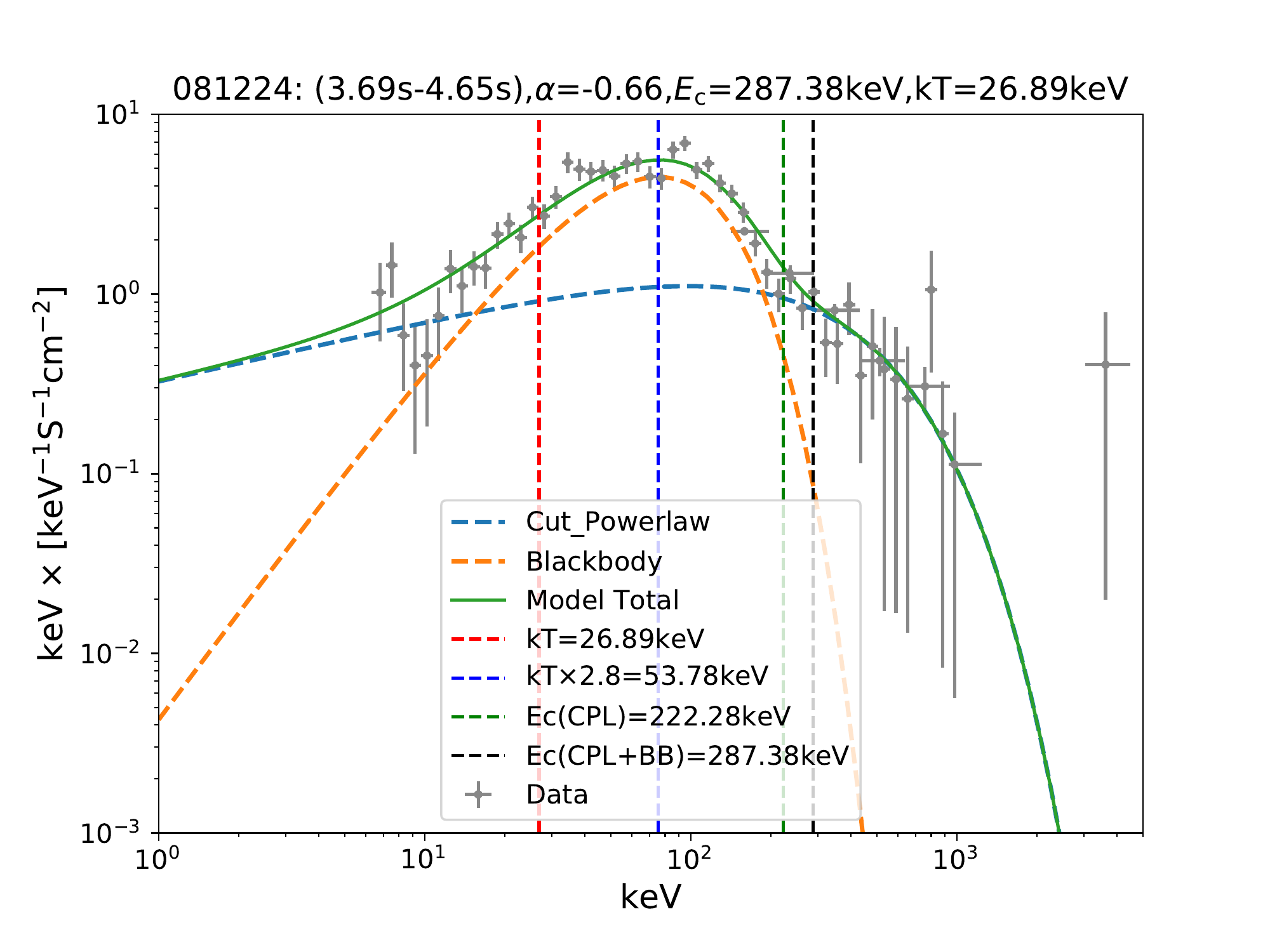}
\includegraphics[angle=0,scale=0.45]{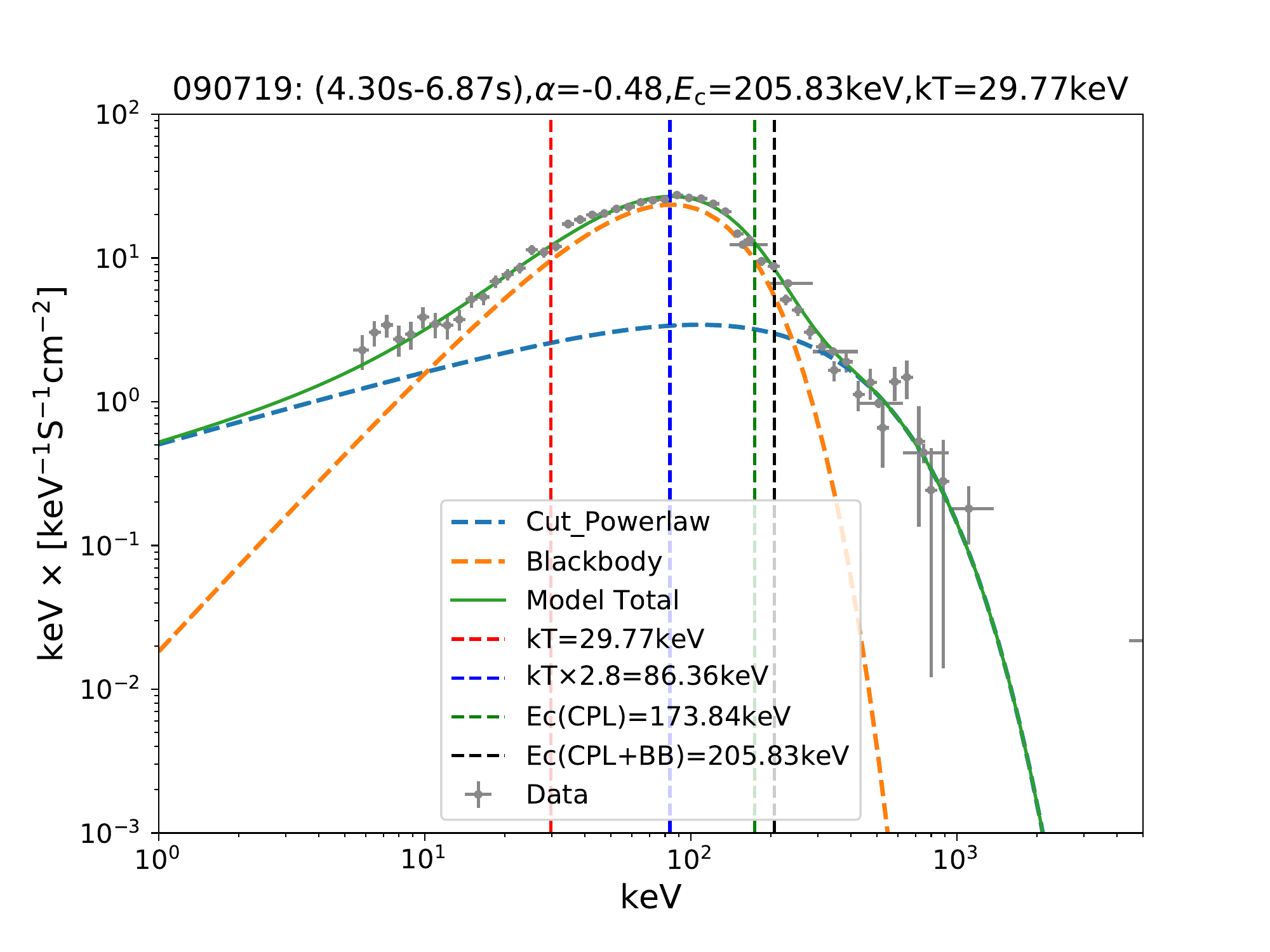}
\includegraphics[angle=0,scale=0.45]{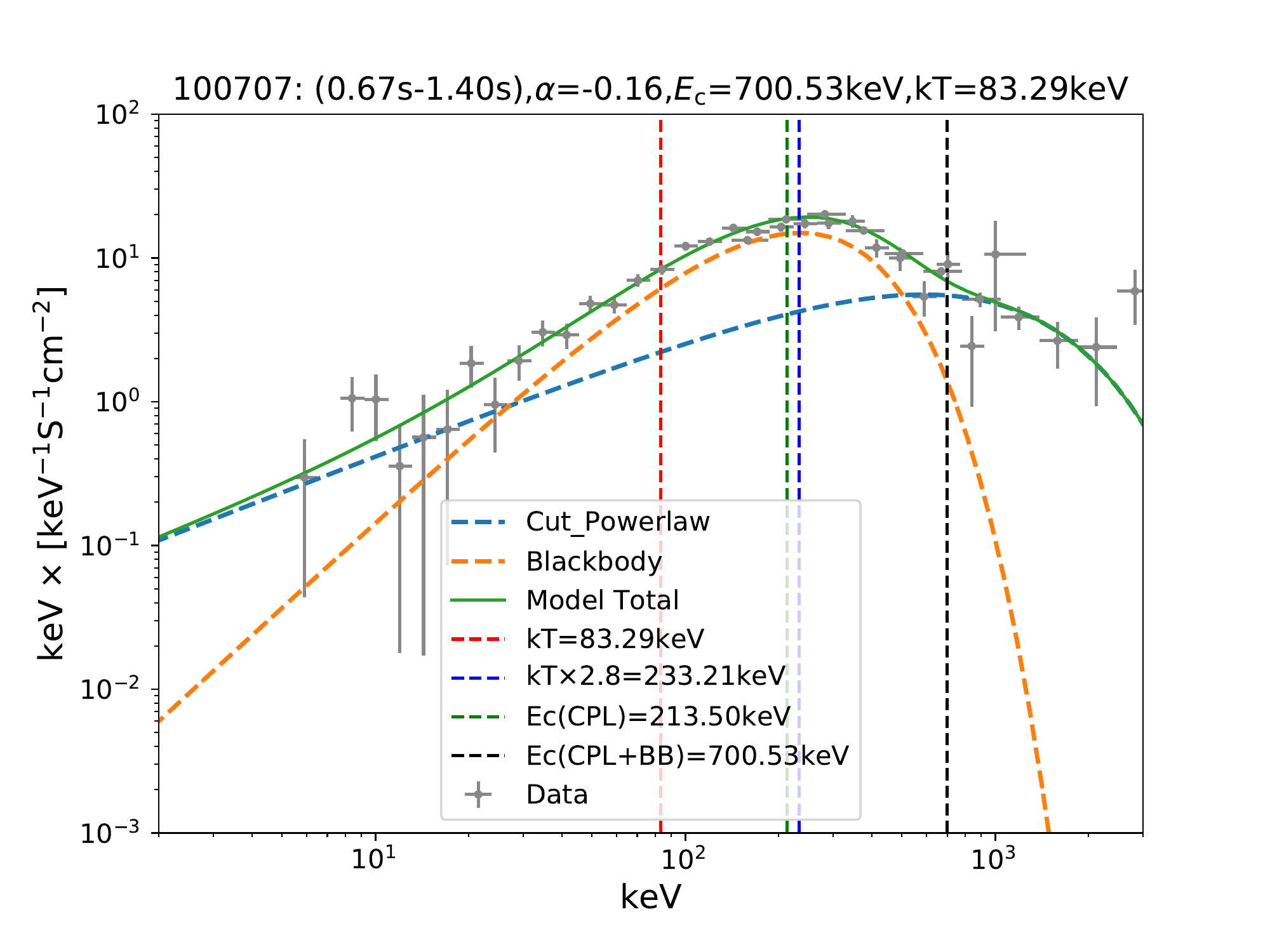}
\includegraphics[angle=0,scale=0.45]{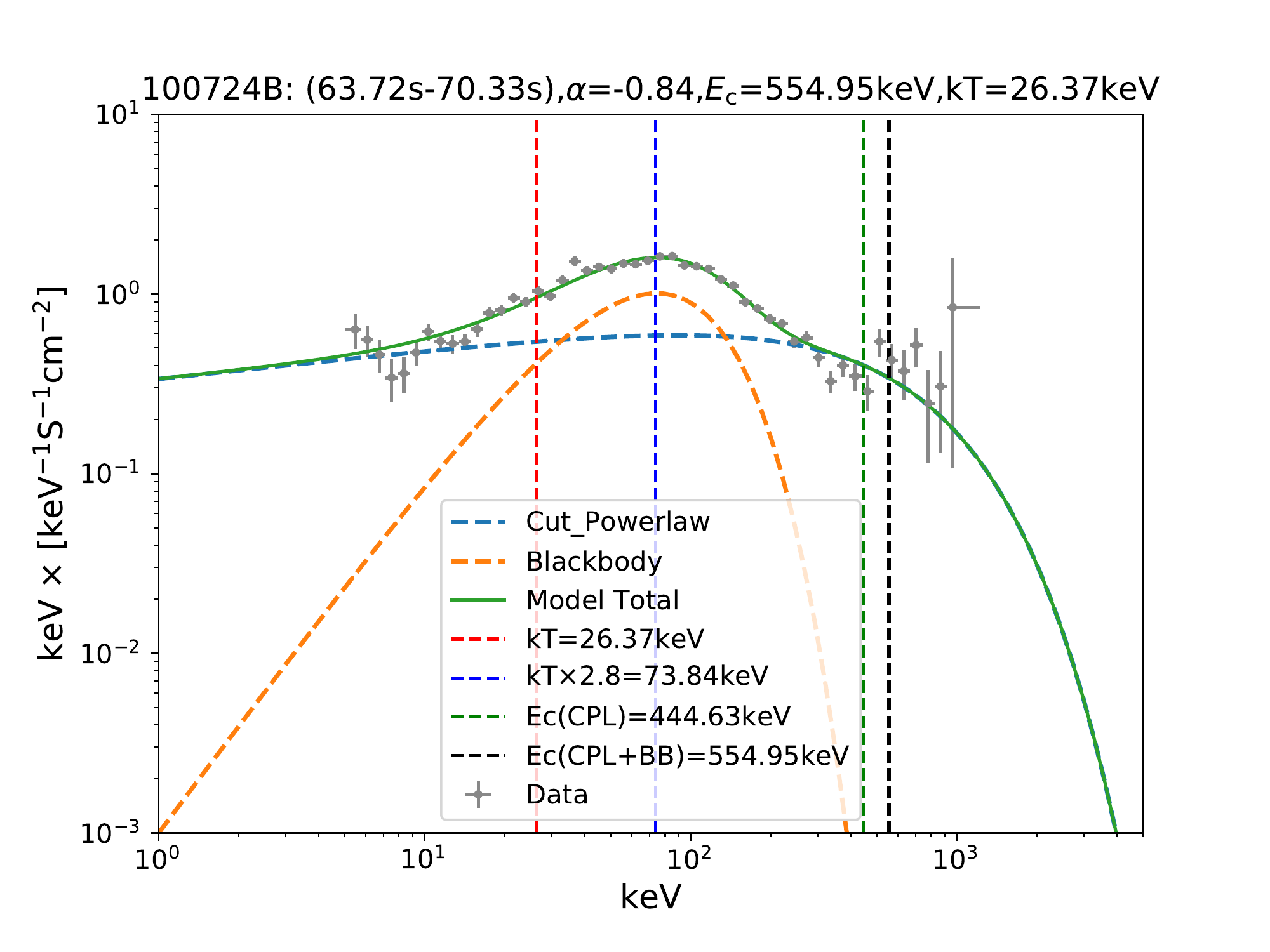}
\includegraphics[angle=0,scale=0.45]{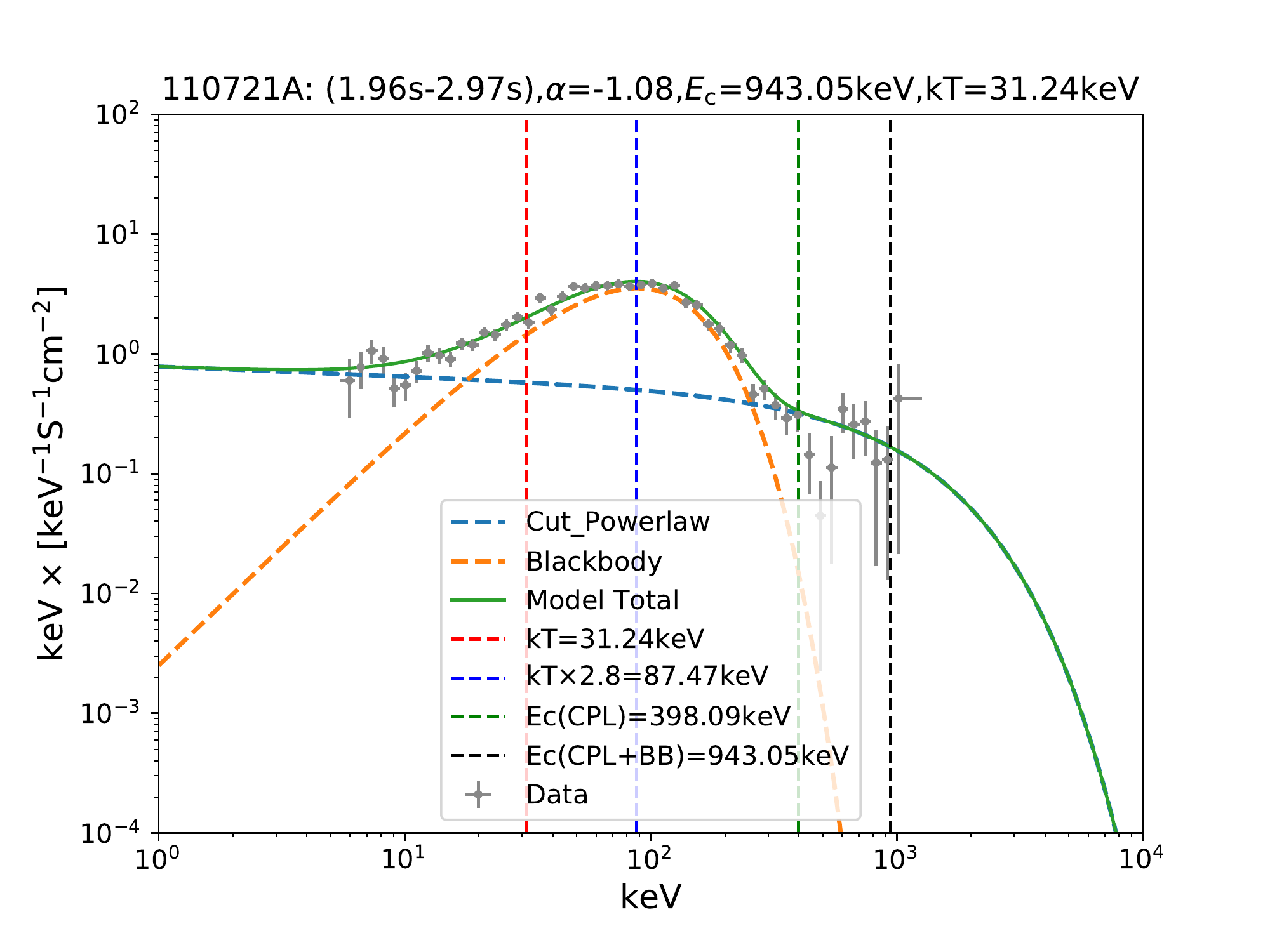}
\includegraphics[angle=0,scale=0.45]{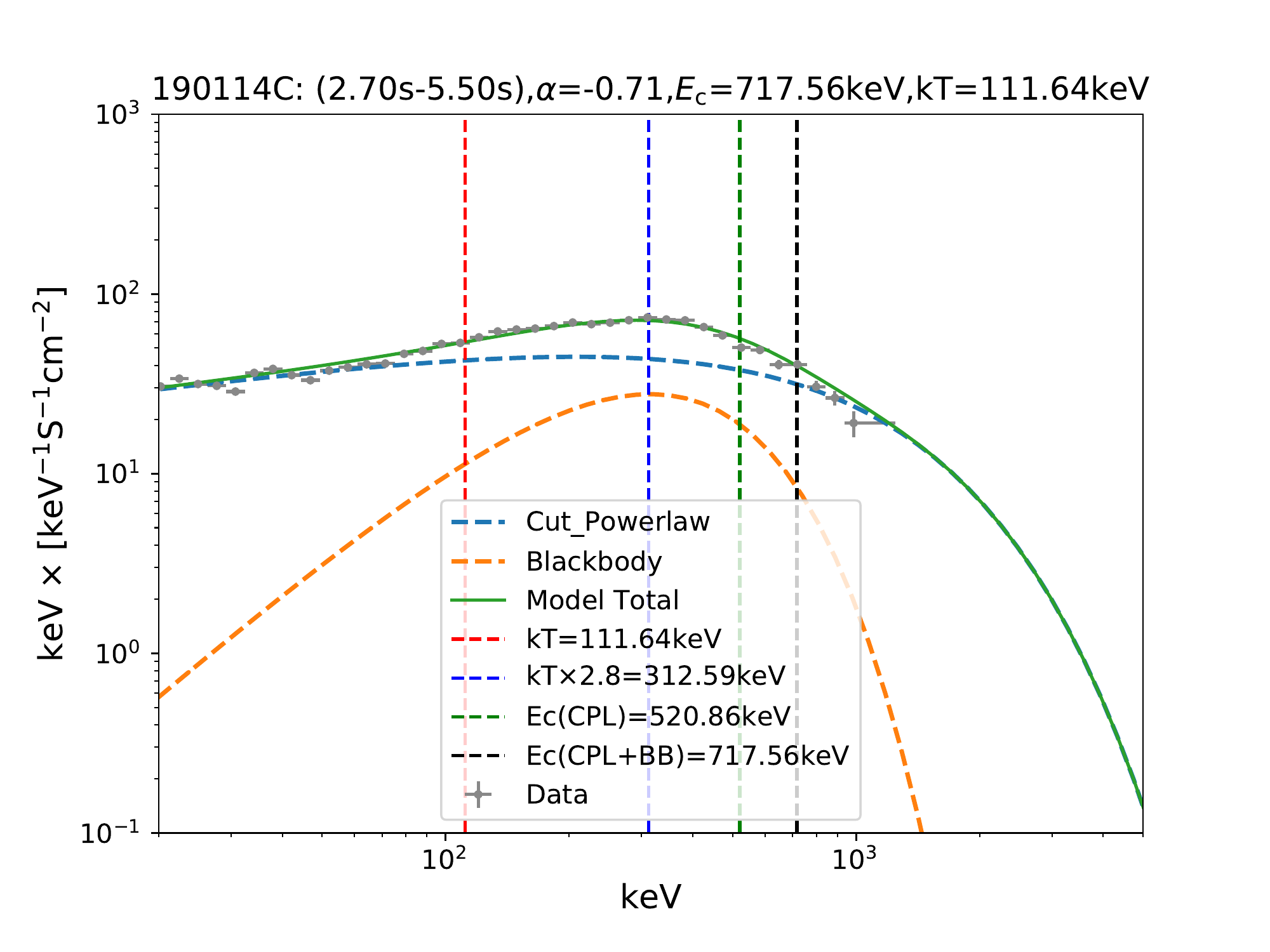}
\caption{Comparison of the time-resolved spectral fit results between CPL+BB model and CPL-alone model in one time bin for each burst. The bins with the biggest negative DIC or the highest thermal ratio are used: (3.69s-4.65s) for GRB081224, (4.30s-6.87s) for GRB090719, (0.42s-1.40s) for GRB100707,  (63.72s-70.33s) for GRB 100724B,(5.74s-6.66s) for GRB 110721A, and (2.7s-5.5s) for GRB 190114C, respectively. The solid line represents the total model while two dash lines represent blackbody component (orange) and CPL component (cyan). Four vertical dashed lines represent the blackbody temperature (red), the peak energy of thermal component (blue), the cutoff peak energy ($E_{\rm c}$) of CPL-alone model (green) and CPL+BB model (black), respectively.}
\end{figure*}\label{Spectrum}

\begin{figure*}
\includegraphics[angle=0,scale=0.45]{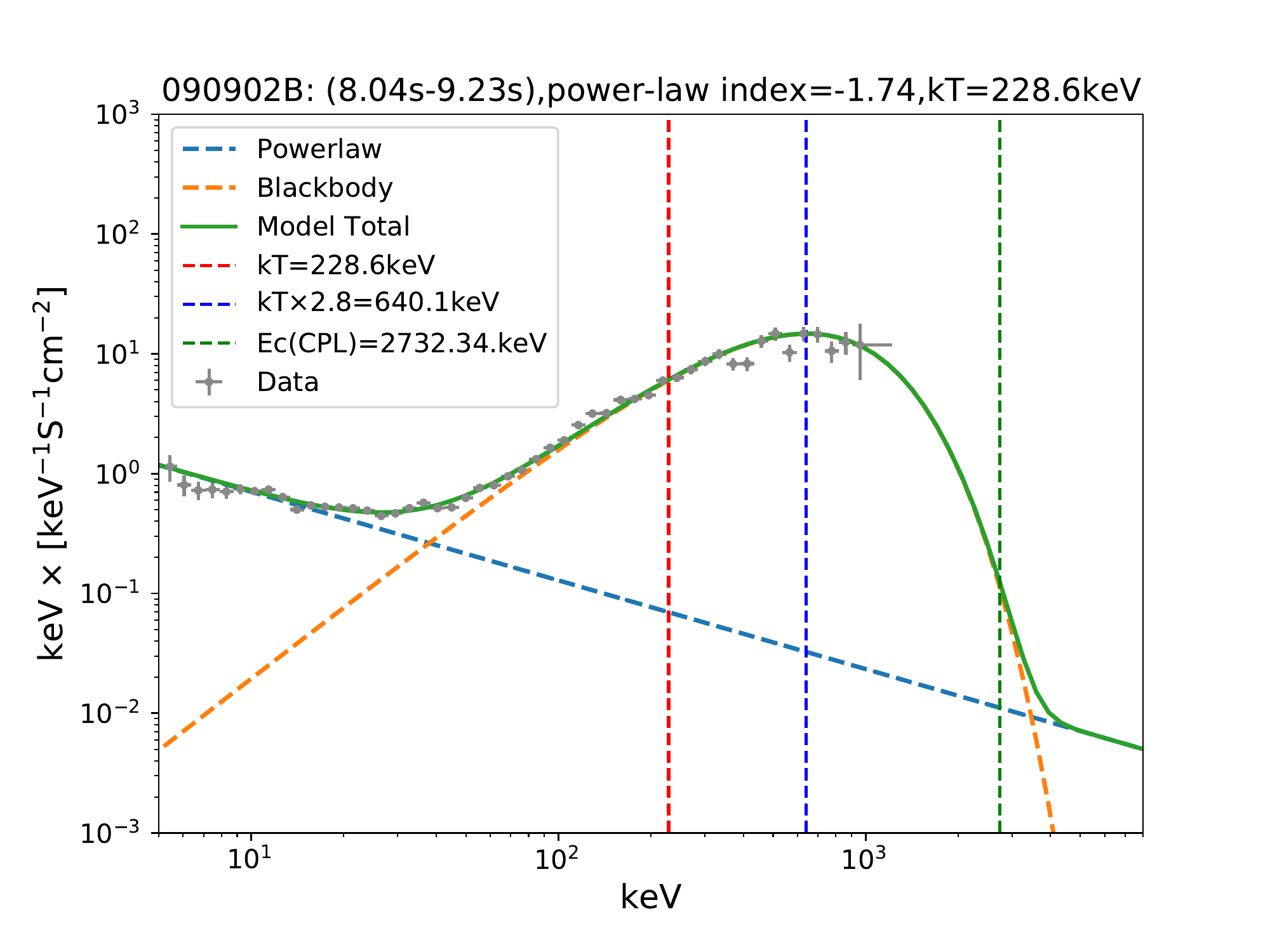}
\includegraphics[angle=0,scale=0.45]{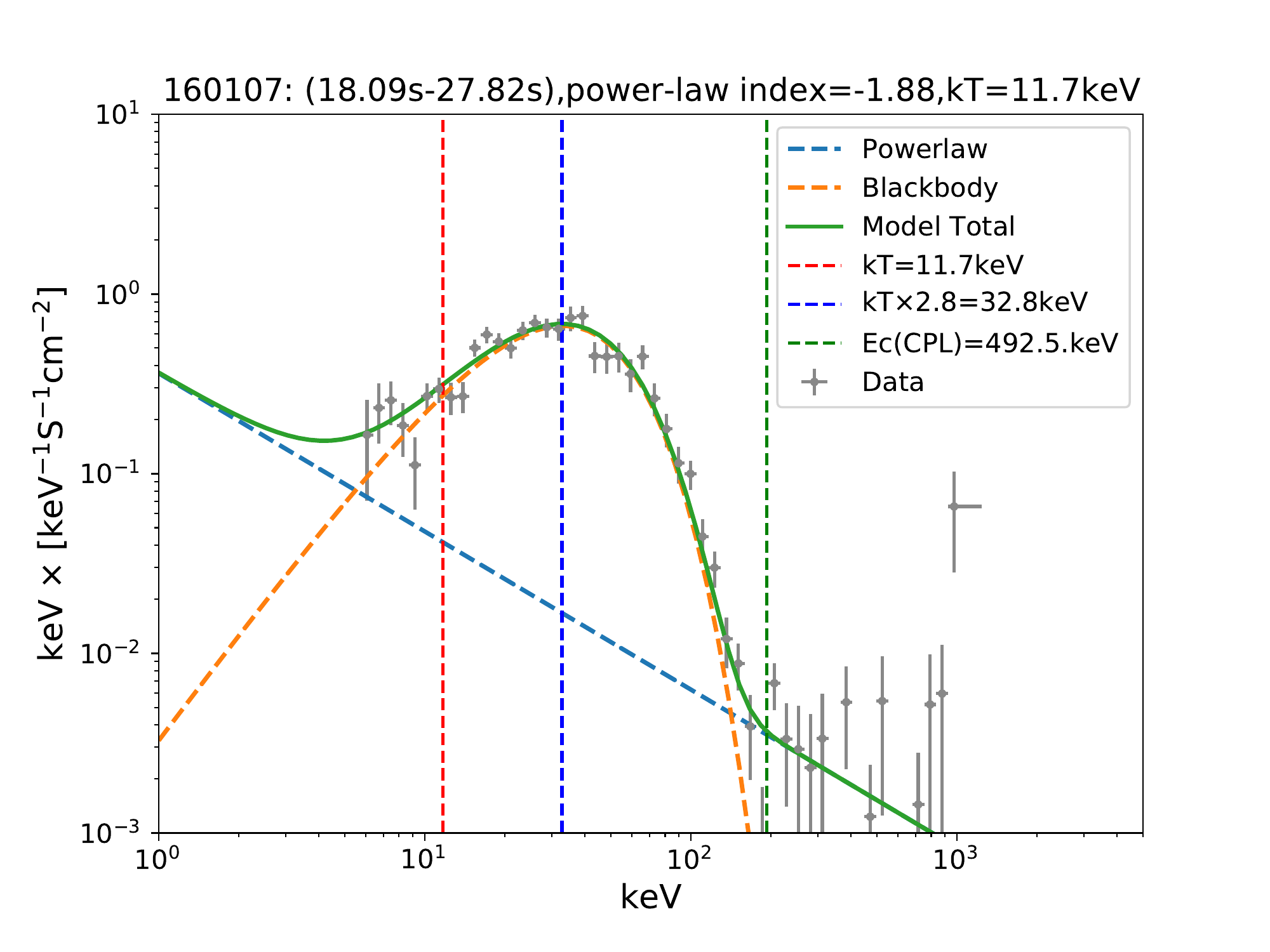}
\caption{Same as Figure \ref{Spectrum}, but for Group II bursts. The bins (8.04s-9.23s) for GRB 090902B  and (18.09s-27.82s) for GRB 160107 are used.}
\end{figure*}\label{SpectrumII}

\clearpage
\appendix

\setcounter{figure}{0}    
\setcounter{section}{0}
\setcounter{table}{0}
\renewcommand{\thesection}{A\arabic{section}}
\renewcommand{\thefigure}{A\arabic{figure}}
\renewcommand{\thetable}{A\arabic{table}}
\renewcommand{\theequation}{A\arabic{equation}}

In this Appendix, we provide additional figures and tables. Additional figures present the temporal evolution of spectral parameters ($E_{\rm p}$ and $\alpha$, Figure \ref{EvolutionTools}) and relation of $E_{\rm p}-\alpha$ of GRB 081224 (Figure \ref{RelationTools}) using different fitting tools and different binning techniques. Additional tables (Appendix) present the results of the time-resolved spectral fits for each burst.

\subsection{Figures}

It is important to address the questions of whether or not the results obtained from various fitting tools and binning techniques are consistent with each other. We, therefore, use GRB 081224 as an example to investigate the difference of evolution of parameters and the relation of parameters by the 3ML and the RMFIT (version 3.3pr7) software packages, and the $S$ and the BBloacks techniques. To compare the results between 3ML and RMFIT, we adopt the same time bin information based on the BBloacks method, which is then applied to both the 3ML and the RMFIT tools. Based on the 3ML tool, we also compare the results from different time bin techniques (S and BBloacks) and different $S$ values ($S$=30 and $S$=40). Temporal evolution of parameters ($\alpha$ and $E_{\rm p}$) and the relation of parameters ($E_{\rm p}$-$\alpha$) are presented in Figure \ref{EvolutionTools} and \ref{RelationTools}, respectively. We find that they all generally present the same results. The slopes of the parameter relation are also almost the same.

\subsection{Tables}
Tables \ref{Table:090719}-\ref{Table:160107} include the results of the time-resolved spectra fits for five Group I bursts (GRB 090719, GRB 100707, GRB 100724B, GRB 110721A, and GRB 190114C) and two Group II bursts (GRB 090902B and GRB 160107).

\clearpage
\begin{figure*}
\includegraphics[angle=0,scale=0.50]{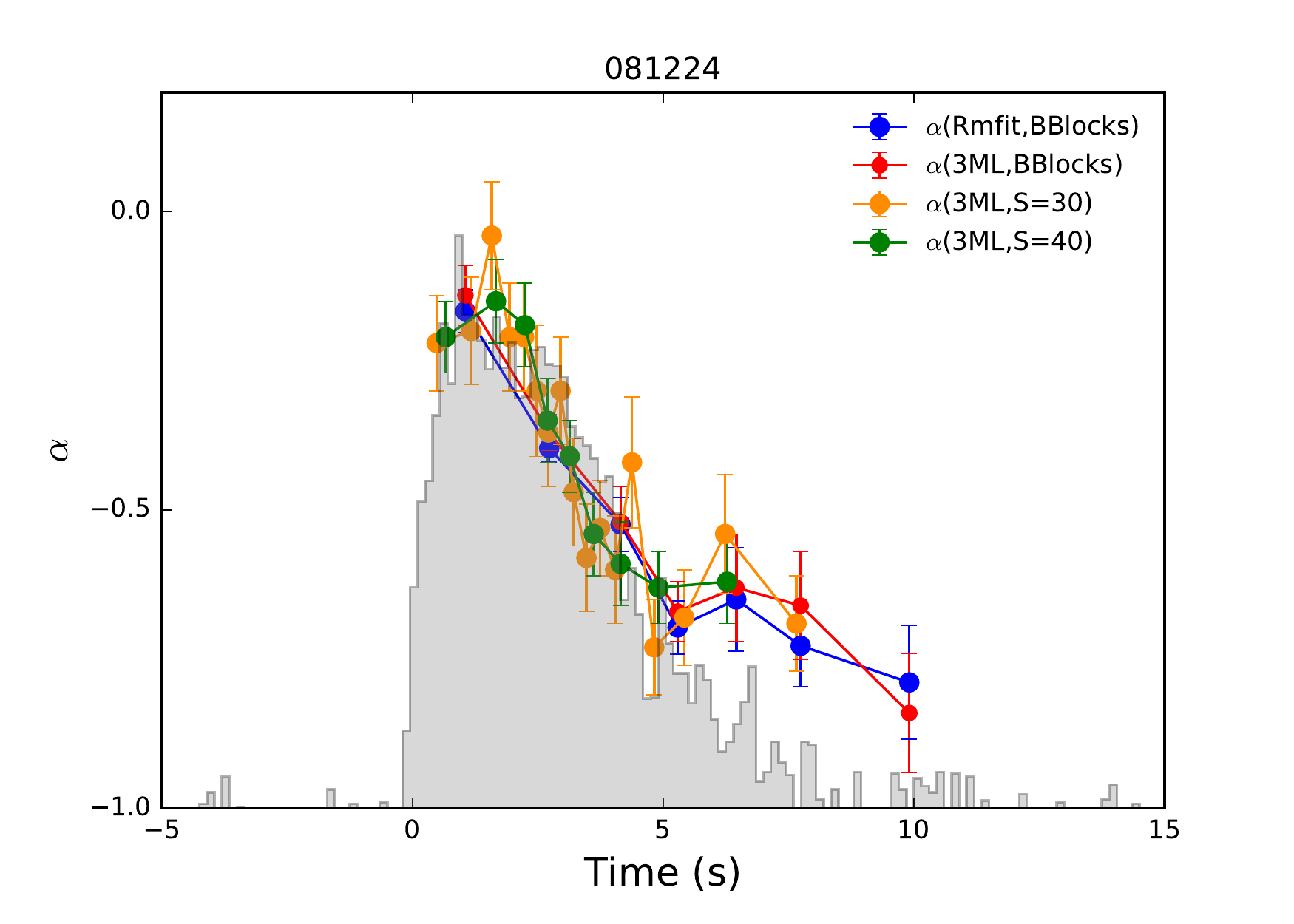}
\includegraphics[angle=0,scale=0.50]{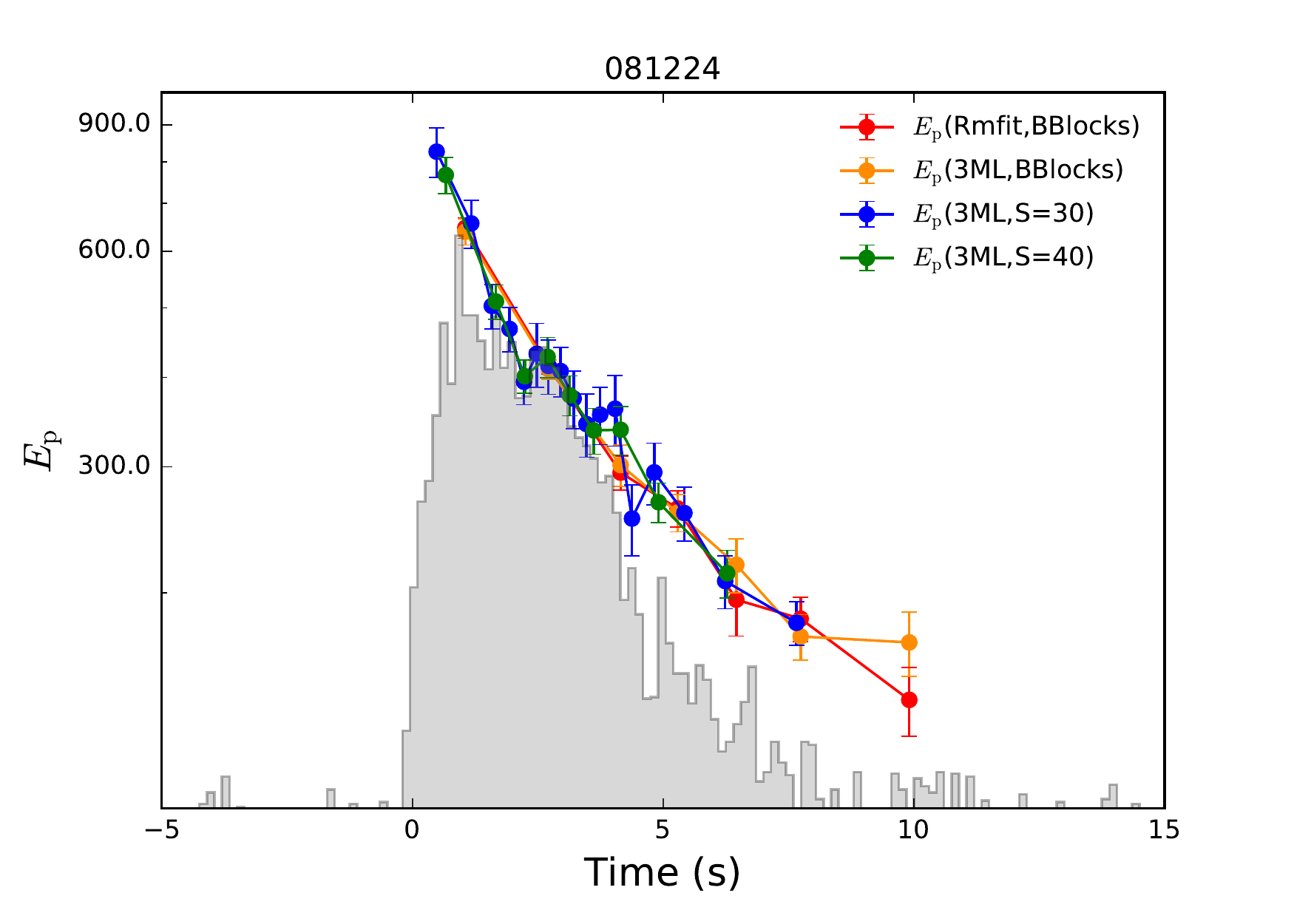}
\caption{Temporal evolution of spectral parameters of GRB081224: $\alpha$ (left panel) and $E_{\rm p}$ (right panel). 
Data points with blue, red, orange, and green indicate the cases of RMIT+BBlocks, 3ML+BBlocks, 3ML+S(=30) and 3ML+S(=40), respectively. Light curves are overlaid in grey.}
\end{figure*}\label{EvolutionTools}

\begin{figure*}
\centering
\includegraphics[angle=0,scale=0.80]{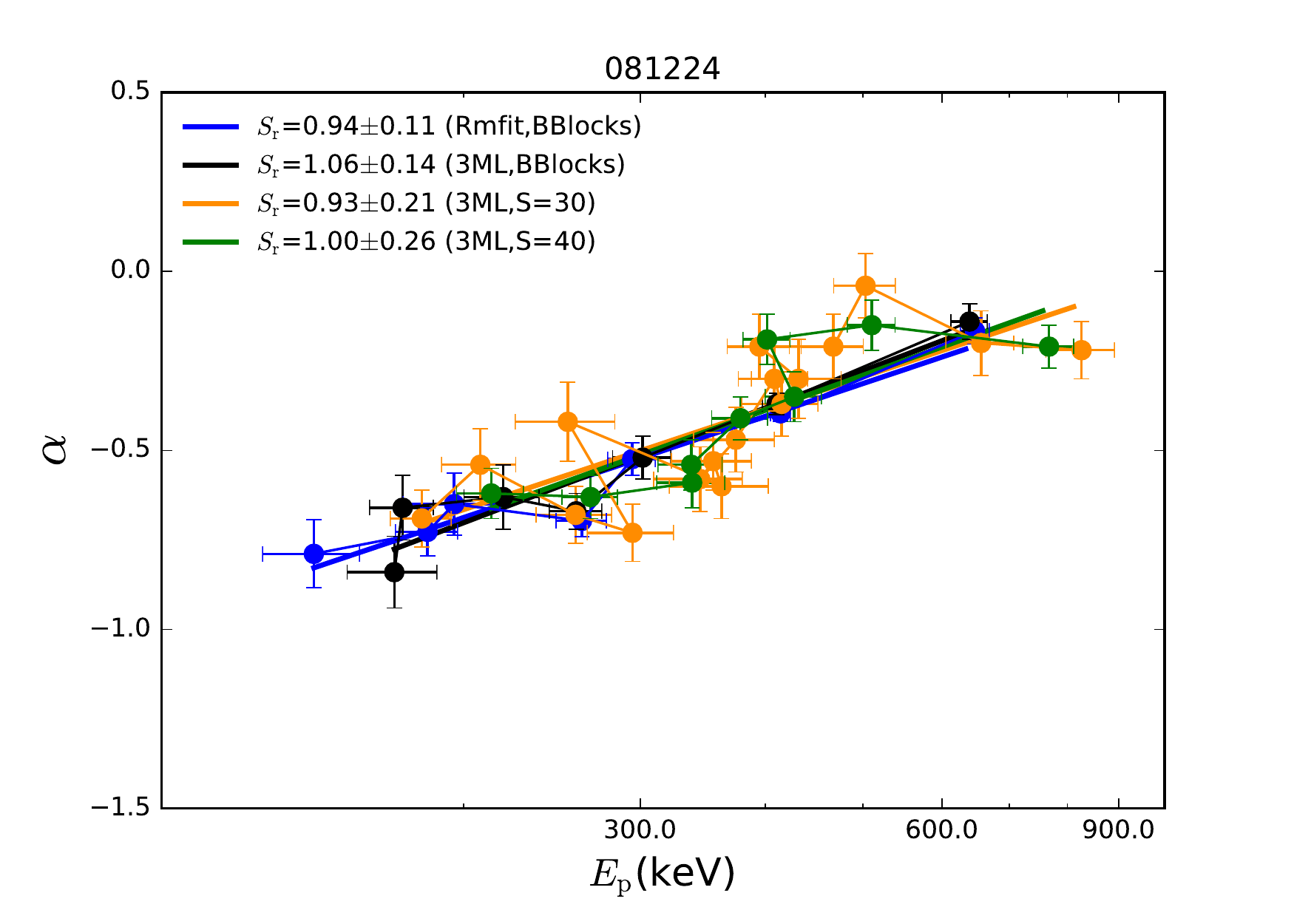}
\caption{Relation of $E_{\rm p}-\alpha$ of GRB 081224. The symbols are the same as Figure \ref{EvolutionTools} but for the relation of $E_{\rm p}-\alpha$. The solid lines represent the slopes of different cases.}
\end{figure*}\label{RelationTools}

\clearpage 
\begin{deluxetable}{cc|cc|ccc|cccc}
\tablewidth{0pt}
\setlength{\tabcolsep}{0.35em}
\tabletypesize{\scriptsize}
\tablecaption{Results of the Time-resolved Spectral Fits of GRB 081224}
\tablehead{
\multirow{3}{0.8cm}{$t_{1}$$\sim$$t_{2}$} 
&\multirow{3}{0.8cm}{$S$} 
&\multicolumn{2}{|c}{Cutoff Power-law Fitting}
&\multicolumn{3}{|c}{Cutoff Power-law Plus Blackbody Fitting}
&\multicolumn{3}{|c}{Difference}\\
\cline{3-11}
&
&\colhead{$\alpha$}
&\colhead{$E_{\rm c}$}
&\colhead{$\alpha$}
&\colhead{$E_{\rm c}$}
&\colhead{$kT$}
&\colhead{$\Delta \alpha^{\rm I}$}
&\colhead{$\Delta E_{\rm c}$}
&\colhead{Ratio$^{\rm I}$}
&\colhead{$\Delta$DIC$^{\rm I}$}\\
(s)
&&&(keV)
&&&(keV)
&
&
&
&
}
\startdata
0.35$\sim$1.76&48.23&-0.13$^{+0.05}_{-0.05}$&342.6$^{+22.2}_{-22.4}$&-0.20$^{+0.11}_{-0.12}$&337.4$^{+54.8}_{-54.8}$&129.1$^{+53.6}_{-74.8}$&-0.07$^{+0.12}_{-0.13}$&-5.2$^{+59.1}_{-59.2}$&0.09$^{+0.51}_{-0.10}$&-55.3\\
3.69$\sim$4.65&45.88&-0.57$^{+0.05}_{-0.05}$&222.3$^{+19.4}_{-19.1}$&-0.66$^{+0.10}_{-0.10}$&287.4$^{+65.7}_{-64.7}$&26.9$^{+9.7}_{-16.0}$&-0.09$^{+0.11}_{-0.11}$&65.1$^{+68.5}_{-67.5}$&0.05$^{+0.15}_{-0.06}$&-378.3\\
4.65$\sim$5.97&40.97&-0.67$^{+0.05}_{-0.05}$&201.7$^{+18.7}_{-18.5}$&-0.67$^{+0.05}_{-0.06}$&198.2$^{+20.1}_{-19.1}$&104.0$^{+67.5}_{-67.5}$&0.00$^{+0.07}_{-0.08}$&-3.5$^{+27.5}_{-26.6}$&0.01$^{+0.03}_{-0.01}$&-16.8\\
5.97$\sim$6.95&28.07&-0.67$^{+0.08}_{-0.08}$&172.9$^{+23.7}_{-23.7}$&-0.67$^{+0.12}_{-0.12}$&187.8$^{+31.1}_{-36.5}$&15.1$^{+12.5}_{-10.5}$&0.00$^{+0.14}_{-0.14}$&15.0$^{+39.2}_{-43.5}$&0.01$^{+0.24}_{-0.01}$&-61.7\\
6.95$\sim$8.54&27.23&-0.70$^{+0.09}_{-0.09}$&140.6$^{+19.0}_{-19.0}$&-0.69$^{+0.17}_{-0.16}$&148.3$^{+28.1}_{-32.4}$&14.5$^{+13.1}_{-9.1}$&0.01$^{+0.19}_{-0.18}$&7.7$^{+33.9}_{-37.6}$&0.01$^{+0.40}_{-0.01}$&-188.4\\
\enddata
\end{deluxetable}\label{Table:081224}

\clearpage 
\begin{deluxetable}{cc|cc|ccc|cccc}
\tablewidth{0pt}
\setlength{\tabcolsep}{0.35em}
\tabletypesize{\scriptsize}
\tablecaption{Results of the Time-resolved Spectral Fits of GRB 090719}
\tablehead{
\multirow{3}{0.8cm}{$t_{1}$$\sim$$t_{2}$} 
&\multirow{3}{0.8cm}{$S$} 
&\multicolumn{2}{|c}{Cutoff Power-law Fitting}
&\multicolumn{3}{|c}{Cutoff Power-law Plus Blackbody Fitting}
&\multicolumn{3}{|c}{Difference}\\
\cline{3-11}
&
&\colhead{$\alpha$}
&\colhead{$E_{\rm c}$}
&\colhead{$\alpha$}
&\colhead{$E_{\rm c}$}
&\colhead{$kT$}
&\colhead{$\Delta \alpha^{\rm I}$}
&\colhead{$\Delta E_{\rm c}$}
&\colhead{Ratio$^{\rm I}$}
&\colhead{$\Delta$DIC$^{\rm I}$}\\
(s)
&&&(keV)
&&&(keV)
&
&
&
&
}
\startdata
0.42$\sim$0.89&21.01&0.03$^{+0.13}_{-0.13}$&197.4$^{+25.0}_{-24.6}$&-0.16$^{+0.26}_{-0.31}$&221.3$^{+50.0}_{-65.3}$&83.4$^{+30.1}_{-37.8}$&-0.19$^{+0.29}_{-0.34}$&23.9$^{+55.9}_{-69.8}$&0.16$^{+0.79}_{-0.19}$&-30.0\\
0.89$\sim$3.16&62.62&0.01$^{+0.05}_{-0.05}$&110.4$^{+5.0}_{-5.0}$&0.25$^{+0.20}_{-0.20}$&74.9$^{+26.9}_{-22.0}$&85.7$^{+21.3}_{-13.4}$&0.24$^{+0.21}_{-0.21}$&-35.5$^{+27.3}_{-22.6}$&0.28$^{+0.81}_{-0.29}$&-99.6\\
3.16$\sim$4.30&53.97&-0.43$^{+0.05}_{-0.05}$&209.8$^{+13.9}_{-13.9}$&-0.45$^{+0.09}_{-0.08}$&228.9$^{+34.1}_{-32.1}$&25.5$^{+16.5}_{-20.4}$&-0.02$^{+0.10}_{-0.09}$&19.1$^{+36.9}_{-35.0}$&0.02$^{+0.50}_{-0.02}$&-3658.8\\
4.30$\sim$6.87&106.00&-0.41$^{+0.02}_{-0.02}$&173.8$^{+5.8}_{-5.8}$&-0.48$^{+0.05}_{-0.05}$&205.8$^{+15.5}_{-15.7}$&29.8$^{+4.1}_{-4.0}$&-0.07$^{+0.05}_{-0.05}$&32.0$^{+16.6}_{-16.7}$&0.06$^{+0.06}_{-0.03}$&-22.1\\
6.87$\sim$7.59&43.30&-0.46$^{+0.06}_{-0.06}$&121.5$^{+8.9}_{-9.0}$&-0.47$^{+0.07}_{-0.07}$&120.3$^{+13.7}_{-11.6}$&38.2$^{+25.5}_{-27.5}$&-0.01$^{+0.09}_{-0.09}$&-1.1$^{+16.4}_{-14.7}$&0.01$^{+1.23}_{-0.02}$&-351.5\\
6.87$\sim$7.59&43.30&-0.46$^{+0.06}_{-0.06}$&121.3$^{+9.2}_{-9.1}$&-0.47$^{+0.07}_{-0.07}$&120.3$^{+13.7}_{-11.6}$&38.2$^{+25.5}_{-27.5}$&-0.01$^{+0.09}_{-0.09}$&-1.0$^{+16.5}_{-14.8}$&0.01$^{+1.23}_{-0.02}$&-351.5\\
7.59$\sim$9.32&41.19&-0.71$^{+0.07}_{-0.07}$&96.8$^{+8.2}_{-8.2}$&-0.72$^{+0.14}_{-0.13}$&95.2$^{+11.2}_{-12.5}$&20.4$^{+17.9}_{-15.6}$&-0.01$^{+0.16}_{-0.15}$&-1.5$^{+13.9}_{-15.0}$&0.02$^{+3.38}_{-0.03}$&-7936.8\\
9.32$\sim$11.98&34.50&-1.07$^{+0.08}_{-0.08}$&98.3$^{+12.1}_{-12.2}$&-0.81$^{+0.28}_{-0.28}$&64.2$^{+34.6}_{-30.4}$&49.4$^{+29.2}_{-35.9}$&0.26$^{+0.29}_{-0.29}$&-34.1$^{+36.7}_{-32.8}$&0.10$^{+2.80}_{-0.11}$&-7169.2\\
\enddata
\end{deluxetable}\label{Table:090719}

\clearpage 
\begin{deluxetable}{cc|cc|ccc|cccc}
\tablewidth{0pt}
\setlength{\tabcolsep}{0.35em}
\tabletypesize{\scriptsize}
\tablecaption{Results of the Time-resolved Spectral Fits of GRB 100707}
\tablehead{
\multirow{3}{0.8cm}{$t_{1}$$\sim$$t_{2}$} 
&\multirow{3}{0.8cm}{$S$} 
&\multicolumn{2}{|c}{Cutoff Power-law Fitting}
&\multicolumn{3}{|c}{Cutoff Power-law Plus Blackbody Fitting}
&\multicolumn{3}{|c}{Difference}\\
\cline{3-11}
&
&\colhead{$\alpha$}
&\colhead{$E_{\rm c}$}
&\colhead{$\alpha$}
&\colhead{$E_{\rm c}$}
&\colhead{$kT$}
&\colhead{$\Delta \alpha^{\rm I}$}
&\colhead{$\Delta E_{\rm c}$}
&\colhead{Ratio$^{\rm I}$}
&\colhead{$\Delta$DIC$^{\rm I}$}\\
(s)
&&&(keV)
&&&(keV)
&
&
&
&
}
\startdata
0.36$\sim$0.70&34.38&0.41$^{+0.11}_{-0.11}$&264.1$^{+25.8}_{-25.8}$&-0.23$^{+0.31}_{-0.30}$&911.0$^{+394.6}_{-424.4}$&111.6$^{+11.0}_{-10.7}$&-0.64$^{+0.33}_{-0.32}$&646.9$^{+395.5}_{-425.1}$&0.35$^{+0.59}_{-0.22}$&-32.4\\
0.70$\sim$1.19&54.98&0.38$^{+0.07}_{-0.07}$&223.9$^{+13.5}_{-13.3}$&0.14$^{+0.27}_{-0.25}$&460.5$^{+107.7}_{-125.3}$&79.3$^{+6.5}_{-6.6}$&-0.24$^{+0.28}_{-0.26}$&236.6$^{+108.5}_{-126.0}$&0.29$^{+0.31}_{-0.15}$&-57.1\\
1.19$\sim$2.77&123.61&0.18$^{+0.03}_{-0.03}$&187.1$^{+5.5}_{-5.6}$&0.07$^{+0.10}_{-0.10}$&295.5$^{+33.3}_{-32.8}$&52.0$^{+2.9}_{-2.8}$&-0.11$^{+0.10}_{-0.10}$&108.4$^{+33.8}_{-33.3}$&0.20$^{+0.09}_{-0.06}$&-120.2\\
2.77$\sim$4.17&100.54&0.09$^{+0.04}_{-0.04}$&106.5$^{+4.1}_{-4.1}$&-0.27$^{+0.12}_{-0.12}$&205.2$^{+31.3}_{-31.9}$&34.4$^{+1.9}_{-1.9}$&-0.36$^{+0.13}_{-0.13}$&98.8$^{+31.6}_{-32.2}$&0.27$^{+0.12}_{-0.09}$&-65.2\\
4.17$\sim$5.04&72.69&-0.12$^{+0.06}_{-0.06}$&91.5$^{+5.2}_{-5.2}$&-0.12$^{+0.06}_{-0.06}$&91.7$^{+5.2}_{-5.5}$&154.5$^{+98.5}_{-99.2}$&0.00$^{+0.08}_{-0.08}$&0.1$^{+7.3}_{-7.5}$&0.01$^{+0.01}_{-0.01}$&-24.3\\
5.04$\sim$6.12&62.16&-0.29$^{+0.06}_{-0.06}$&99.4$^{+6.6}_{-6.6}$&-0.19$^{+0.21}_{-0.20}$&107.8$^{+20.8}_{-18.2}$&16.3$^{+4.3}_{-4.4}$&0.10$^{+0.22}_{-0.21}$&8.4$^{+21.8}_{-19.4}$&0.07$^{+0.23}_{-0.06}$&-57.7\\
6.12$\sim$10.00&80.36&-0.27$^{+0.05}_{-0.05}$&85.0$^{+4.1}_{-4.2}$&-0.46$^{+0.19}_{-0.17}$&113.5$^{+23.1}_{-25.1}$&22.0$^{+4.6}_{-6.6}$&-0.19$^{+0.20}_{-0.18}$&28.5$^{+23.5}_{-25.5}$&0.11$^{+0.15}_{-0.10}$&-42.5\\
10.00$\sim$12.69&55.57&-0.31$^{+0.07}_{-0.07}$&69.5$^{+4.7}_{-4.7}$&-0.31$^{+0.07}_{-0.07}$&69.7$^{+4.5}_{-4.6}$&155.3$^{+98.2}_{-98.1}$&0.00$^{+0.10}_{-0.10}$&0.3$^{+6.5}_{-6.5}$&0.01$^{+0.01}_{-0.01}$&-20.8\\
12.69$\sim$14.21&35.75&-0.70$^{+0.10}_{-0.10}$&92.5$^{+12.5}_{-12.4}$&-0.68$^{+0.12}_{-0.12}$&87.7$^{+14.6}_{-14.8}$&60.4$^{+32.6}_{-37.9}$&0.02$^{+0.16}_{-0.16}$&-4.8$^{+19.2}_{-19.3}$&0.01$^{+0.38}_{-0.01}$&-20.7\\
14.21$\sim$16.81&37.36&-0.77$^{+0.11}_{-0.11}$&73.3$^{+10.2}_{-10.2}$&-0.84$^{+0.16}_{-0.16}$&82.1$^{+20.4}_{-20.1}$&42.1$^{+38.4}_{-29.8}$&-0.07$^{+0.19}_{-0.19}$&8.8$^{+22.8}_{-22.6}$&0.04$^{+7.01}_{-0.04}$&-7076.7\\
16.81$\sim$20.00&31.74&-1.17$^{+0.13}_{-0.13}$&76.6$^{+15.3}_{-15.2}$&-1.12$^{+0.17}_{-0.16}$&66.4$^{+17.1}_{-20.0}$&48.7$^{+23.1}_{-29.4}$&0.05$^{+0.21}_{-0.21}$&-10.2$^{+22.9}_{-25.1}$&0.02$^{+0.56}_{-0.03}$&-103.2\\
\enddata
\end{deluxetable}\label{Table:100707}

\clearpage 
\begin{deluxetable}{cc|cc|ccc|ccccc}
\tablewidth{0pt}
\setlength{\tabcolsep}{0.35em}
\tabletypesize{\scriptsize}
\tablecaption{Results of the Time-resolved Spectral Fits of GRB 100724B}
\tablehead{
\multirow{3}{0.8cm}{$t_{1}$$\sim$$t_{2}$} 
&\multirow{3}{0.8cm}{$S$} 
&\multicolumn{2}{|c}{Cutoff Power-law Fitting}
&\multicolumn{3}{|c}{Cutoff Power-law Plus Blackbody Fitting}
&\multicolumn{3}{|c}{Difference}\\
\cline{3-11}
&
&\colhead{$\alpha$}
&\colhead{$E_{\rm c}$}
&\colhead{$\alpha$}
&\colhead{$E_{\rm c}$}
&\colhead{$kT$}
&\colhead{$\Delta \alpha^{\rm I}$}
&\colhead{$\Delta E_{\rm c}$}
&\colhead{Ratio$^{\rm I}$}
&\colhead{$\Delta$DIC$^{\rm I}$}\\
(s)
&&&(keV)
&&&(keV)
&
&
&
&
}
\startdata
1.19$\sim$6.72&25.69&-1.02$^{+0.03}_{-0.03}$&1880.0$^{+404.9}_{-404.6}$&-0.99$^{+0.05}_{-0.05}$&2119.6$^{+431.1}_{-434.4}$&26.9$^{+4.8}_{-5.2}$&0.03$^{+0.06}_{-0.06}$&239.5$^{+591.4}_{-593.6}$&0.01$^{+0.03}_{-0.01}$&-21.8\\
6.72$\sim$8.32&22.18&-0.86$^{+0.06}_{-0.06}$&920.1$^{+243.3}_{-241.6}$&-0.90$^{+0.09}_{-0.09}$&1219.9$^{+451.4}_{-452.1}$&38.4$^{+24.5}_{-24.5}$&-0.04$^{+0.11}_{-0.11}$&299.8$^{+512.8}_{-512.6}$&0.01$^{+0.19}_{-0.01}$&-112.9\\
8.32$\sim$10.73&39.04&-0.90$^{+0.06}_{-0.06}$&1074.7$^{+330.8}_{-306.5}$&-1.00$^{+0.04}_{-0.04}$&2946.0$^{+485.3}_{-479.2}$&38.3$^{+3.4}_{-3.4}$&-0.10$^{+0.07}_{-0.07}$&1871.4$^{+587.3}_{-568.8}$&0.04$^{+0.03}_{-0.02}$&-63.9\\
10.73$\sim$11.92&36.78&-0.90$^{+0.05}_{-0.05}$&1243.2$^{+302.6}_{-312.0}$&-1.02$^{+0.04}_{-0.04}$&2933.3$^{+627.6}_{-622.9}$&48.7$^{+6.8}_{-6.9}$&-0.12$^{+0.06}_{-0.06}$&1690.1$^{+696.7}_{-696.7}$&0.04$^{+0.05}_{-0.02}$&-41.9\\
11.92$\sim$14.49&41.85&-0.82$^{+0.04}_{-0.04}$&625.7$^{+78.9}_{-79.4}$&-0.95$^{+0.05}_{-0.05}$&1705.8$^{+384.9}_{-392.5}$&34.8$^{+3.1}_{-3.2}$&-0.13$^{+0.06}_{-0.06}$&1080.1$^{+392.9}_{-400.5}$&0.05$^{+0.03}_{-0.02}$&-44.9\\
14.49$\sim$18.34&65.31&-0.97$^{+0.02}_{-0.02}$&1737.3$^{+220.4}_{-223.5}$&-1.05$^{+0.02}_{-0.02}$&3502.9$^{+378.4}_{-376.5}$&48.5$^{+3.8}_{-3.8}$&-0.08$^{+0.03}_{-0.03}$&1765.5$^{+437.9}_{-437.8}$&0.04$^{+0.02}_{-0.02}$&-130.2\\
18.34$\sim$25.16&68.80&-0.88$^{+0.02}_{-0.02}$&653.0$^{+55.3}_{-55.7}$&-1.04$^{+0.03}_{-0.03}$&1860.3$^{+374.4}_{-370.1}$&37.1$^{+3.2}_{-3.2}$&-0.16$^{+0.04}_{-0.04}$&1207.3$^{+378.4}_{-374.2}$&0.04$^{+0.02}_{-0.02}$&-48.9\\
25.16$\sim$26.98&27.72&-0.90$^{+0.06}_{-0.06}$&541.3$^{+103.6}_{-102.2}$&-0.90$^{+0.06}_{-0.07}$&588.5$^{+133.6}_{-136.9}$&14.6$^{+11.2}_{-11.6}$&0.00$^{+0.08}_{-0.09}$&47.2$^{+169.1}_{-170.9}$&0.01$^{+0.56}_{-0.01}$&-2108.9\\
26.98$\sim$32.17&34.57&-1.04$^{+0.04}_{-0.04}$&604.6$^{+109.5}_{-110.9}$&-1.24$^{+0.04}_{-0.05}$&4271.4$^{+1687.0}_{-1711.8}$&29.4$^{+3.5}_{-3.4}$&-0.20$^{+0.06}_{-0.06}$&3666.8$^{+1690.5}_{-1715.4}$&0.04$^{+0.03}_{-0.02}$&-22.3\\
32.17$\sim$38.35&27.64&-1.07$^{+0.06}_{-0.06}$&465.8$^{+100.7}_{-103.1}$&-1.28$^{+0.06}_{-0.07}$&3000.1$^{+1671.9}_{-1673.7}$&26.7$^{+2.9}_{-2.9}$&-0.21$^{+0.08}_{-0.09}$&2534.3$^{+1674.9}_{-1676.9}$&0.06$^{+0.06}_{-0.04}$&-27.1\\
38.35$\sim$40.27&41.93&-0.89$^{+0.04}_{-0.04}$&493.3$^{+64.8}_{-64.4}$&-0.95$^{+0.06}_{-0.06}$&846.0$^{+197.3}_{-208.7}$&25.3$^{+3.2}_{-3.2}$&-0.06$^{+0.07}_{-0.07}$&352.7$^{+207.7}_{-218.4}$&0.05$^{+0.05}_{-0.03}$&-29.8\\
40.27$\sim$41.57&27.68&-0.86$^{+0.06}_{-0.06}$&381.1$^{+68.2}_{-68.8}$&-1.05$^{+0.10}_{-0.10}$&1077.9$^{+462.1}_{-480.7}$&31.3$^{+4.5}_{-4.4}$&-0.19$^{+0.12}_{-0.12}$&696.8$^{+467.1}_{-485.6}$&0.08$^{+0.09}_{-0.05}$&-22.5\\
41.57$\sim$45.13&32.65&-0.89$^{+0.05}_{-0.05}$&272.1$^{+37.1}_{-37.3}$&-0.95$^{+0.08}_{-0.09}$&429.8$^{+104.0}_{-106.7}$&19.9$^{+3.0}_{-3.1}$&-0.06$^{+0.09}_{-0.10}$&157.8$^{+110.4}_{-113.1}$&0.06$^{+0.09}_{-0.04}$&-26.3\\
45.13$\sim$48.00&43.13&-0.90$^{+0.04}_{-0.04}$&424.0$^{+50.7}_{-49.7}$&-0.95$^{+0.07}_{-0.07}$&563.3$^{+159.8}_{-148.3}$&26.2$^{+9.9}_{-14.1}$&-0.05$^{+0.08}_{-0.08}$&139.3$^{+167.6}_{-156.4}$&0.03$^{+0.10}_{-0.03}$&-56.2\\
48.00$\sim$54.09&37.12&-0.98$^{+0.05}_{-0.05}$&340.7$^{+49.3}_{-49.4}$&-1.24$^{+0.08}_{-0.08}$&1433.2$^{+816.5}_{-741.0}$&29.9$^{+2.8}_{-2.8}$&-0.26$^{+0.09}_{-0.09}$&1092.5$^{+818.0}_{-742.7}$&0.09$^{+0.08}_{-0.04}$&-27.3\\
54.09$\sim$58.33&66.51&-0.83$^{+0.02}_{-0.02}$&446.2$^{+31.8}_{-31.9}$&-0.83$^{+0.04}_{-0.04}$&562.3$^{+63.0}_{-63.0}$&23.6$^{+2.6}_{-2.5}$&0.00$^{+0.04}_{-0.04}$&116.1$^{+70.6}_{-70.6}$&0.04$^{+0.03}_{-0.02}$&-40.6\\
58.33$\sim$59.48&48.64&-0.80$^{+0.04}_{-0.04}$&457.7$^{+45.1}_{-45.4}$&-0.85$^{+0.05}_{-0.05}$&597.6$^{+103.5}_{-106.4}$&29.3$^{+5.5}_{-5.2}$&-0.05$^{+0.06}_{-0.06}$&139.9$^{+112.8}_{-115.7}$&0.04$^{+0.05}_{-0.03}$&-18.2\\
59.48$\sim$60.99&63.64&-0.79$^{+0.03}_{-0.03}$&386.6$^{+30.1}_{-30.1}$&-0.84$^{+0.06}_{-0.06}$&523.0$^{+84.1}_{-84.8}$&29.1$^{+5.0}_{-5.0}$&-0.05$^{+0.07}_{-0.07}$&136.4$^{+89.3}_{-90.0}$&0.04$^{+0.06}_{-0.03}$&-25.0\\
60.99$\sim$62.17&46.49&-0.74$^{+0.04}_{-0.04}$&371.4$^{+37.1}_{-37.0}$&-0.81$^{+0.06}_{-0.06}$&631.7$^{+114.6}_{-116.5}$&27.9$^{+2.4}_{-2.4}$&-0.07$^{+0.07}_{-0.07}$&260.3$^{+120.5}_{-122.2}$&0.07$^{+0.05}_{-0.03}$&-43.4\\
62.17$\sim$63.72&70.70&-0.82$^{+0.03}_{-0.03}$&608.0$^{+51.1}_{-50.6}$&-0.92$^{+0.04}_{-0.04}$&1239.9$^{+247.0}_{-250.8}$&32.2$^{+3.1}_{-3.1}$&-0.10$^{+0.05}_{-0.05}$&631.9$^{+252.2}_{-255.9}$&0.04$^{+0.03}_{-0.02}$&-51.9\\
63.72$\sim$70.33&107.98&-0.81$^{+0.02}_{-0.02}$&444.6$^{+19.3}_{-19.5}$&-0.84$^{+0.02}_{-0.02}$&555.0$^{+41.0}_{-41.1}$&26.4$^{+2.5}_{-2.4}$&-0.03$^{+0.03}_{-0.03}$&110.3$^{+45.3}_{-45.5}$&0.03$^{+0.02}_{-0.01}$&-53.1\\
70.33$\sim$73.24&52.97&-0.82$^{+0.03}_{-0.03}$&369.6$^{+32.2}_{-32.4}$&-0.96$^{+0.06}_{-0.06}$&758.0$^{+162.6}_{-165.2}$&29.0$^{+2.7}_{-2.6}$&-0.14$^{+0.07}_{-0.07}$&388.3$^{+165.8}_{-168.3}$&0.07$^{+0.05}_{-0.03}$&-45.2\\
73.24$\sim$74.33&44.68&-0.76$^{+0.04}_{-0.04}$&457.1$^{+45.8}_{-46.5}$&-0.74$^{+0.06}_{-0.06}$&509.0$^{+71.4}_{-71.4}$&22.8$^{+5.6}_{-5.0}$&0.02$^{+0.07}_{-0.07}$&51.9$^{+84.8}_{-85.2}$&0.02$^{+0.06}_{-0.02}$&-19.1\\
73.24$\sim$74.33&44.68&-0.76$^{+0.04}_{-0.04}$&454.7$^{+44.6}_{-44.7}$&-0.74$^{+0.06}_{-0.06}$&509.0$^{+71.4}_{-71.4}$&22.8$^{+5.6}_{-5.0}$&0.02$^{+0.07}_{-0.07}$&54.3$^{+84.2}_{-84.2}$&0.02$^{+0.06}_{-0.02}$&-18.9\\
74.33$\sim$76.18&72.29&-0.77$^{+0.02}_{-0.02}$&462.5$^{+30.5}_{-31.2}$&-0.80$^{+0.04}_{-0.04}$&649.7$^{+77.9}_{-77.2}$&27.7$^{+2.4}_{-2.4}$&-0.03$^{+0.04}_{-0.04}$&187.2$^{+83.7}_{-83.3}$&0.05$^{+0.03}_{-0.02}$&-54.7\\
76.18$\sim$78.39&54.59&-0.87$^{+0.03}_{-0.03}$&404.1$^{+37.4}_{-37.8}$&-0.87$^{+0.05}_{-0.05}$&484.5$^{+64.9}_{-67.2}$&21.4$^{+3.6}_{-3.4}$&0.00$^{+0.06}_{-0.06}$&80.4$^{+74.9}_{-77.1}$&0.03$^{+0.05}_{-0.02}$&-21.6\\
78.39$\sim$80.02&31.16&-0.91$^{+0.06}_{-0.06}$&343.5$^{+57.4}_{-58.9}$&-1.16$^{+0.13}_{-0.12}$&1322.5$^{+1084.5}_{-791.9}$&32.9$^{+5.2}_{-4.6}$&-0.25$^{+0.14}_{-0.13}$&979.0$^{+1086.0}_{-794.1}$&0.09$^{+0.14}_{-0.07}$&-24.8\\
80.02$\sim$87.06&27.17&-1.04$^{+0.06}_{-0.06}$&456.6$^{+113.1}_{-113.6}$&-1.32$^{+0.06}_{-0.06}$&5473.5$^{+3753.6}_{-3310.6}$&31.2$^{+3.3}_{-3.4}$&-0.28$^{+0.08}_{-0.08}$&5016.9$^{+3755.3}_{-3312.5}$&0.06$^{+0.05}_{-0.03}$&-22.5\\
87.06$\sim$111.46&20.87&-1.03$^{+0.08}_{-0.08}$&406.9$^{+116.0}_{-115.2}$&-1.29$^{+0.07}_{-0.06}$&4054.7$^{+2267.9}_{-2249.9}$&26.5$^{+2.5}_{-2.5}$&-0.26$^{+0.11}_{-0.10}$&3647.8$^{+2270.8}_{-2252.9}$&0.07$^{+0.06}_{-0.04}$&-26.9\\
117.17$\sim$121.70&22.60&-1.03$^{+0.10}_{-0.10}$&153.3$^{+29.7}_{-30.2}$&-0.87$^{+0.33}_{-0.31}$&56.0$^{+36.3}_{-31.7}$&41.6$^{+4.9}_{-4.7}$&0.16$^{+0.34}_{-0.33}$&-97.3$^{+46.9}_{-43.8}$&0.44$^{+0.62}_{-0.33}$&-78.5\\
121.70$\sim$123.12&21.20&-1.01$^{+0.10}_{-0.10}$&190.8$^{+43.0}_{-43.8}$&-1.20$^{+0.24}_{-0.23}$&418.2$^{+288.3}_{-246.4}$&20.7$^{+8.3}_{-14.3}$&-0.19$^{+0.26}_{-0.25}$&227.4$^{+291.4}_{-250.3}$&0.10$^{+0.60}_{-0.11}$&-2049.3\\
126.92$\sim$130.80&41.82&-0.91$^{+0.06}_{-0.06}$&142.0$^{+15.6}_{-15.9}$&-0.91$^{+0.07}_{-0.07}$&136.8$^{+19.6}_{-18.4}$&81.1$^{+64.7}_{-53.3}$&0.00$^{+0.09}_{-0.09}$&-5.2$^{+25.1}_{-24.3}$&0.02$^{+1.38}_{-0.02}$&-382.2\\
130.80$\sim$135.52&23.19&-1.32$^{+0.12}_{-0.13}$&217.3$^{+75.6}_{-81.0}$&-1.62$^{+0.12}_{-0.13}$&1693.0$^{+1484.9}_{-1269.8}$&16.3$^{+3.8}_{-3.1}$&-0.30$^{+0.17}_{-0.18}$&1475.6$^{+1486.8}_{-1272.3}$&0.08$^{+0.16}_{-0.07}$&-12.0\\
\enddata
\end{deluxetable}\label{Table:100724B}

\clearpage 
\begin{deluxetable}{cc|cc|ccc|cccc}
\tablewidth{0pt}
\setlength{\tabcolsep}{0.35em}
\tabletypesize{\scriptsize}
\tablecaption{Results of the Time-resolved Spectral Fits of GRB 110721A}
\tablehead{
\multirow{3}{0.8cm}{$t_{1}$$\sim$$t_{2}$} 
&\multirow{3}{0.8cm}{$S$} 
&\multicolumn{2}{|c}{Cutoff Power-law Fitting}
&\multicolumn{3}{|c}{Cutoff Power-law Plus Blackbody Fitting}
&\multicolumn{3}{|c}{Difference}\\
\cline{3-11}
&
&\colhead{$\alpha$}
&\colhead{$E_{\rm c}$}
&\colhead{$\alpha$}
&\colhead{$E_{\rm c}$}
&\colhead{$kT$}
&\colhead{$\Delta \alpha^{\rm I}$}
&\colhead{$\Delta E_{\rm c}$}
&\colhead{Ratio$^{\rm I}$}
&\colhead{$\Delta$DIC$^{\rm I}$}\\
(s)
&&&(keV)
&&&(keV)
&
&
&
&
}
\startdata
0.00$\sim$0.47&33.73&-0.95$^{+0.02}_{-0.02}$&6669.0$^{+761.9}_{-753.4}$&-0.98$^{+0.03}_{-0.03}$&8054.5$^{+1259.2}_{-1271.7}$&128.8$^{+43.3}_{-39.5}$&-0.03$^{+0.04}_{-0.04}$&1385.6$^{+1471.7}_{-1478.1}$&0.02$^{+0.09}_{-0.02}$&-24.2\\
0.47$\sim$1.96&77.74&-0.94$^{+0.01}_{-0.01}$&1864.5$^{+127.5}_{-127.2}$&-0.94$^{+0.02}_{-0.02}$&2397.4$^{+186.0}_{-182.1}$&39.5$^{+3.0}_{-3.0}$&0.00$^{+0.02}_{-0.02}$&532.9$^{+225.5}_{-222.1}$&0.03$^{+0.01}_{-0.01}$&-94.8\\
1.96$\sim$2.97&74.04&-0.90$^{+0.03}_{-0.03}$&397.9$^{+31.4}_{-31.4}$&-1.08$^{+0.04}_{-0.04}$&958.0$^{+171.4}_{-177.4}$&31.4$^{+2.3}_{-2.3}$&-0.18$^{+0.05}_{-0.05}$&560.1$^{+174.3}_{-180.1}$&0.08$^{+0.04}_{-0.03}$&-62.0\\
2.97$\sim$3.70&49.77&-1.13$^{+0.04}_{-0.04}$&479.3$^{+66.5}_{-67.7}$&-1.22$^{+0.06}_{-0.06}$&899.6$^{+246.5}_{-265.7}$&23.5$^{+4.0}_{-4.2}$&-0.09$^{+0.07}_{-0.07}$&420.3$^{+255.3}_{-274.1}$&0.05$^{+0.08}_{-0.03}$&-28.1\\
3.70$\sim$4.90&45.72&-1.28$^{+0.03}_{-0.03}$&817.1$^{+159.3}_{-164.4}$&-1.29$^{+0.05}_{-0.05}$&1060.3$^{+287.6}_{-300.2}$&17.8$^{+5.0}_{-5.3}$&-0.01$^{+0.06}_{-0.06}$&243.2$^{+328.8}_{-342.3}$&0.02$^{+0.08}_{-0.02}$&-31.7\\
4.90$\sim$6.25&36.04&-1.28$^{+0.04}_{-0.04}$&1015.4$^{+275.3}_{-288.4}$&-1.33$^{+0.09}_{-0.11}$&3271.4$^{+1297.8}_{-2408.6}$&25.6$^{+16.8}_{-14.6}$&-0.05$^{+0.10}_{-0.12}$&2256.0$^{+1326.7}_{-2425.8}$&0.01$^{+0.21}_{-0.02}$&-231.3\\
6.25$\sim$9.03&33.51&-1.23$^{+0.05}_{-0.05}$&603.5$^{+134.1}_{-136.1}$&-1.46$^{+0.09}_{-0.09}$&10532.3$^{+3487.8}_{-9275.5}$&34.6$^{+4.6}_{-4.3}$&-0.23$^{+0.10}_{-0.10}$&9928.8$^{+3490.4}_{-9276.5}$&0.06$^{+0.07}_{-0.04}$&-27.4\\
9.03$\sim$13.87&22.49&-1.21$^{+0.06}_{-0.07}$&609.7$^{+181.9}_{-189.1}$&-1.35$^{+0.14}_{-0.17}$&6061.0$^{+780.8}_{-5518.4}$&29.7$^{+12.2}_{-15.5}$&-0.14$^{+0.15}_{-0.18}$&5451.3$^{+801.7}_{-5521.6}$&0.04$^{+0.22}_{-0.05}$&-125.5\\
\enddata
\end{deluxetable}\label{Table:110721A}

\clearpage 
\begin{deluxetable}{cc|cc|ccc|cccc}
\tablewidth{0pt}
\setlength{\tabcolsep}{0.35em}
\tabletypesize{\scriptsize}
\tablecaption{Results of the Time-resolved Spectral Fits of GRB 190114C}
\tablehead{
\multirow{3}{0.8cm}{$t_{1}$$\sim$$t_{2}$} 
&\multirow{3}{0.8cm}{$S$} 
&\multicolumn{2}{|c}{Cutoff Power-law Fitting}
&\multicolumn{3}{|c}{Cutoff Power-law Plus Blackbody Fitting}
&\multicolumn{3}{|c}{Difference}\\
\cline{3-11}
&
&\colhead{$\alpha$}
&\colhead{$E_{\rm c}$}
&\colhead{$\alpha$}
&\colhead{$E_{\rm c}$}
&\colhead{$kT$}
&\colhead{$\Delta \alpha^{\rm I}$}
&\colhead{$\Delta E_{\rm c}$}
&\colhead{Ratio$^{\rm I}$}
&\colhead{$\Delta$DIC$^{\rm I}$}\\
(s)
&&&(keV)
&&&(keV)
&
&
&
&
}
\startdata
0.70$\sim$1.58&240.02&-0.61$^{+0.01}_{-0.01}$&383.8$^{+8.1}_{-8.2}$&-0.52$^{+0.03}_{-0.03}$&236.2$^{+28.7}_{-28.9}$&230.3$^{+18.6}_{-19.5}$&0.09$^{+0.03}_{-0.03}$&-147.5$^{+29.8}_{-30.0}$&0.35$^{+0.81}_{-0.24}$&-44.3\\
1.58$\sim$1.71&111.95&-0.44$^{+0.02}_{-0.02}$&478.6$^{+19.8}_{-19.7}$&-0.66$^{+0.06}_{-0.05}$&621.1$^{+66.5}_{-66.8}$&155.8$^{+10.2}_{-10.5}$&-0.22$^{+0.06}_{-0.05}$&142.5$^{+69.4}_{-69.7}$&0.28$^{+0.14}_{-0.10}$&-367.7\\
2.64$\sim$2.88&125.08&-0.49$^{+0.02}_{-0.02}$&515.0$^{+18.8}_{-19.0}$&-0.58$^{+0.05}_{-0.05}$&460.3$^{+84.1}_{-88.2}$&197.4$^{+16.5}_{-16.4}$&-0.09$^{+0.05}_{-0.05}$&-54.7$^{+86.2}_{-90.2}$&0.35$^{+0.18}_{-0.14}$&-622.0\\
2.88$\sim$3.09&95.22&-0.53$^{+0.02}_{-0.02}$&632.4$^{+30.9}_{-30.6}$&-0.65$^{+0.04}_{-0.04}$&656.8$^{+79.0}_{-73.7}$&187.6$^{+21.5}_{-21.9}$&-0.12$^{+0.04}_{-0.04}$&24.4$^{+84.8}_{-79.8}$&0.23$^{+0.52}_{-0.13}$&-12.4\\
3.09$\sim$3.21&87.32&-0.43$^{+0.02}_{-0.03}$&784.8$^{+44.0}_{-43.8}$&-0.73$^{+0.05}_{-0.06}$&1474.5$^{+205.2}_{-192.5}$&162.1$^{+12.2}_{-7.6}$&-0.30$^{+0.05}_{-0.07}$&689.8$^{+209.9}_{-197.4}$&0.25$^{+0.5}_{-0.18}$&-19.7\\
3.21$\sim$3.60&137.87&-0.36$^{+0.02}_{-0.02}$&605.2$^{+18.4}_{-18.7}$&-0.62$^{+0.04}_{-0.04}$&939.7$^{+93.4}_{-93.5}$&149.2$^{+7.4}_{-7.6}$&-0.26$^{+0.04}_{-0.04}$&334.4$^{+95.2}_{-95.4}$&0.23$^{+0.44}_{-0.21}$&-36.7\\
3.60$\sim$3.74&72.47&-0.34$^{+0.03}_{-0.03}$&593.5$^{+34.2}_{-34.0}$&-0.68$^{+0.09}_{-0.08}$&1068.1$^{+218.2}_{-232.0}$&151.0$^{+11.2}_{-11.1}$&-0.34$^{+0.09}_{-0.09}$&474.7$^{+220.9}_{-234.5}$&0.27$^{+0.61}_{-0.18}$&-20.0\\
3.74$\sim$3.96&131.51&-0.18$^{+0.02}_{-0.02}$&528.9$^{+17.2}_{-16.9}$&-0.68$^{+0.05}_{-0.05}$&1421.1$^{+174.1}_{-178.5}$&140.4$^{+4.6}_{-4.5}$&-0.50$^{+0.05}_{-0.05}$&892.2$^{+174.9}_{-179.3}$&0.32$^{+0.57}_{-0.29}$&-62.7\\
4.10$\sim$4.44&159.08&-0.41$^{+0.02}_{-0.02}$&347.1$^{+10.4}_{-10.4}$&-0.60$^{+0.04}_{-0.04}$&421.3$^{+29.5}_{-29.6}$&114.8$^{+7.0}_{-7.2}$&-0.19$^{+0.04}_{-0.04}$&74.2$^{+31.3}_{-31.4}$&0.20$^{+0.09}_{-0.07}$&-921.9\\
4.44$\sim$4.51&60.65&-0.67$^{+0.04}_{-0.04}$&322.4$^{+28.9}_{-28.8}$&-0.92$^{+0.11}_{-0.11}$&445.3$^{+98.1}_{-90.8}$&96.6$^{+14.8}_{-13.2}$&-0.25$^{+0.12}_{-0.12}$&122.9$^{+102.3}_{-95.3}$&0.22$^{+0.30}_{-0.15}$&-207.4\\
4.51$\sim$4.77&132.57&-0.64$^{+0.02}_{-0.02}$&486.0$^{+18.9}_{-19.0}$&-1.02$^{+0.04}_{-0.04}$&751.0$^{+101.7}_{-104.0}$&111.1$^{+5.1}_{-5.0}$&-0.38$^{+0.04}_{-0.04}$&265.0$^{+103.4}_{-105.7}$&0.41$^{+0.73}_{-0.35}$&-132.9\\
4.77$\sim$4.95&125.55&-0.47$^{+0.02}_{-0.02}$&352.6$^{+13.5}_{-13.5}$&-0.85$^{+0.05}_{-0.05}$&582.4$^{+66.1}_{-69.9}$&89.9$^{+3.9}_{-4.0}$&-0.38$^{+0.05}_{-0.05}$&229.9$^{+67.5}_{-71.2}$&0.33$^{+0.61}_{-0.28}$&-53.9\\
4.95$\sim$5.45&169.35&-0.61$^{+0.01}_{-0.01}$&412.9$^{+11.9}_{-11.8}$&-1.01$^{+0.03}_{-0.03}$&735.0$^{+70.4}_{-71.5}$&91.0$^{+2.7}_{-2.7}$&-0.40$^{+0.03}_{-0.03}$&322.1$^{+71.4}_{-72.5}$&0.35$^{+0.59}_{-0.33}$&-176.2\\
5.45$\sim$5.51&74.54&-0.43$^{+0.04}_{-0.04}$&314.7$^{+20.8}_{-20.9}$&-0.89$^{+0.09}_{-0.16}$&660.2$^{+193.7}_{-174.4}$&81.5$^{+5.4}_{-4.9}$&-0.46$^{+0.10}_{-0.16}$&345.5$^{+194.8}_{-175.7}$&0.40$^{+1.03}_{-0.28}$&-92.9\\
5.51$\sim$5.69&101.53&-0.49$^{+0.03}_{-0.03}$&181.4$^{+8.6}_{-8.5}$&-1.12$^{+0.08}_{-0.09}$&577.4$^{+148.8}_{-148.9}$&47.9$^{+1.4}_{-1.5}$&-0.63$^{+0.09}_{-0.09}$&396.0$^{+149.0}_{-149.1}$&0.34$^{+0.65}_{-0.32}$&-47.1\\
\enddata
\end{deluxetable}\label{Table:190114C}

\clearpage 
\begin{deluxetable}{cc|cc|cc|ccc}
\tablewidth{0pt}
\setlength{\tabcolsep}{0.35em}
\tabletypesize{\scriptsize}
\tablecaption{Results of the Time-resolved Spectral Fits of GRB 090902B}
\tablehead{
\multirow{3}{0.8cm}{$t_{1}$$\sim$$t_{2}$} 
&\multirow{3}{0.8cm}{$S$} 
&\multicolumn{2}{|c}{Cutoff Power-law Fitting}
&\multicolumn{2}{|c}{Power-law Plus Blackbody Fitting}
&\multicolumn{2}{|c}{Difference}\\
\cline{3-9}
&
&\colhead{$\alpha$}
&\colhead{$E_{\rm c}$}
&\colhead{Power-law Index}
&\colhead{$kT$}
&\colhead{$\Delta \alpha^{\rm II}$}
&\colhead{Ratio$^{\rm II}$}
&\colhead{$\Delta$DIC$^{\rm II}$}\\
(s)
&&&
(keV)
&&
(keV)
&
&
&
}
\startdata
5.94$\sim$6.04&26.05&0.05$^{+0.12}_{-0.12}$&377.8$^{+51.0}_{-51.8}$&-1.54$^{+0.12}_{-0.11}$&146.0$^{+6.8}_{-6.7}$&-1.59$^{+0.17}_{-0.16}$&0.81$^{+0.32}_{-0.24}$&-2.4\\
6.04$\sim$6.70&44.89&-0.16$^{+0.06}_{-0.06}$&499.4$^{+39.6}_{-39.8}$&-1.53$^{+0.04}_{-0.04}$&161.6$^{+4.4}_{-4.4}$&-1.37$^{+0.07}_{-0.07}$&0.76$^{+0.14}_{-0.13}$&-1.9\\
6.70$\sim$7.24&51.25&-0.59$^{+0.03}_{-0.03}$&1047.7$^{+82.9}_{-83.5}$&-1.59$^{+0.04}_{-0.04}$&187.3$^{+5.5}_{-5.5}$&-1.00$^{+0.05}_{-0.05}$&0.71$^{+0.14}_{-0.13}$&-114.7\\
7.24$\sim$7.75&59.81&-0.72$^{+0.02}_{-0.02}$&1279.4$^{+90.5}_{-90.7}$&-1.62$^{+0.03}_{-0.03}$&199.3$^{+6.1}_{-6.0}$&-0.90$^{+0.04}_{-0.04}$&0.71$^{+0.15}_{-0.13}$&-185.9\\
7.75$\sim$8.04&59.62&-0.98$^{+0.02}_{-0.02}$&2429.5$^{+214.5}_{-218.4}$&-1.65$^{+0.03}_{-0.03}$&253.4$^{+10.6}_{-10.5}$&-0.67$^{+0.04}_{-0.04}$&0.67$^{+0.20}_{-0.15}$&-222.7\\
8.04$\sim$9.23&133.37&-1.14$^{+0.01}_{-0.01}$&2736.4$^{+137.7}_{-138.0}$&-1.74$^{+0.01}_{-0.01}$&228.6$^{+4.5}_{-4.6}$&-0.60$^{+0.01}_{-0.01}$&0.65$^{+0.08}_{-0.08}$&-1306.4\\
9.23$\sim$9.74&98.08&-1.25$^{+0.01}_{-0.01}$&3163.4$^{+280.8}_{-281.2}$&-1.76$^{+0.02}_{-0.02}$&180.0$^{+5.4}_{-5.4}$&-0.51$^{+0.02}_{-0.02}$&0.54$^{+0.10}_{-0.09}$&-447.5\\
9.74$\sim$9.81&51.14&-1.09$^{+0.03}_{-0.03}$&1635.5$^{+256.2}_{-255.6}$&-1.82$^{+0.06}_{-0.05}$&149.9$^{+7.0}_{-6.9}$&-0.73$^{+0.07}_{-0.06}$&0.62$^{+0.21}_{-0.15}$&-132.0\\
9.81$\sim$10.07&76.08&-1.16$^{+0.02}_{-0.02}$&1727.8$^{+177.4}_{-177.1}$&-1.78$^{+0.03}_{-0.03}$&144.3$^{+4.9}_{-4.9}$&-0.62$^{+0.04}_{-0.04}$&0.55$^{+0.12}_{-0.11}$&-218.0\\
10.07$\sim$10.92&107.41&-1.15$^{+0.01}_{-0.01}$&2554.4$^{+160.6}_{-160.6}$&-1.78$^{+0.02}_{-0.02}$&201.0$^{+4.8}_{-4.8}$&-0.63$^{+0.02}_{-0.02}$&0.65$^{+0.11}_{-0.09}$&-821.7\\
10.92$\sim$11.09&61.85&-0.98$^{+0.02}_{-0.02}$&1516.1$^{+155.1}_{-157.2}$&-1.70$^{+0.04}_{-0.04}$&169.7$^{+6.2}_{-6.3}$&-0.72$^{+0.04}_{-0.04}$&0.61$^{+0.17}_{-0.13}$&-160.2\\
11.09$\sim$11.42&67.86&-1.12$^{+0.02}_{-0.02}$&1904.0$^{+196.6}_{-196.6}$&-1.80$^{+0.04}_{-0.04}$&161.5$^{+5.7}_{-5.7}$&-0.68$^{+0.04}_{-0.04}$&0.62$^{+0.15}_{-0.13}$&-246.9\\
11.42$\sim$12.11&81.20&-1.10$^{+0.02}_{-0.02}$&1770.6$^{+152.2}_{-151.7}$&-1.75$^{+0.03}_{-0.03}$&151.9$^{+4.1}_{-4.2}$&-0.65$^{+0.04}_{-0.04}$&0.57$^{+0.10}_{-0.09}$&-323.3\\
12.11$\sim$12.22&22.09&-1.50$^{+0.12}_{-0.12}$&467.0$^{+245.4}_{-224.1}$&-1.89$^{+0.07}_{-0.07}$&27.3$^{+3.7}_{-3.7}$&-0.39$^{+0.14}_{-0.14}$&0.13$^{+0.15}_{-0.07}$&-11.0\\
12.51$\sim$12.99&51.63&-1.17$^{+0.07}_{-0.07}$&135.9$^{+17.5}_{-17.4}$&-1.85$^{+0.03}_{-0.03}$&20.2$^{+1.0}_{-1.0}$&-0.68$^{+0.08}_{-0.08}$&0.14$^{+0.05}_{-0.04}$&-14.9\\
15.21$\sim$15.69&76.43&-0.64$^{+0.03}_{-0.03}$&840.0$^{+53.6}_{-53.8}$&-1.57$^{+0.02}_{-0.02}$&156.4$^{+3.5}_{-3.5}$&-0.93$^{+0.04}_{-0.04}$&0.64$^{+0.10}_{-0.09}$&-6.5\\
21.00$\sim$21.16&35.59&-0.99$^{+0.08}_{-0.08}$&214.0$^{+36.2}_{-37.2}$&-1.57$^{+0.03}_{-0.03}$&24.8$^{+1.8}_{-1.8}$&-0.58$^{+0.09}_{-0.09}$&0.11$^{+0.05}_{-0.04}$&-1.2\\
23.32$\sim$24.04&23.29&-1.56$^{+0.13}_{-0.14}$&166.6$^{+51.8}_{-59.2}$&-2.08$^{+0.07}_{-0.07}$&14.7$^{+2.3}_{-2.4}$&-0.52$^{+0.15}_{-0.16}$&0.10$^{+0.16}_{-0.06}$&-1.2\\
24.04$\sim$24.24&30.08&-0.86$^{+0.09}_{-0.09}$&278.9$^{+54.2}_{-55.0}$&-1.59$^{+0.05}_{-0.05}$&44.4$^{+3.1}_{-3.1}$&-0.73$^{+0.10}_{-0.10}$&0.20$^{+0.10}_{-0.07}$&-14.6\\
\enddata
\end{deluxetable}\label{Table:090902B}

\clearpage 
\begin{deluxetable}{cc|cc|cc|ccc}
\tablewidth{0pt}
\setlength{\tabcolsep}{0.35em}
\tabletypesize{\scriptsize}
\tablecaption{Results of the Time-resolved Spectral Fits of GRB 160107}
\tablehead{
\multirow{3}{0.8cm}{$t_{1}$$\sim$$t_{2}$} 
&\multirow{3}{0.8cm}{$S$} 
&\multicolumn{2}{|c}{Cutoff Power-law Fitting}
&\multicolumn{2}{|c}{Power-law Plus Blackbody Fitting}
&\multicolumn{2}{|c}{Difference}\\
\cline{3-9}
&
&\colhead{$\alpha$}
&\colhead{$E_{\rm c}$}
&\colhead{Power-law Index}
&\colhead{$kT$}
&\colhead{$\Delta \alpha^{\rm II}$}
&\colhead{Ratio$^{\rm II}$}
&\colhead{$\Delta$DIC$^{\rm II}$}\\
(s)
&&&
(keV)
&&(keV)
&
&
&
}
\startdata
0.10$\sim$4.73&38.19&-1.40$^{+0.05}_{-0.05}$&322.5$^{+62.4}_{-64.2}$&-1.77$^{+0.03}_{-0.03}$&24.0$^{+2.2}_{-2.3}$&-0.37$^{+0.06}_{-0.06}$&0.07$^{+0.05}_{-0.03}$&-3.3\\
4.73$\sim$18.08&45.68&-1.62$^{+0.05}_{-0.05}$&283.3$^{+55.0}_{-54.9}$&-1.96$^{+0.03}_{-0.03}$&20.3$^{+1.8}_{-1.8}$&-0.34$^{+0.06}_{-0.06}$&0.07$^{+0.04}_{-0.03}$&-4.9\\
18.09$\sim$27.82&29.50&-1.72$^{+0.08}_{-0.08}$&492.5$^{+238.8}_{-228.6}$&-1.89$^{+0.04}_{-0.05}$&12.2$^{+4.0}_{-3.4}$&-0.17$^{+0.09}_{-0.09}$&0.03$^{+0.14}_{-0.03}$&-42.0\\
18.09$\sim$27.82&29.50&-1.72$^{+0.08}_{-0.08}$&492.5$^{+238.8}_{-228.6}$&-1.88$^{+0.04}_{-0.05}$&11.7$^{+3.5}_{-3.4}$&-0.16$^{+0.09}_{-0.09}$&0.03$^{+0.15}_{-0.02}$&-43.6\\
28.28$\sim$28.75&21.62&-1.30$^{+0.11}_{-0.11}$&295.6$^{+103.5}_{-109.4}$&-1.72$^{+0.06}_{-0.06}$&23.0$^{+4.4}_{-4.5}$&-0.42$^{+0.13}_{-0.13}$&0.07$^{+0.13}_{-0.05}$&-6.3\\
28.75$\sim$31.47&27.10&-1.68$^{+0.08}_{-0.08}$&532.4$^{+247.7}_{-238.4}$&-1.98$^{+0.06}_{-0.06}$&24.1$^{+3.5}_{-3.4}$&-0.30$^{+0.10}_{-0.10}$&0.07$^{+0.09}_{-0.04}$&-5.2\\
33.31$\sim$33.85&20.08&-1.06$^{+0.14}_{-0.14}$&181.6$^{+53.5}_{-57.8}$&-1.70$^{+0.06}_{-0.06}$&23.4$^{+2.7}_{-2.7}$&-0.64$^{+0.15}_{-0.15}$&0.12$^{+0.10}_{-0.06}$&-2.4\\
74.40$\sim$75.08&24.50&-1.68$^{+0.08}_{-0.08}$&741.1$^{+444.2}_{-377.7}$&-1.86$^{+0.06}_{-0.07}$&17.8$^{+11.6}_{-12.7}$&-0.18$^{+0.10}_{-0.11}$&0.03$^{+0.65}_{-0.03}$&-1722.1\\
75.08$\sim$77.13&27.40&-1.71$^{+0.09}_{-0.10}$&223.9$^{+72.2}_{-76.7}$&-2.10$^{+0.06}_{-0.05}$&15.9$^{+2.2}_{-2.2}$&-0.39$^{+0.11}_{-0.11}$&0.08$^{+0.11}_{-0.04}$&-7.9\\
\enddata
\end{deluxetable}\label{Table:160107}

\end{document}